\documentclass[smallextended]{svjour3}       % onecolumn (second format)
\smartqed  % flush right qed marks, e.g. at end of proof

\usepackage[nocompress]{cite}

\usepackage{graphicx}
\usepackage{algpseudocode}
\usepackage{multirow}
\usepackage{wrapfig}
\usepackage{float} 
\usepackage{color}

\usepackage{amssymb,amsfonts,amsmath,subfigure,url}
\usepackage{wrapfig}
\usepackage{enumitem}
\usepackage{times}
\usepackage{eurosym}
\usepackage{pifont}
\usepackage{array,multirow,adjustbox}

\usepackage{longtable}\usepackage[lined,algoruled,commentsnumbered,linesnumbered,longend,noresetcount]{algorithm2e}
\usepackage{textcomp}

\usepackage{rotating}

\usepackage{color}
\usepackage{pdf14}
\usepackage{hyperref}
\usepackage{ulem}

\definecolor{darkgreen}{rgb}{0.3,.48,0.225}

\newcommand{\RED}[1]{\textcolor{red}{#1}}%
\newcommand{\CHANGED}[1]{\textcolor{black}{#1}}%
\newcommand{\emphx}[1]{\textit{#1}}

\newcommand{\CH}{P1}
\newcommand{\SGM}{P2}
\newcommand{\DM}{P3}
\newcommand{\GAC}{P4}
\newcommand{\VW}{P5}

\usepackage{marginnote}
\usepackage[backgroundcolor=white,bordercolor=blue,linecolor=blue]{todonotes}
\newcommand{\MREVISION}[2]{\textcolor{black}{#2}}

\usepackage{wrapfig}

\usepackage{url}

\usepackage{pdf14}

\begin{document}

\pagestyle{plain}

\title{Automating System Test Case Classification and Prioritization for Use Case-Driven Testing in Product Lines}

%\author{\IEEEauthorblockN{Ines Hajri, Arda Goknil, Fabrizio Pastore, Lionel C. Briand}
%\IEEEauthorblockA{
%SnT Centre for Security, Reliability and Trust, University of Luxembourg, Luxembourg\\
%\sf\{hajri, goknil, pastore, briand\}@svv.lu}

%\author{\IEEEauthorblockN{Ines Hajri$^\dagger$, Arda Goknil$^\dagger$, Fabrizio Pastore$^\dagger$, Lionel C. Briand$^\dagger$$^\ddagger$}
%\IEEEauthorblockA{
%$^\dagger$SnT Centre for Security, Reliability and Trust, University of Luxembourg, Luxembourg\\
%$^\ddagger$University of Ottawa, Canada\\
%\sf\{hajri, goknil, pastore, briand\}@svv.lu}

%}

\author{Ines Hajri         \and
        Arda Goknil \and Fabrizio Pastore \and Lionel C. Briand %etc.
}


\institute{I. Hajri, A. Goknil, F. Pastore, L. C. Briand \at
              SnT Centre, University of Luxembourg, Luxembourg \\
%              Tel.: +123-45-678910\\
%              Fax: +123-45-678910\\
              \email{ines.hajri.svv@gmail.com, ar.goknil@gmail.com, fabrizio.pastore@uni.lu, lionel.briand@uni.lu}           %  \\
           \and
           L. C. Briand \at
              University of Ottawa
}

\date{Received: date / Accepted: date}

\maketitle

\begin{abstract}
Product Line Engineering (PLE) is a crucial practice in many software development environments where software systems are complex and developed for multiple customers with varying needs. At the same time, many development processes are use case-driven and this strongly influences their requirements engineering and system testing practices. In this paper, we propose, apply, and assess an automated system test case classification and prioritization approach specifically targeting system testing in the context of use case-driven development of product families. %In our previous work, we developed Product line Use case modeling Method (PUM) to support variability modeling in Product Line (PL) use case diagrams and specifications. We also developed a use case configurator, PUMConf, which interactively receives configuration decisions from users to generate Product Specific (PS) use case models from PL use case models. Our approach is built upon PUM and PUMConf. 
Our approach provides: (i) automated support to classify, for a new product in a product family, relevant and valid system test cases associated with previous products, and (ii) automated prioritization of system test cases using multiple risk factors such as fault-proneness of requirements and requirements volatility in a product family. %distribution of failures and the number of products in which the test case failed. 
%Our tool support is an extension of PUMConf integrated with IBM Doors. 
Our evaluation was performed in the context of an industrial product family in the automotive domain. Results provide empirical evidence that we propose a practical and beneficial way to classify and prioritize system test cases for industrial product lines. % in industrial settings.

%\begin{IEEEkeywords}
%Product Line Engineering, Use Case Driven Development, Regression Testing.
%\end{IEEEkeywords}


\end{abstract}

%\IEEEpeerreviewmaketitle

% !TEX root =  Main.tex
\section{Introduction}  
\label{sec:introduction}

%FABRIZIO-25.10: removed "is becoming a", it's already a well known practice
Product Line Engineering (PLE) is a common practice in many domains such as automotive and avionics to enhance product quality, to reduce development costs, and to improve time-to-market~\cite{Pohl05}. In such domains, many development processes are use case-driven and this strongly influences their requirements engineering and system testing practices~\cite{Nebut2006, Nebut06, Wang2015a, Wang2015b}. For example, IEE S.A. (in the following ``IEE'')~\cite{IEE}, a leading supplier of embedded software and hardware systems in the automotive domain and the case study supplier in this paper, develops automotive sensing systems enhancing safety and comfort in vehicles for multiple major car manufacturers worldwide. The current development and testing practice at IEE is use case-driven, and IEE, like many other development environments, follows the common product line testing strategy referred to as \textit{opportunistic reuse of test assets}~\cite{Neto2011}. A new product line is typically started with a first product from an initial customer. Analysts elicit requirements as use case specifications and then derive system test cases from these specifications. For each subsequent customer for that product, the analysts start from the current use case specifications, and negotiate variabilities with the customer to produce new specifications. They then manually choose and prioritize, from the existing test suite of the previous product(s), test cases that can and need to be rerun to ensure existing, unmodified functionalities are still working correctly in the new product. 
%FABRIZIO-25.10: The following sentence does not goes straight to the point. TRacing commonalities is more a modelling problem, which has been solved by PUM, so I would give it for done and go directly to the problem of tetsing
%With such practice, however, it is not easy for analysts to keep track of commonalities and variabilities across products. As a result, for each new customer, analysts need to evaluate the entire use cases and the system test cases derived from them. %This form of test reuse is not performed systematically, which means that 
With this form of test reuse, there is no structured, automated method that supports the activity of classifying and prioritizing test cases. It is fully manual, error-prone and time-consuming, which leads to ad-hoc change management for use case models (use case diagrams and specifications) and system test cases in product lines. Therefore, product line test case classification and prioritization techniques, based on a dedicated use case modeling methodology, are needed to automate the reuse of system test cases in the context of use case-driven development. %and product families.

The need for supporting PLE for the purpose of test automation has already been acknowledged and many product line testing approaches have been proposed in the literature~\cite{Neto2011, do2014strategies, Engstrom2011, Lee-SPLTestingSurvey-SPLC-2012, Runeson2012}. %~\cite{johansen2011survey}~\cite{oster2011survey}~\cite{tevanlinna2004product}
Most of the existing approaches follow the product line testing strategy \textit{design test assets for reuse}~\cite{Neto2011} in which test assets, e.g., abstract test cases or behavioral models, are created in advance for the entire product family, including common and reusable parts. When a new product is developed, test assets are selected to be reused, extended, and refined into product-specific test cases. Due to deadline pressures and limited resources, many companies, including IEE, find the upfront creation of test assets to be impractical because of the large amount of manual effort required before there are (enough) customers to justify it.

\MREVISION{R3.4}{Lity et al.~\cite{Lity2016, Lochau2014, Lity2012} propose a test case selection approach which follows an alternative product line testing strategy, i.e., \textit{incremental testing of product lines}~\cite{Neto2011}. In this strategy, the initial product is tested individually and the following products are tested using regression testing techniques, i.e., test case selection and prioritization. %Lity et al. apply incremental model slicing to determine the impact of changes on a system model, e.g., finite state machines, and to select the scenarios to be retested with the new product. 
The approach does not require the entire test suite of the product family to be generated in advance since the test cases of the new product are selected and derived incrementally from the test suites of the previous product(s). Its main limitation is the need for detailed behavioral models, e.g., finite state machines and sequence diagrams, which rarely exist in industrial practice since software development and testing are typically driven by requirements in Natural Language (NL) and behavioral models are typically specified only for a limited set of critical system features~\cite{Larman-Applying-2002}.
To evaluate the applicability of behavioral modeling in practice, we asked IEE engineers to specify \MREVISION{R3.3}{System Sequence Diagrams (SSDs)} for some of the use cases in one of their projects, at a level of detail that was appropriate for our objectives. For example, the SSD for one of the mid-size use cases included 74 messages, 19 nested blocks, and 24 references to other SSDs that had to be derived. This was considered too complex for the engineers and required significant help from the authors of this paper, and many iterations and meetings. \MREVISION{R2.4}{The main problem with sequence diagrams is the nested blocks (loops that cover alternative flows backwarding) for loops and references to other sequence diagrams. With these structures, it was not feasible to follow the execution flow visually for the engineers.} Our conclusion is that the adoption of behavioral modeling, at the level of detail required for automated test case selection, is typically not practical for system test automation unless detailed behavioral models are already used for other purposes, e.g., software design.}
Many approaches for test case classification and prioritization require the source code of the system under test together with code coverage information~\cite{Yoo2012}. %We deliberately avoid the use of source code information because 
However, this information is often partially available in industrial contexts. Indeed, when system testing is outsourced to companies or independent test teams, 
%or to departments that are located in different parts of the world, 
the source code of the system under test is often partially or not available. For example, test teams may have access only to the source code of a single product, not the entire product line. In addition, structural coverage information is often unavailable in the case of embedded systems. Indeed,
traditional compiler-based methods used to collect coverage data~\cite{Yang2006ASO} cannot be applied when test cases need to be run on dedicated hardware. These are the main motivations in this paper to rely on a requirements-driven approach to test case classification and prioritization.
%The collection of code coverage information is often infeasible because dedicated tracing tools~\cite{lauterbach} do not work with all the embedded hardware and, in addition, their license is seldom available to all the departments in the company.
%ARDA: I think the following sentence is repeating the sentence ``''
%To successfully prioritize system test cases despite the absence of coverage data and source code, we rely on the use of machine learning; more precisely, we leverage logistic regression models to capture how likely changes in requirements, size of use case scenarios and other information derived from specifications impact on the failure likelihood of each test case.}
% thus learning the domain information that has shown to be effective

In our previous work~\cite{Hajri2015}, we proposed the \MREVISION{R3.5}{Product line Use case modeling Method (PUM)}, which supports variability modeling in Product Line (PL) use case diagrams and specifications in NL, intentionally avoiding any reliance on feature models or behavioral models such as activity and sequence diagrams. %thus avoiding modeling and traceability overhead. 
%PUM adopts the existing PL extensions of use case diagrams in the work of Halmans and Pohl~\cite{Halmans2003a}. In order to model variability in use case specifications written in natural language, we introduced new product line extensions for the Restricted Use Case Modeling method (RUCM)~\cite{Yue2013}. 
PUM relies on the Restricted Use Case Modeling method (RUCM)~\cite{Yue2013}, which introduces a template with keywords and restriction rules to reduce ambiguity and to enable automated analysis of use case specifications. RUCM has been successfully applied in many domains (e.g.,~\cite{Zhang2018, Wang2015a, Hajri2015, Mai2018, Mai2018b, Hajri2017b, Mai2019, wang2019automatic}).
Based on PUM, we developed a use case-driven configuration approach~\cite{Hajri2016c, Hajri2016b} guiding engineers in making configuration decisions %(e.g., checking consistency of a decision with prior decisions)
 and automatically generating \MREVISION{R2.6, R3.6}{Product Specific (PS)} use case models. %from the PL use case models and configuration decisions. 
It is supported by a tool, \textit{PUMConf (Product line Use case Model Configurator)}, integrated with IBM DOORS.

In this paper, we propose, apply and assess an approach for the definition, selection, and prioritization of test cases in product lines, based on our use case-driven modeling and configuration techniques~\cite{Hajri2015, Hajri2016c}. Our goal is to rely, to the largest extent possible, on common practices, including the ones at IEE \MREVISION{R2.7}{(e.g., use case modeling and requirements traceability)}, to achieve widespread applicability. Our approach supports the incremental testing of new products of a product family where requirements are captured as use case specifications.
%In our context, we aim to 
Consistent with the strategy referred to as ``incremental testing of product lines'', we automate the definition of system test cases by reusing test cases that belong to existing products.
% which are impacted by changes in configuration decisions when a new product is configured.
%Our approach follows the strategy \textit{incremental testing of product lines}. 
After the initial product is tested individually, new test cases might be needed and some of the existing test
cases may need to be modified for new products, while some existing test cases are simply reused verbatim.  
The definition of test cases for new products is based on the classification and selection of existing test cases in the product line and on the identification of new, untested scenarios for new products under test. 
%We automatically identify system test cases from existing products which are impacted by changes in configuration decisions when a new product is configured. 
Test case prioritization is based on prediction models trained using product line historical data.

%Our approach supports the classification and prioritization of system test cases for new products in a product line.
%First, system test cases are derived for the new product. This is done by reusing system test cases executed in previous product(s), and by identifying use case scenarios of the new 
%System test cases for a new product are derived by reusing system test cases for previous product(s), and by identifying use case scenarios of the new product that have not been tested so far in the product family. 
To reuse the existing system test cases, our approach automatically classifies them as \textit{obsolete}, \textit{retestable}, and \textit{reusable}. An \textit{obsolete} test case cannot be executed on the new product as the corresponding use case scenarios are not selected for the new product. 
A \textit{retestable} test case is still valid but needs to be rerun to determine the possible impact of changes whereas a \textit{reusable} test case is also valid but does not need to be rerun for the new product. 
%Fabrizio-25.10: tried to clarify what is a decision
%To do so, we implemented a model differencing pipeline which identifies decision changes to be used in the classification of system test cases. 
We implemented a model differencing pipeline which identifies changes in the decisions made to configure a product (e.g., selecting a variant use case).
%There are two sets of decisions: (i) the set of previously made decisions used to generate the PS use case models for the previous product(s) and (ii) the set of decisions made to generate the PS use case models for the new product. 
%Fabrizio-25.10: using specifications  instead of 'models', I found very misleading the keyword model, I understand what you mean but we could use a higher level concept
There are two sets of decisions: (i) decisions made to generate the PS use case specifications for the previous product(s) and (ii) decisions made to generate the PS use case specifications for the new product.
Our approach compares the two sets to classify the decisions as \textit{new}, \textit{deleted} and \textit{updated}, and to identify the impacted parts of the use case models of the previous product(s). \MREVISION{R2.8}{Our approach needs traceability links between use cases and system test cases. These links can be manually assigned by engineers, or automatically generated as a side-product of the automated test case generation approaches (e.g., \cite{Nebut2006, Wang2015a, wang2019automatic}).} By using the traceability links from the impacted parts of the use case models to the system test cases, we automatically classify the existing system test cases to be reused for testing the new product. In addition, we automatically identify the use case scenarios of the new product that have not been tested before, and provide information on how to modify existing system test cases to cover these new, untested use case scenarios, i.e., the impact of use case changes on existing system test cases. Note that we do not address evolving PL use case models, which need to be treated in a separate approach. %\textcolor{red}{The test case classification our approach provides is based on the impact of requirements changes on system test cases. For the final decision of the selection of the system test cases, the analyst may also need to consider some implementation and hardware changes, e.g., code refactoring or replacing some hardware with less expensive technology, which are simply design decisions not necessarily driven by requirements changes. For instance, a reusable test case might need to be selected to be rerun because part of the source code verified by the test case is refactored. In addition, in order to ensure the safety for highly safety-critical systems, the test execution strategy the company follows may enforce all the reusable test cases to be rerun.}
%Lionel: What about new test cases? Can we provide information to help with that?
%Arda: Yes, we can provide the information about that. We updated the paragraph above.
%The classification is an early requirements analysis output indicating the impact of requirements changes on system test cases. 
%Lionel: Interesting, this is important. 
%During the development of the new product, for the final decision of the execution of the system test cases, the analyst may also need to consider implementation and hardware changes which are simply design decisions not necessarily driven by requirements changes. 
%ARDA: We added the red colored sentences above to emphasize that our approach is only based on requirements changes. For reusable and retestable test cases, the analyst may need to consider some other changes like code refactoring and hardware changes which are not mainly driven by requirements changes. This statement is important because we would like to discuss the results of the evaluation based on this fact.
%Fabrizio-25.10: already said before that this is the second, so I remove 'Second'
%Second, the system test cases are automatically prioritized based on multiple risk factors 

System test cases are automatically prioritized based on multiple risk factors 
such as fault-proneness of requirements and requirements volatility in the product line. %, and implementation complexity of requirements. %distribution of failures, the number of products in which the test case has failed, and the number of products in which the test case has been executed. 
%To this end, we compute a prioritization score for each system test case based on these factors building a prediction model based on historical data. 
To this end, we rely on prediction models; more precisely, we leverage logistic regression models that capture how likely changes in these risk factors impact the failure likelihood of each test case.
To support these activities, we extended \textit{PUMConf}. %, to automatically classify and prioritize system test cases for a new product %based on evolving configuration decisions 
%in a product family.  
We have evaluated the effectiveness of the proposed approach by applying it to classify and prioritize the test cases of five software products belonging to a product line in the automotive domain. \MREVISION{R4.3}{In our evaluation, we have answered the following research questions (RQs):}

\begin{itemize}
\item \textit{RQ1. Does the proposed approach provide correct test case classification results?} With RQ1, we have evaluated the precision and recall of the procedure adopted to classify the test cases developed for previous products.

\item \textit{RQ2. Does the proposed approach accurately identify new scenarios that are relevant for testing a new product?} With RQ2, we have evaluated the precision and recall of the approach in identifying new scenarios to be tested for a new product (i.e., new requirements not covered by existing test cases).

\item \textit{RQ3. Does the proposed approach successfully prioritize test cases?} 
With RQ3, we have evaluated whether the approach is able to effectively prioritize system test cases that trigger failures and thus can help minimize testing effort while retaining maximum fault detection power.

\item \textit{RQ4. Can the proposed approach significantly reduce testing costs compared to current industrial practice?}
With RQ4, we have evaluated to what extent the proposed approach can help significantly reduce the cost of defining and executing system test cases. 

\end{itemize}

To summarize, the contributions of this paper are:

\begin{itemize}

\item a test case classification and prioritization approach that is specifically tailored to the use case-driven development of product families, that does not rely on behavioral system models, and that guides engineers in testing new products in a product family;

\item a publicly available tool\footnote{For accessing the tool, see:~\url{https://sites.google.com/site/pumconf/}.} integrated with IBM DOORS as a plug-in, which automatically selects and prioritizes system test cases when a new product is configured in a product family;

\item an industrial case study demonstrating the applicability and benefits of our approach.

\end{itemize}

This paper is structured as follows. Section~\ref{sec:background} provides the background on PUM and PUMConf on which this paper builds the proposed approach. %Section~\ref{sec:context} introduces the industrial context of our case study to illustrate the practical motivations for our approach. 
Section~\ref{sec:related} discusses the related work. %in light of the industrial needs identified in Section~\ref{sec:context}. 
In Section~\ref{sec:overview}, we provide an overview of the approach. Sections~\ref{sec:selection} and~\ref{sec:prioritization} provide the details of its core technical parts. %In Section~\ref{sec:tool}, we present our tool while 
\MREVISION{R4.16}{Section~\ref{sec:tool} presents an overview of the provided tool support.} Section~\ref{sec:evaluation} reports on our evaluation in an industrial setting, involving an embedded system called Smart Trunk Opener (STO). In Section~\ref{sec:conclusion}, we conclude the paper.

%\MREVISION{R3.1}{DENEME}
%
% !TEX root =  Main.tex
\section{Background}
\label{sec:background}
In this section we give the background regarding the elicitation of PL use case models (see Section~\ref{subsec:elicitationVariability}), and our configuration approach (see Section~\ref{subsec:configuration}). \MREVISION{R4.5}{We also provide a glossary for the main terminology used in the paper (see Section~\ref{subsec:glossary}).}  

%This is needed for understanding our change impact analysis approach in the rest of the paper. Section~\ref{subsec:elicitationVariability} provides a brief description of PL use case and domain modeling proposed in our previous work~\cite{Hajri15}, while we illustrate the basics of our configuration approach~\cite{Hajri16} through example models in Section~\ref{subsec:configuration}.  

In the rest of the paper, we use Smart Trunk Opener (STO) as a case study, to motivate, illustrate and assess our approach. STO is a real-time automotive embedded system developed by IEE. It provides automatic, hands-free access to a vehicle's trunk, in combination with a keyless entry system. In possession of the vehicle's electronic remote control, the user moves her leg in a forward and backward direction at the vehicle's rear bumper. STO recognizes the movement and transmits a signal to the keyless entry system, which confirms that the user has the remote. This allows the trunk controller to open the trunk automatically.

\subsection{Elication of Variability in PL Use Cases with PUM}
\label{subsec:elicitationVariability}

Elicitation of PL use cases is based on the Product line Use case modeling Method (PUM)~\cite{Hajri2015}. In this section, we give a brief description of the PUM artifacts.

%\vspace*{-1.0em}
\begin{figure}[ht]%[width=0.1]
%\begin{wrapfigure}{h}{0.52\linewidth}
%\begin{figure}[h]
%\vspace*{-0.9em}
       %\centerline{\includegraphics[width=\linewidth]{images/productlineDiagram}}
        \centerline{\includegraphics[width=0.74000\linewidth]{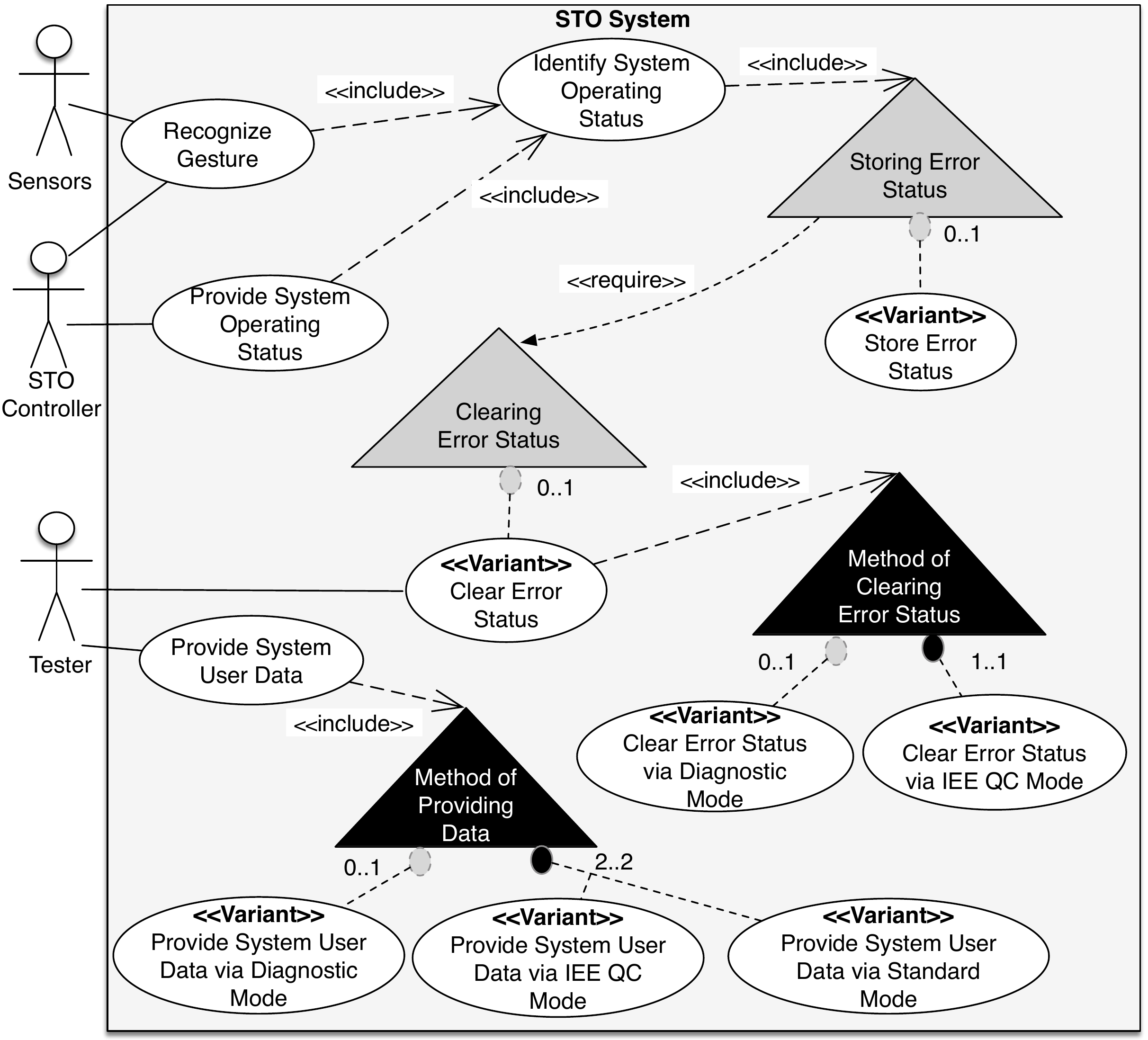}}
%\vspace*{-0.8em}
      \caption{Part of the Product Line Use Case Diagram for STO}
      \label{fig:productlineDiagram}
%\vspace*{-1.1em}
\end{figure}
%\vspace*{-0.9em}

\subsubsection{Use Case Diagram with PL Extensions}
\label{subsec:diagram}

For use case diagrams, we employ the PL extensions proposed by Halmans and Pohl~\cite{Halmans2003a, Buhne2003} since they support explicit representation of variants, variation points, and their dependencies (see Fig.~\ref{fig:productlineDiagram}).

A use case is either \textit{Essential} or \textit{Variant}. Variant use cases are distinguished from essential use cases, i.e., mandatory for all the products in a product family, by using the stereotype \textit{Variant}. A variation point given as a triangle is associated to one, or more than one use case using the relation \textit{include}. A mandatory variation point indicates where the customer has to make a selection (the black triangles in Fig.~\ref{fig:productlineDiagram}). A `tree-like' relation, containing a cardinality constraint, is used to express relations between variants and variation points, which are called \textit{variability relations}. The relation uses a [min..max] notation in which $min$ and $max$ define the minimum and maximum numbers of variants that can be selected in the variation point.

% !TEX root =  Main.tex
%\begin{table}[h]
\begin{table}[t]
%\scriptsize
%\vspace*{-1.3em}
\footnotesize
\caption{Some STO Use Cases in the extended RUCM}
%\vspace*{-1.6em}
\begin{center}
\begin{tabular}{ |p{0.13cm} | p{8.5cm} |}
1	& \textbf{USE CASE} Recognize Gesture\\
2 	& \textbf{1.1 Basic Flow (BF)}\\
3	& 1. The system REQUESTS move capacitance FROM the sensors.\\
4 	& 2. INCLUDE USE CASE Identify System Operating Status.\\
5	& 3. The system VALIDATES THAT the operating status is valid.\\
6	& 4. The system VALIDATES THAT the movement is a valid kick.\\
7	& 5. The system SENDS the valid kick status TO the STO Controller.\\
%9	& 6. The system waits for next execution cycle\\
%10	& Postcondition: The occupant class for airbag control and the occupant class for seat belt reminder have been sent.\\
8	& \textbf{1.2 \textless OPTIONAL\textgreater Bounded Alternative Flow (BAF1)}\\
9	& RFS 1-4\\
10	& 1. IF voltage fluctuation is detected THEN\\
%13	& 2. The system sends previous occupant class for airbag control to AirbagControlUnit. \\
%14	& 3. The system sends previous occupant class for seat belt reminder to AirbagControlUnit. \\
%16	& 4. The system sends the DTC to AirbagControlUnit.\\
%14	& 2. The system resets classification filters.\\
11	& 2. ABORT. \\ %RESUME STEP 1.\\
12	& 3. ENDIF\\
%17	& Postcondition: Classification filters have been reset.\\

13 	& \textbf{1.3 Specific Alternative Flow (SAF1)}\\
14	& RFS 3\\
%20	& 1. IF the occupant class for airbag control is not valid and the occupant class for seat belt reminder is not valid THEN\\
%21	& 2. The system SENDS the previous occupant class for airbag control TO AirbagControlUnit.\\
%22	& 3. The system SENDS the previous occupant class for seat belt reminder TO SeatBeltControlUnit.\\
%	& 4. The system sends the DTC to AirbagControlUnit.
15	& 1. ABORT.\\
%24	& 4. ENDIF\\
%25	& Postcondition: The previous occupant classes for airbag control and seat belt reminder has been sent to AirbagControlUnit and to SeatBeltControlUnit respectively.\\

16	& \textbf{1.4 Specific Alternative Flow (SAF2)}\\
17	& RFS 4\\
%28 	& 1. IF the occupant class for seat belt reminder is not valid THEN\\
18	& 1. The system increments OveruseCounter by the increment step.\\
%19	& 2. The system VALIDATES THAT the OveruseCounter is smaller than the OveruseCounter limit.\\ 
%4. The system sends the DTC to AirbagControlUnit.
19	& 2. ABORT.\\
%32	& 4. ENDIF\\
%33	& Postcondition: The occupant class for airbag control has been sent to AirbagControlUnit and the previous occupant class for seat belt reminder has been sent to SeatBeltControlUnit.\\
20     & \\
21	& \textbf{USE CASE} Identify System Operating Status\\
22 	& \textbf{1.1 Basic Flow (BF)}\\
23 	& 1. The system VALIDATES THAT the watchdog reset is valid.\\
24	& 2. The system VALIDATES THAT the RAM is valid.\\
25	& 3. The system VALIDATES THAT the sensors are valid.\\
26	& 4. The system VALIDATES THAT there is no error detected.\\
%8	& 5. The system SENDS the valid kick status TO STOController.\\
27	& \textbf{1.5 Specific Alternative Flow (SAF4)}\\
28	& RFS 4\\
%28 	& 1. IF the occupant class for seat belt reminder is not valid THEN\\
29	& 1. INCLUDE \textless VARIATION POINT: Storing Error Status\textgreater.\\
%19	& 2. The system VALIDATES THAT the OveruseCounter is smaller than the OveruseCounter limit.\\ 
%4. The system sends the DTC to AirbagControlUnit.
30	& 2. ABORT.\\
31     & \\
32	& \textbf{USE CASE} Provide System User Data\\
33 	& \textbf{1.1 Basic Flow (BF)}\\
34 	& 1. The tester SENDS the system user data request TO the system.\\
35	& 2. INCLUDE \textless VARIATION POINT : Method of Providing Data\textgreater.\\
36     & \\
37	& \textbf{\textless VARIANT\textgreater USE CASE} Provide System User Data via Standard Mode\\
38 	& \textbf{1.1 Basic Flow (BF)}\\
39 	& V1. \textless OPTIONAL\textgreater The system SENDS calibration TO the tester.\\
40	& V2. \textless OPTIONAL\textgreater The system SENDS sensor data TO the tester.\\
41	& V3. \textless OPTIONAL\textgreater The system SENDS trace data TO the tester.\\
42	& V4. \textless OPTIONAL\textgreater The system SENDS error data TO the tester.\\
43	& V5. \textless OPTIONAL\textgreater The system SENDS error trace data TO the tester.\\

\end{tabular}
\end{center}
\label{tab:useCaseRUCM}
\vspace*{-2.3em}
\end{table}%

A variability relation is optional where ($min = 0$) or ($min > 0$ and $max < n$); \textit{n} is the number of variants in a variation point. It is mandatory where ($min = max = n$). Optional and mandatory relations are depicted with light-grey and black filled circles, respectively (see Fig.~\ref{fig:productlineDiagram}). For instance, the essential use case \textit{Provide System User Data}  has to support multiple methods of providing data where the methods of providing data via IEE QC mode and Standard mode are mandatory. In addition, the method of providing data via diagnostic mode can be selected. 
%In STO, the customer may decide the system does not store the errors determined while the operating status is being identified (see the `Storing Error Status' optional variation point in Fig.~\ref{fig:productlineDiagram}). 
It can be decided that the STO system should not store the errors determined during the identification of the operating state (see the optional variation point \textit{Storing Error Status}). %in Fig.~\ref{fig:productlineDiagram}). 
The extensions support the dependencies \textit{require} and \textit{conflict} among variation points and variant use cases~\cite{Buhne2003}. With \textit{require} in Fig.~\ref{fig:productlineDiagram}, the selection of the variant use case in \textit{Storing Error Status} implies the selection of the variant use case in \textit{Clearing Error Status}.

%Some further variability information is given in PL use case specifications. For instance, only PL use case specifications indicate in which flows of events a variation point is included.     

\subsubsection{Restricted Use Case Modeling (RUCM) with PL Extensions}
\label{subsec:rucm}

This section introduces the RUCM (Restricted Use Case Modeling) template and its PL extensions which we proposed in our previous work~\cite{Hajri2015}. RUCM is a use case modeling method with restriction rules and keywords constraining the use of NL~\cite{Yue2013}. %We employ RUCM in the elicitation of PL specifications since it was designed to make specifications more precise and analyzable, while preserving their readability.
%For details, the reader is referred to~\cite{Yue13}.  
%Controlled experiment results show that the restriction rules in RUCM are overall applicable and beneficial~\cite{Yue13}. 
%Since it was not originally designed for product line modeling of embedded systems, we introduce some extensions into RUCM (\textit{Challenge 2}).
Since it was not designed for PL modeling, we introduced some PL extensions (see Table~\ref{tab:useCaseRUCM}). In RUCM, use cases have basic and alternative flows (Lines 2, 8, 13, 16, 22, 27, 33 and 38). In Table~\ref{tab:useCaseRUCM}, we omit some  alternative flows and basic information such as actors and pre/post conditions.

A basic flow describes a main successful path that satisfies stakeholder interests (Lines 3-7, 23-26 and 39-43). It contains use case steps and a postcondition. A step can be a system-actor interaction: an actor sends a request or data to the system (Line 34); the system replies to an actor with a result (Line 7). In addition, the system validates a request or data (Line 5), or it alters its internal state (Line 18).
%The use case inclusion is given in a step with the keyword `\textit{INCLUDE USE CASE}' (Line 4). 
Other use cases are included with the keyword `\textit{INCLUDE USE CASE}' (Line 4). The keywords are in capital letters. 
%In addition, the inclusion of another use case is specified as a step. This is the case of Line 3, as denoted by the keyword `\textit{INCLUDE USE CASE}'. All keywords are written in capital letters for readability. 
`\textit{VALIDATES THAT}' (Line 5) indicates a condition that must be true to take the next step, otherwise an alternative flow is taken. %In Table~\ref{tab:useCaseRUCM}, the system proceeds to Step 3 (Line 5) if the operating status is valid (Line 4).
%The keyword \emphx{Post-condition} indicates a post-condition for the current use case flow (Line 10).

%Alternative flows describe all the other scenarios or branches, both success and failure. An alternative flow always depends on a condition in a specific step, in a flow of reference, referred to as \textit{reference flow}, and that reference flow is either the basic flow or the same alternative flow. In RUCM, there are three types of alternative flows: \textit{specific}, \textit{bounded} and \textit{global}. A specific alternative flow refers to a step in the reference flow (Lines 19 and 27). A bounded alternative flow refers to more than one step in the reference flow (Line 12) while a global alternative flow refers to any step in the reference flow.

An alternative flow describes other scenarios, %or branches, 
both success and failure. It depends on a condition in a specific step in a flow of reference, referred to as \textit{reference flow}, and that reference flow is either the
basic flow or another alternative flow.

RUCM has \textit{specific}, \textit{bounded} and \textit{global} alternative flows. A specific alternative flow refers to a step in a reference flow (Lines 13, 16, and 27). A bounded alternative flow refers to more than one step in a reference flow (Line 8), while a global flow refers to any step in a reference flow. `\textit{RFS}' is used to refer to reference flow steps (Lines 9, 14, 17, and 28). Bounded and global alternative flows begin with `\textit{IF .. THEN}' for the conditions under which they are taken (Line 10). Specific alternative flows do not necessarily begin with `\textit{IF .. THEN}' since a guard condition is already indicated in their reference flow steps (Line 5). 
%It is possible to have composite conditions further refined in multiple alternative flows which are considered to be executed when the condition in the reference flow step evaluates to false. This case is not covered by RUCM. Therefore, PUM suggests to use the `\textit{IF .. THEN}' keyword also in specific alternative flows in such cases. The alternative flows are evaluated in the order they appear in the use case.   

%The keywords ease automatic identification of steps for these interactions.
%We introduce extensions into RUCM regarding the usage of `\textit{IF}' conditions and the way input/output messages are expressed.

%PUM follows the guidelines that suggest not to use multiple branches within the same use case path~\cite{Larman02}, thus enforcing the usage of `\textit{IF}' conditions only as a means to specify guard conditions for alternative flows.

%Our extensions are
PUM extensions to RUCM include (i) new keywords for modeling interactions in embedded systems and (ii) new keywords for modeling variability. The keywords `\emphx{SENDS .. TO}' and `\emphx{REQUESTS .. FROM}' capture system-actor interactions (Lines 3, 7, 34, and 39-43). For instance, Step 1 (Line 3) indicates an input message from sensors to the system. %, while Step 5 (Line 7) contains an output message from the system to the STO Controller. 
%According to our experience, in embedded systems, system-actor interactions are specified in terms of messages. More keywords can be defined for other types of systems. 
%We introduce the notion of variation point and variant use case, complementary to the extensions in Section~\ref{subsec:diagram}, into RUCM. 
For consistency with PL use case diagrams, PUM introduces into RUCM the notion of variation point and variant use case. Variation points can be included in basic or alternative flows with the keyword `\textit{INCLUDE \textless VARIATION POINT : ... \textgreater}' (Lines 29 and 35). Variant use cases are given with the keyword `\textit{\textless VARIANT \textgreater}' (Line 37). %The same keyword is also used for variant actors related to a variation point given in the use case diagram.    
%
%Some variability cannot be captured in PL diagrams due to the required level of granularity for product configuration. To model such variability, as part of our extensions, we introduce optional steps, optional alternative flows and a variant order of steps.
To capture variability that cannot be modeled in use case diagrams because of their coarse granularity, PUM introduces optional steps, optional alternative flows and a variant order of steps. Optional steps and alternative flows begin with the keyword `\textit{\textless OPTIONAL\textgreater}' (Lines 8 and 39-43). The keyword `V' is used before step numbers to express variant step order (Lines 39-43). \MREVISION{R2.11}{A variant order occurs with optional and/or mandatory steps. %It is important because variability in the system behavior can be given with multiple execution orders of the same steps. 
For instance, the steps in the basic flow of \textit{Provide System User Data via Standard Mode} are optional, while their execution order varies.}

\subsection{Configuration of PS Use Case Models}
\label{subsec:configuration}
% !TEX root =  Main.tex
%\section{Configuration of Product Specific Use Case}
%\label{sec:configuration}

%PUMConf relies on variability information given in the PL use case models. 
PUMConf supports users in making configuration decisions and automatically generating PS use cases from PL use cases.

The user selects (1) variant use cases in the PL use case diagram and (2) optional use case elements in the PL use case specifications, to generate PS use case diagram and specifications. %The reader is referred to our prior work~\cite{Hajri16,Hajri16b} for the details of our configuration approach. 
For instance, the user makes decisions for the variation points in Fig.~\ref{fig:productlineDiagram}. A decision is about selecting, for the product, variant use cases in the variation point. The user selects \textit{Store Error Status} and \textit{Clear Error Status} in the variation points \textit{Storing Error Status} and \textit{Clearing Error Status}, respectively. She also unselects \textit{Clear Error Status via Diagnostic Mode} in the variation point \textit{Method of Clearing Error Status}, while \textit{Clear Error Status via IEE QC Mode} is automatically selected by PUMConf because of the mandatory variability relation. Finally, the user unselects \textit{Provide System User Data via Diagnostic Mode} in the variation point \textit{Method of Providing Data}.

%\begin{figure}[h]
%\begin{wrapfigure}{h}{0.64\linewidth}
\begin{figure}[ht]%[width=0.1]
%\vspace*{-1.8em}
        \centerline{\includegraphics[width=0.840\linewidth]{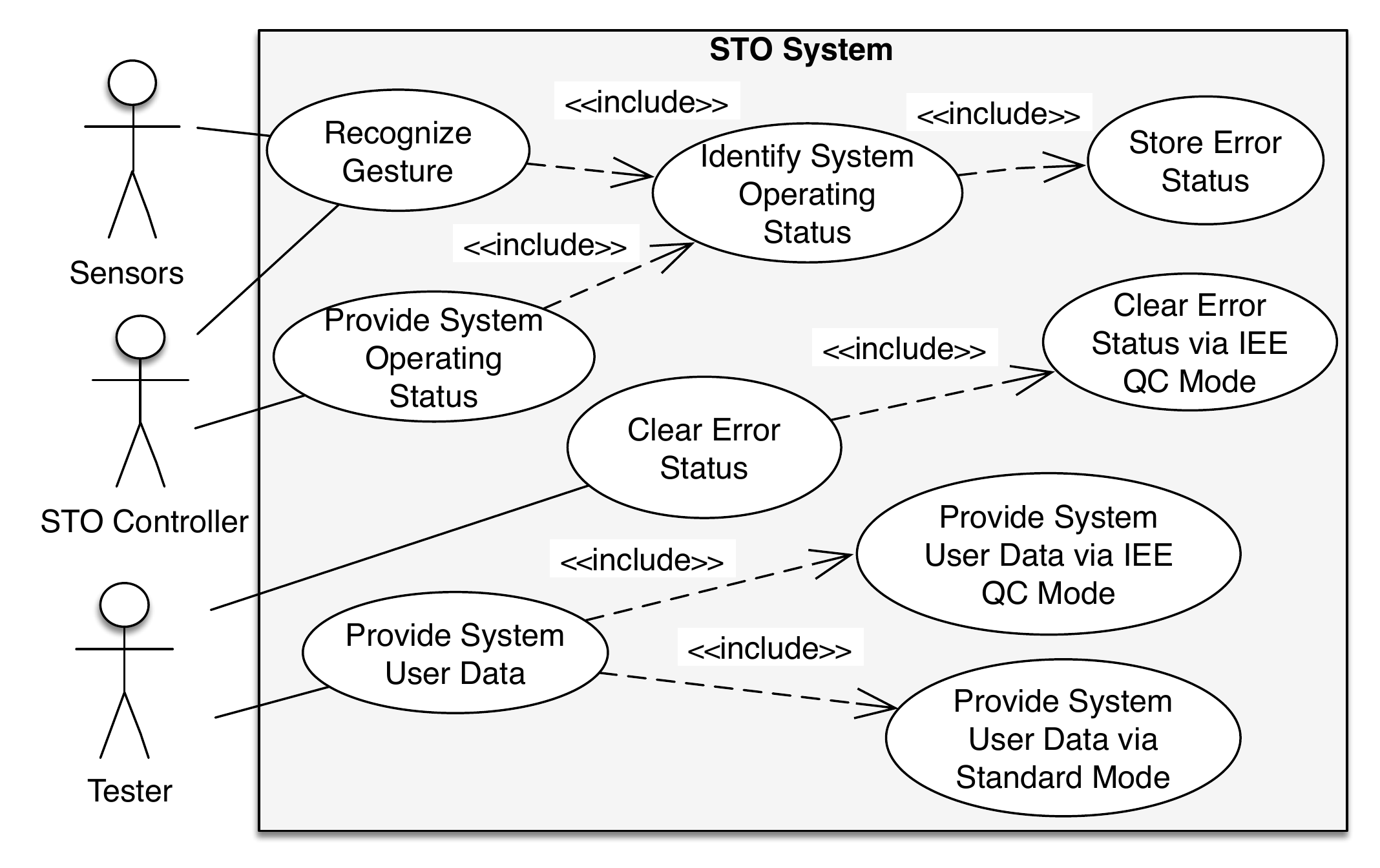}}
        %\centerline{\includegraphics[width=\linewidth]{images/productspecificDiagramExample}}
%\vspace*{-0.7em}
      \caption{Generated Product Specific Use Case Diagram for STO}
      \label{fig:ExamplePSUseCaseDiagram}
\vspace*{-1.2em}
\end{figure}

Given the configuration decisions, PUMConf automatically generates the PS use case diagram from the PL diagram (see Fig.~\ref{fig:ExamplePSUseCaseDiagram} generated from Fig.~\ref{fig:productlineDiagram}). %For instance, the user unselects \textit{Provide System User Data via Diagnostic Mode} for \textit{Method of Providing Data} in Fig.~\ref{fig:productlineDiagram}. 
For instance, for the decision for the variation point \textit{Method of Providing Data}, PUMConf creates the use cases \textit{Provide System User Data via IEE QC Mode} and \textit{Provide System User Data via Standard Mode}, and two \textit{include} relations in Fig.~\ref{fig:ExamplePSUseCaseDiagram}.

%\vspace*{-1.3em}
% !TEX root =  Main.tex
%\begin{table}[h]
\begin{table}[t]
%\begin{wraptable}{r}{9.00cm}
%\scriptsize
\footnotesize
%\vspace*{-1.5em}
\caption{Some of the Generated Product Specific Use Case Specifications}
%\vspace*{-0.8em}
\begin{center}
\begin{tabular}{| p{0.13cm} | p{8.5cm} |}
1	& \textbf{USE CASE} Recognize Gesture\\
2 	& \textbf{1.1 Basic Flow (BF)}\\
3	& 1. The system REQUESTS the move capacitance FROM the sensors.\\
4 	& 2. INCLUDE USE CASE Identify System Operating Status.\\
5	& 3. The system VALIDATES THAT the operating status is valid.\\
6	& 4. The system VALIDATES THAT the movement is a valid kick.\\
7	& 5. The system SENDS the valid kick status TO the STO Controller.\\
8 	& \textbf{1.2 Specific Alternative Flow (SAF1)}\\
9	& RFS 3\\
10	& 1. ABORT.\\
11	& \textbf{1.3 Specific Alternative Flow (SAF2)}\\
12	& RFS 4\\
13	& 1. The system increments the OveruseCounter by the increment step.\\
14	& 2. ABORT.\\

15     & \\
16	& \textbf{USE CASE} Identify System Operating Status\\
17 	& \textbf{1.1 Basic Flow (BF)}\\
18 	& 1. The system VALIDATES THAT the watchdog reset is valid.\\
19	& 2. The system VALIDATES THAT the RAM is valid.\\
20	& 3. The system VALIDATES THAT the sensors are valid.\\
21	& 4. The system VALIDATES THAT there is no error detected.\\
22	& \textbf{1.5 Specific Alternative Flow (SAF4)}\\
23	& RFS 4\\
24	& 1. INCLUDE USE CASE Store Error Status.\\
25	& 2. ABORT.\\
26     & \\
27	& \textbf{USE CASE} Provide System User Data\\
28 	& \textbf{1.1 Basic Flow (BF)}\\
29 	& 1. The tester SENDS the system user data request TO the system.\\
30     & 2. The system VALIDATES THAT `Precondition of Provide System User Data via Standard Mode'.\\
31	& 3. INCLUDE USE CASE Provide System User Data via Standard Mode.\\
32	& \textbf{1.2 Specific Alternative Flow (SAF1)}\\
33	& RFS 2\\
34	& 1. INCLUDE USE CASE Provide System User Data via IEE QC Mode.\\
35	& 2. ABORT.\\
36     & \\
37	& \textbf{USE CASE} Provide System User Data via Standard Mode\\
38 	& \textbf{1.1 Basic Flow (BF)}\\
39 	& 1. The system SENDS the trace data TO the tester.\\
40	& 2. The system SENDS the calibration data TO the tester.\\
41	& 3. The system SENDS the error trace data TO the tester.\\

\end{tabular}
\end{center}
\label{tab:ExamplePSUseCase}
\vspace*{-2.2em}
\end{table}%

%\vspace*{-0.3em}

%The decision-making is an iterative process. We devised an algorithm to check the consistency of configuration decisions~\cite{Hajri2016c}. In the case of contradicting configuration decisions, such as two decisions resulting in selecting variant use cases violating some dependency constraints, the algorithm automatically detects and reports them. The user must then backtrack and revise the decisions to resolve the contradictions. Assume that the user first makes a decision in \textit{Clearing Error Status}, which is unselecting \textit{Clear Error Status}. No contradiction is identified since there is no prior decision. The user proceeds with \textit{Storing Error Status} and selects \textit{Store Error Status}. Our algorithm identifies a contradiction with the decision in \textit{Clearing Error Status} since the selection of \textit{Store Error Status} implies the selection of \textit{Clear Error Status} via the \textit{requires} dependency (see Fig.~\ref{fig:productlineDiagram}). The user is asked to resolve the contradiction by updating one of the decisions for \textit{Storing Eror Status} and \textit{Clearing Error Status}. The user selects \textit{Clear Error Status} to resolve the contradiction. %At this point, there is no contradiction identified.

%Next, the user makes decisions for the PL specifications. 
After identifying variant use cases to be included in the PS diagram, the user makes decisions based on the PL specifications. In Table~\ref{tab:useCaseRUCM}, there are two variation points (Lines 29 and 35), one variant use case (Lines 37-43), five optional steps (Lines 39-43), one optional alternative flow (Lines 8-12), and one variant order group (Lines 39-43). The decisions for the variation points are already made in the PL diagram. Three optional steps are selected with the order \textit{V3}, \textit{V1}, and \textit{V5}. The optional alternative flow is unselected.

%For instance, based on the decision for \textit{Method of Providing Data} included by \textit{Provide System User Data} in Fig.~\ref{fig:productlineDiagram}, PUMConf creates \textit{Provide System User Data via IEE QC Mode} and \textit{Provide System User Data via Standard Mode} with the two \textit{include} relations in Fig.~\ref{fig:ExamplePSUseCaseDiagram}.

%For instance, for \textit{Provide System User Data}, \textit{Method of Providing Data} and the selected variant use cases in Fig.~\ref{fig:productlineDiagram}, the algorithm creates \textit{Provide System User Data}, \textit{Provide System User Data via IEE QC Mode} and \textit{Provide System User Data via Standard Mode} with the two \textit{include} relations in Fig.~\ref{fig:ExamplePSUseCaseDiagram}.

%The approach saves all configuration decisions in a decision model~\cite{Hajri16}. Such a decision model is important since the analyst/customer may need to update decisions to reconfigure the PS models for the same product. The decision model conforms to a decision metamodel, described in~\cite{Hajri16}. 

PUMConf automatically generates PS use case specifications from PL specifications, diagram decisions and specification decisions. 
%(see Table~\ref{tab:ExamplePSUseCase} generated from Table~\ref{tab:useCaseRUCM}). %First, the diagram decisions are considered for the variation points in the PL specifications (Lines 29 and 35 in Table~\ref{tab:useCaseRUCM}). 
Table~\ref{tab:ExamplePSUseCase} shows a PS use case specification generated from Table~\ref{tab:useCaseRUCM}, where selected optional steps are generated with the order decided in the PS specifications (Lines 39-41).
 %First, the diagram decisions are considered for the variation points in the PL specifications (Lines 29 and 35 in Table~\ref{tab:useCaseRUCM}). 
For multiple variants selected for the same variation point, PUM introduces validation checks to select the variation point to be used, based on their preconditions.
For instance, based on the diagram decision for \textit{Method of Providing Data} in Fig.~\ref{fig:productlineDiagram}, PUMConf creates two include statements for \textit{Provide System User Data via Standard Mode} and \textit{via IEE QC Mode} (Lines 31 and 34 in Table~\ref{tab:ExamplePSUseCase}), and a validation step (Line 30) that checks if the precondition of \textit{Provide System User Data via Standard Mode} holds. If it holds, \textit{Provide System User Data via Standard Mode} is executed in the basic flow (Line 31). If not, \textit{Provide System User Data via IEE QC Mode} is executed (Lines 32-35).

\subsection{\MREVISION{R4.5}{Glossary}}
\label{subsec:glossary}

An \emphx{actor} specifies a type of role played by an entity interacting with the system (e.g., by exchanging signals and data), but which is external to the system (see Section~\ref{subsec:elicitationVariability}).

A \emphx{use case} is a list of actions or event steps typically defining the interactions between an actor and a system to achieve a goal. It is generally named with a phrase that summarizes the story description, e.g., \emphx{Recognize Gesture} (see Section~\ref{subsec:elicitationVariability}).

A \emphx{use case specification} is a textual document that captures the specific details of a use case. Use case specifications provide a way to document the functional requirements of a system. They generally follow a template (see Section~\ref{subsec:elicitationVariability}).

A \emphx{use case scenario model} is a graph representation of a use case specification (see Section~\ref{subsubsec:usecase_model_generation}).

A \emphx{use case flow} is a sequence of interactions between actors and the system captured by a use case specification. A use case specification may include multiple alternative use case flows (see Section~\ref{subsec:elicitationVariability}).

A \emphx{use case scenario} is a sequence of interactions between actors and the system. It represents a single use case execution. It is a possible path through a use case specification. It may include multiple use case flows (see Section~\ref{subsubsec:usecase_model_generation}).

A \emphx{finite state machine} is an abstract machine, i.e., a theoretical model of a system used in automata theory, that can be in exactly one of a finite number of states at any given time (see Section~\ref{sec:related}). 

A \emphx{sequence diagram} shows interactions between objects in a sequential order, i.e., the order in which these interactions take place (see Section~\ref{sec:related}).
   
A \emphx{system sequence diagram (SSD)} is a sequence diagram that shows, for a particular scenario of a use case, the events that actors generate, their order, and possible inter-system events (see Section~\ref{sec:related}).

\emphx{Model slicing} allows for a model reduction by abstracting from model elements not influencing a selected element, e.g., a state transition, used as a slicing criterion (see Section~\ref{sec:related}). A reduced model is a slice that preserves the execution semantics compared to the original model with respect to the slicing criterion.

\emphx{Fault proneness of requirements} allows engineers to identify the requirements which have had reported failures (see Sections~\ref{sec:related}, \ref{sec:overview} and \ref{sec:prioritization}). As the system evolves into several versions, engineers can use the data collected from prior versions to identify requirements that are likely to be error prone~\cite{Srikanth2005}.
 
\emphx{Requirements volatility} is a measure of how much a system's requirements change during the development of the system (see Sections~\ref{sec:related} and~\ref{sec:overview}). Projects for which the requirements change greatly have a high volatility, while projects whose requirements are relatively stable have a low volatility~\cite{Malaiya1999, Henry93}.
 
A \emphx{single-product setting} is an experiment setting where the new product is compared to only one previous product in the product line at once (see Section~\ref{sec:evaluation}). Assume that there are $\mathit{N}$ previous products ($\mathit{P_{1}}$, $\mathit{P_{2}}$, ..., and $\mathit{P_{N}}$) in a product line. In a single-product setting, the new product ($\mathit{P_{new}}$) is compared to each previous product distinctly at $N$ times ($\mathit{P_{new}}$ - $\mathit{P_{1}}$, $\mathit{P_{new}}$ - $\mathit{P_{2}}$, ..., $\mathit{P_{new}}$ - $\mathit{P_{N}}$).
 
A \emphx{whole-line setting} is an experiment setting where the new product is compared to all the previous products in the product line at once (see Section~\ref{sec:evaluation}). In a whole-line setting, the new product ($\mathit{P_{new}}$) is compared to all the $N$ previous products at the same time ($\mathit{P_{new}}$ - $\{\mathit{P_{1}}$, $\mathit{P_{2}}$, ..., $\mathit{P_{N}}\}$).  

An \emphx{executable test case} is a sequence of executable instructions (i.e., invocations of test driver function) that trigger the system under test, thus simulating the interactions between one or more actors and the system (see Sections \ref{sec:related} and~\ref{sec:selection}).

%
% !TEX root =  Main.tex
\section{Related Work}

\label{sec:related}

\MREVISION{R2.12}{We cover the related work across three categories: (i) \textit{testing of product lines}, (ii) \textit{test case classification and selection}, and (iii) \textit{test case prioritization}. The last two categories cover the features our approach addresses in the context of product lines. In the first category, we present existing product line testing strategies and discuss how our approach is related to specific testing strategies and activities such as test case generation and execution.}

\subsection{Testing of Product Lines}
\label{subsec:PLTesting}
\MREVISION{R2.13}{Various product line testing strategies have been proposed in the literature \cite{do2014strategies, Neto2011, Lee-SPLTestingSurvey-SPLC-2012, Engstrom2011, Runeson2012, oster2011survey, tevanlinna2004product, johansen2011survey}. Neto et al.~\cite{Neto2011} present a comprehensive survey, including \textit{testing product by product}, \textit{opportunistic reuse of test assets}, \textit{design test assets for reuse}, \textit{division of responsibilities}, and \textit{incremental testing of product lines}. The strategy \textit{testing product by product} does not attempt to reuse test cases developed for previous products, while the strategy \textit{opportunistic reuse of test assets} focuses on the reuse of test assets across products without considering any systematic reuse method. The strategy \textit{design test assets for reuse} enforces the creation of test assets early in product line development, under the assumption that product lines and configuration choices are exhaustively modeled before the release of any product. This assumption does not hold when product lines and configuration choices are refined during product configuration, which is a common industry practice. The strategy \textit{division of responsibilities} is about defining testing phases that facilitate test reuse. Our approach follows the strategy referred to as  \textit{incremental testing of product lines}, which relies on regression testing techniques, i.e., test case selection and prioritization. We are the first to support incremental testing of product lines through test case selection and prioritization in the context of use case-driven development.}

Product line testing covers two separate but closely related test engineering activities: domain testing and application testing. Domain testing verifies and validates reusable components in a product line while application testing does so for a specific product in the product line. %Lee et al.~\cite{Lee-SPLTestingSurvey-SPLC-2012} present another survey of product line testing techniques in which research contributions are classified in five categories within the context of domain testing and application testing: \textit{test case creation}, \textit{test case selection}, \textit{test execution for absent variants}, \textit{variability binding in testing}, and \textit{application-specific tests}. Our approach falls under the category of \textit{test case selection}. 
\MREVISION{R1.4}{Domain test cases can be created either directly from domain artifacts or through domain test models (derived from domain artifacts). Application test cases can be created directly from domain test cases by using variability binding information in products. A test case can be executed before or after variability binding in products, and the variability binding can occur during the development phase, at compile time, or at runtime. 
Our approach currently supports application testing, but can be adapted to classify domain test cases. More specifically, the scenario generation and impact analysis algorithms in Section~\ref{subsubsec:impact_identification} can be adapted to identify scenarios of the variant requirements and eventually to determine test cases examining those scenarios.}
%For each new product in a product family, %PS use case models are generated from PL use case models and configuration decisions in PUMConf. 
%we \textcolor{red}{classify} and prioritize system test cases of previous product(s). %(\textit{Challenges 1} and \textit{2}). %and \textit{3}). 
For each new product, our approach can be used to classify and prioritize domain test cases derived from PL use case models. 

There are various product line testing approaches that support %\textit{test case creation} and \textit{application-specific tests} 
test case generation and execution (e.g.,~\cite{Nebut06, Reuys2006, Kamsties2004, Geppert2004, Uzuncaova2010, Uzuncaova2008, Arrieta2017}). Some of them generate system test cases from use case models in a product family. %~\cite{McGregor2001} %by following the strategy \textit{design test assets for reuse}.
However, they require detailed behavioral models (e.g., sequence or activity diagrams) %, which limits the adoption of these techniques in industrial settings.  
which engineers tend to avoid because of the costs related to their development and maintenance.
\MREVISION{R2.13}{Among these works generating system test cases from use cases, the ScenTED approach proposed by 
Reuys et al.~\cite{Reuys2005, Reuys2006} %~\cite{Pohl2006}~\cite{Kamsties2004}, 
is a representative approach in terms of its reliance on behavioral models for test case generation and execution in product lines.} It is 
based on the systematic refinement of PL use case scenarios to PL system and integration test scenarios. ScenTED requires activity diagrams capturing activities described in use case specifications together with variants of the product family. %These activity diagrams are used to derive domain test cases that include variants, and, at the same time, that guarantee the coverage of all the branches in the activity diagrams. Application specific test cases are then derived by selecting the variants in the domain test cases. 
%Stricker et al.~\cite{Stricker2010} propose the ScenTED-DF approach to avoid redundant testing during the execution of those derived application specific test cases. The approach builds on data flow-based testing techniques for single systems and extends those techniques to consider product line variability. 
Extensions of ScenTED include the ScenTED-DF approach~\cite{Stricker2010} which relies on data-flow analysis to avoid redundant execution of test cases derived with ScenTED. %Another approach for generating system test cases from PL models is that of Nebut et al.~\cite{Nebut06}. The approach requires that PL requirements be specified using UML use case models, %for the product family, 
%extended with parameters and contracts to specify variability. %The user makes configuration decisions on those use case models to generate PS use cases. %The approach then needs that 
%PS use cases are generated from those use case models.
%Detailed sequence diagrams are needed for each PS use case to generate system test cases. 
A methodology that does not rely on detailed behavioral models is PLUTO (Product Lines Use Case Test Optimization)~\cite{Bertolino2003}. PLUTO automatically derives test scenarios from PL use cases with some special tags for variability, but executable system test cases need to be manually derived from test scenarios. %(i.e., optional, alternative and parametric). 
%Variations are explicitly enclosed into the sections of use case specifications by means of these tags that indicate those parts of the PL requirements to be instantiated for configuring a product~\cite{Bertolino2003}.
%PLUTO includes special tags for capturing variability (i.e., \textit{optional}, \textit{alternative} and \textit{parametric}), which are used to indicate those PL requirements that need to be instantiated for configuring a product in the product line. 
%Test scenarios are automatically generated from PS use cases, but executable system test cases need to be manually derived from test scenarios.  

This paper complements the approaches above by providing a mechanism for selecting and prioritizing test cases, that have already been generated and executed in previous products.

\subsection{Test Case Classification and Selection}
\label{subsec:rel_selection}

When defining a product in a product family for a new customer, the changed parts of the new product need to be tested, as well as the other parts to detect regression faults. %there is a need not only for testing the changed parts of the product but also for testing other parts for regression. %As the product grows, not all test cases can be rerun 
In most practical contexts, given the number of test cases and their execution time, not all of them can be rerun for regression due to limited resources. %Therefore, there is a need to select test cases and minimize the test suite for the new product in the product family. 
Test case selection is a strategy commonly adopted by regression testing techniques to reduce testing costs~\cite{Engstrom2010, Yoo2012, Do2016}. \MREVISION{R2.14}{Therefore, we investigate test case classification and selection approaches under two categories: (i) the selection of regression test cases for a single product and (ii) the selection of test cases for each product in a product line.}

Regression test selection techniques aim to reduce testing costs by selecting a subset of test cases from an existing test suite~\cite{Rothermel1996}. Most of them are code-based and use code changes and code coverage information to guide test selection (e.g.,~\cite{Kung1995, Binkley1997, Rothermel1997, Rothermel2000, Harrold2001, Qu2011, Nardo2015}). Other techniques use different artifacts such as requirements specifications (e.g.,~\cite{Vaysburg2002, Mirarab2008, Dukaczewski2013}), architecture models (e.g.,~\cite{Mayrhauser1999, Muccini2006, Muccini2007}), %~\cite{Muccini2003} 
or UML diagrams (e.g.,~\cite{Briand2009, Chen2002, Hemmati2010}). %~\cite{Zech2017}%Most approaches do not address the selection of test cases in the context of use case-driven development and testing in product lines. 
For instance, Briand et al.~\cite{Briand2009} present an approach for automating regression test selection based on UML diagrams and traceability information linking UML models to test cases. They propose a formal mapping between changes on UML diagrams (i.e., class and sequence diagrams) and a classification of regression test cases into three categories (i.e., \textit{reusable}, \textit{retestable}, and \textit{obsolete}). %Chen at al.~\cite{Chen2002} propose another approach for changes on the UML activity diagrams that are traced to test cases. %used to describe system requirements. 
%The approach requires that the activity diagrams be traced to test cases. 
%Test cases are selected by comparing the activity diagrams with their modified version. 

%The above approaches do not address classifying and selecting test cases in the context of use case-driven development of product lines. %(\textit{Challenges 1} and \textit{2}). %do not consider products evolving in a product family. 
The approaches mentioned above require detailed design artifacts (e.g., finite state machines and sequence diagrams), rather than requirements in NL, such as use case specifications.  %  which is not very applicable in industrial settings. 
Further, they compare a system artifact with its modified version to select test cases from a test suite in the context of a single system, not in the context of a product line. %where multiple test suites exist for a family of products. %(\textit{Challenge 2}). 
%In our approach, after the initial product is tested, the following products in the product family are tested using regression testing techniques, i.e., test case selection and prioritization based on configuration decision changes between the previous product(s) and the new product to be tested. Our approach can compare the new product configuration with multiple existing configurations in the product family in order to identify the impact of use case changes on system test cases (\textit{Challenge 1}) and classify test cases from multiple test suites. %(\textit{Challenge 2}).

%To evaluate RTS techniques, analytic and empirical evaluation can be employed. The complexity and the safety of algorithms are measured for the analytic evaluation. The number of test cases selected or test case reduction rate and the time required to select the subset of test suite are used to empirically evaluate the RTS techniques.

\MREVISION{R2.15}{There are several product line test case selection approaches~\cite{Wang2016, Wang2017, Cabral2010, Knapp2014, Schurr2010, Runeson2012, Engstrom2013}.} %~\cite{Engstrom2013}. 
%The testing strategy that our approach follows falls under the category \textit{Incremental Testing of Product Lines}~\cite{Neto2011}. 
%Wang et al.~\cite{Wang2016}~\cite{Wang2017} propose a method which relies on feature models and component family models to support test case selection for product lines.
Wang et al.~\cite{Wang2016, Wang2017} 
propose a product line test case selection method using feature models.  
%Feature models capture configuration decisions that affect test case selection; 
%These decisions include the selection of the execution states in which the software can be tested, the identification of the features to be tested, and the selection of the configuration parameters to be used during testing. 
%The feature model is used also to capture dependencies between the features to be tested.
%component family models capture how test cases are grouped into test suites, and associate each test case with effectiveness attributes (i.e., fault detection capability, average execution time, and execution frequency). In addition, these models capture the dependencies between features and test cases by means of restrictions. %
The method works in three steps: (i) software engineers indicate features that need to be tested; (ii) a toolset is used to check the consistency between features included in a program; and (iii) test cases are automatically selected so that all the test cases associated with a feature to be tested will be executed. %The main limitation of the approach is that the test selection algorithm relies only upon trace links between features and test cases, and it does not incrementally reuse the test cases of the previous products in the product family. 
%One limitation of the approach is that the test selection algorithm relies upon additional trace links between features and test cases. 
The main limitation is that all the test cases of the product family need to be derived upfront %even if some of them may never be executed 
and that the scope of the product family must be defined in advance. There are other similar approaches suffering from the same limitation~\cite{Cabral2010, Knapp2014, Schurr2010}. %~\cite{Kahsai2008}. 
In contrast, our approach requires that only test cases for the initial product be available in advance. %Test cases of a new product are derived incrementally from the existing test suites (\textit{Challenge 1}). %by comparing the configuration of the new product with the previous configuration(s) in the product family (\textit{Challenge 1}). %and \textit{2}). 
A test case selection approach that does not require early generation of test cases for the product family is that of Lity et al.~\cite{Lity2016, Lochau2014, Lity2012}, which is based on model slicing for incremental product line testing. Lity et al. apply incremental model slicing to determine the impact of changes on a test model, e.g., finite state machines, and to reason about their potential retest. %To do so, the approach requires (i) a set of slicing criteria (e.g., test goals used to guide the test case generation process) suitable to investigate the impact of model changes on test cases, (ii) data and control flow dependencies between model elements, and (iii) model deltas specifying test model changes between subsequent products under test. 
%Differences between test models of products in a product family are specified by delta modeling~\cite{Clarke2010}. %After testing the first product in the product family, the test model changes expressed in deltas are used to reduce the number of test cases that are executed when testing the subsequent products. 
%These differences are used to reduce the number of test cases that are executed for new products. 
The approach first computes test model regression deltas between the previous and new products. Based on a structural coverage criterion and the computed regression deltas, a set of impacted test goals, i.e., structural test model elements, is identified from the test model. The impacted test goals are analyzed to identify obsolete test cases for the new product. For each product, the approach computes a test model slice, which comprises test model elements influencing a test goal based on control and data dependencies between elements. Reusable and retestable test cases are identified based on the changes between the test model slices of the previous and new products. The approach needs detailed behavioral models, e.g., finite state machines and message sequence diagrams, which rarely exist in contexts where requirements are mostly captured in NL. In complex industrial systems, behavioral models that are precise enough to enable test case selection are so complex that their specification cost is prohibitive and the task is often perceived as overwhelming by engineers. In contrast, our approach does not require that detailed behavioral models be provided by engineers. With the help of NLP, it automatically extracts behavioral information from use case specifications compliant with RUCM (see Sections~\ref{subsubsec:usecase_model_generation} and \ref{subsubsec:scenario_generation}). Lity et al. do not address how to trace from impacted test goals to their corresponding test cases while our approach provides a detailed traceability method required for test case classification (see Section \ref{subsubsec:matching_scenarios_testcases}). We automatically identify all tested use case scenarios and derive new, untested scenarios from the tested scenarios (see Sections \ref{subsubsec:scenario_generation} and \ref{subsubsec:matching_scenarios_testcases}). To classify test cases, we directly identify the impact of configuration decision changes on the tested scenarios. Therefore, in contrast to the work by Lity et al., our approach does not need model slices and a retest coverage criterion, i.e., a structural coverage criterion, for the retest decision. In addition, Lity et al. do not support the definition of test cases for new requirements while our approach identifies use case scenarios that have not been tested before, and provides information on how to modify existing test cases to cover those new, untested scenarios (see Section \ref{subsubsec:impact_identification}). %(\textit{Challenge 1}). 
%Fabrizio, was Dukaczewski et al.~\cite{Dukaczewski2013} discuss how 
%An approach that does not require detailed test models is that of Dukaczewski et al.~\cite{Dukaczewski2013} who apply incremental SPL testing strategies at the requirements level. 
Dukaczewski et al.~\cite{Dukaczewski2013} briefly discuss how to apply the incremental product line testing strategy %at the requirements level. 
to NL requirements. 
%They propose a delta-oriented incremental testing approach based on NL requirements and associated test cases. %In order to apply the approach, the requirements of the core system have to be separated from the requirements of the possible variants. 
They do not provide any method to model variability in requirements; it is only suggested that a requirement is split into several requirements, one for each possible product variant. Also, there is no reported systematic approach supported by a tool. To the best of our knowledge, our work is the first systematic and automated approach for supporting the incremental product line testing strategy for NL requirements. %Obsolete and new test cases are identified by adding and removing requirements for the subsequent product. 

\subsection{Test Case Prioritization}
\label{subsec:rel_prioritization}
Test case prioritization techniques schedule test cases in an order that increases their effectiveness in meeting some performance goals (e.g., rate of fault detection and number of test cases required to discover all the faults)~\cite{Rothermel2001, Khatibsyarbini2017, Yoo2012}. They mostly use information about previous executions of test cases (e.g.,~\cite{Rothermel2001, Li2007, Engstrom2011b, Sanchez2011, Lachmann2016b, Hemmati2017}),   %~\cite{Wong1997}
human knowledge (e.g.,~\cite{Srikanth2014, Srikanth2016, Srikanth2012, Arafeen2013, Krishnamoorthi2009, Tonella2006}), %~\cite{Srikanth2005b}
 or a model of the system under test (e.g.,~\cite{Haidry2013, Kundu2009, Tahat2012, Korel2008}). %~\cite{Korel2005} %to prioritize test cases for a single system in the context of regression testing. %The rate of fault detection is frequently employed to evaluate TCP techniques. %Most of the approaches in the context of SPL solely order product variants in a product family, but do not consider the prioritization of test cases per product. %Regarding test case minimization techniques, the objective of these techniques is to reduce the size of the test suite by removing the obsolete and redundant test cases. Since TSM problems are NP-complete, the TSM techniques use heuristics to produce an approximate test set.
%For instance, Tahat et al.~\cite{Tahat2012} present and evaluate two model-based test case prioritization methods using finite state machine models of the system under test. %These methods assume that modifications are made both on the system under test and its model. The existing test suite is executed on the system model and information about this execution is used to prioritize test cases. %Execution of the model is inexpensive compared to that of the system; therefore, the cost of test prioritization is relatively small. 
For instance, Shrikanth et al.~\cite{Srikanth2005} propose a test case prioritization approach that takes into consideration customer-assigned priorities of requirements, developer-perceived implementation complexity, requirements volatility, %(i.e., the amount of times a single requirement changed during development), 
and fault proneness of requirements. %(i.e., the number of faults affecting the source code that implements a single requirement). %Our approach uses information about test cases from their previous executions (e.g., fault proneness of requirements and requirements volatility). %to sort test cases for their execution on a new product in a product family (\textit{Challenge 2}). 
%The main difference between our approach and the aforementioned approaches is 
Tonella et al.~\cite{Tonella2006} propose a test case prioritization technique using user knowledge through a machine learning algorithm (i.e., Case-Based Ranking). Lachmann et al.~\cite{Lachmann2016b} propose another test case prioritization technique for system-level regression testing based on supervised machine learning. They consider test case history and natural language test case descriptions for prioritization. Since they consider the next version of a single system, their approach does not take into account variability and the classification of test cases in a product line for test case prioritization.
In contrast to the aforementioned approaches, we do aim at prioritizing test cases for a new product in a product family, not for the next version of a single system. Our approach considers multiple factors (i.e., test execution history, requirements variability, the classification of test cases and the size of scenarios exercised by test cases) in a product line, identifies their impact on the test case prioritization for the previous products in the product line, and prioritizes test cases for a new product accordingly. %(\textit{Challenge 2}). %For instance, the volatility of a requirement is calculated by analyzing the configuration decisions of the previous products while, for each new product in the product family, the customer needs to assign the priority of product requirements.

%~\cite{Fazeli2017}

%Alves et al.~\cite{Alves2016}

%Tahat et al.~\cite{Tahat2012} propose two model based criteria for the prioritization of regression test cases that rely on structural coverage and dependency analysis. The proposed structural coverage criterion simply gives priority to the set of test cases that cover modified model elements (edges) over the set of test that cover only unmodified part of the models, but do not further sort the test cases in the two sets. The dependency based criterion instead further prioritize test cases based on the impact that the changes had on the data and control dependency among the different nodes of the model covered by the test, test cases are sorted according to the amount of data and control dependencies that have been impacted by the changes. 

%There are some approaches for test case prioritization in the context of product lines
There are approaches that address test case prioritization in product lines (e.g.,~\cite{Runeson2012, Engstrom2013, Hajiaji2014, Hajjaji2017, Baller2014, Henard2014, Ensan2011, Devroey2017, Devroey2014, Hajjaji2017c, Hajjaji2017b, Lity2017}). 
%Most of them (e.g.,~\cite{Hajiaji2014}~\cite{Baller2014}~\cite{Henard2014}~\cite{Ensan2011}~\cite{Devroey2017} \cite{Devroey2014}~\cite{Hajjaji2017c}~\cite{Hajjaji2017b}~\cite{Lity2017}) sort product variants in a product family to maximize a certain testing criterion (e.g., the failure detection rate), but do not consider the prioritization of test cases per product. 
\MREVISION{R3.9}{For instance, to increase feature interaction coverage during product-by-product testing, similarity-based prioritization techniques incrementally select the most diverse products in terms of features to be tested~\cite{Henard2014, Hajiaji2014, Hajjaji2017}.} %selects the next product to be tested based on the similarities with previously tested product(s) in the product family. 
%Fabrizio-25.10: the following is somehow a "detail" not useful for discussion here
%The next product has always the minimum similarity with all the previously tested products with respect to feature selection. 
Baller et al.~\cite{Baller2014} propose an approach to prioritize products in a product family based on the selection of test suites with regard to cost/profit objectives. %The proposed approach uses an incremental heuristic for deriving a sequence of the products to be tested for approaching optimal profits under reduced costs. %Sanchez et al.~\cite{Sanchez2014} propose five different prioritization criteria based on common metrics of feature models to maximize the rate of early fault detection of an SPL test suite. The set of products to be tested are prioritized according to the proposed criteria. It is concluded that there are significant differences in the rate of early fault detection provided by different prioritization criteria. 
%In contrast to the aforementioned techniques applied to prioritize products to be tested, our prioritization approach prioritizes test cases for a new product in a product family based on multiple risk factors (\textit{Challenge 2}).
The aforementioned techniques prioritize the products to be tested, which is not useful in our context since products are seldom developed in parallel.
In contrast, our approach prioritizes the test cases of a new product to support early detection of software faults based on multiple risk factors. %(\textit{Challenge 2}).
%, which is a common need for companies developing product lines.}
%in a product family 

There are search-based approaches for multi-objective test case prioritization in product lines (e.g.,~\cite{Wang2014, Parejo2016, Arrieta2016, Arrieta2019}). For instance, Parejo et al.~\cite{Parejo2016} model test case prioritization as a multi-objective optimization problem and implement a search-based algorithm to solve it based on the NSGA-II evolutionary algorithm. Arrieta et al.~\cite{Arrieta2019} propose another approach that cost-effectively optimizes product line test process. None of them work based on information at the level of NL requirements. %None of these works consider test case classification and NL requirements as a factor to prioritize test cases.}

% !TEX root =  Main.tex
% !TEX root =  ../Main.tex
%\begin{table*}[tb]
\begin{sidewaystable}
\scriptsize
\caption{Summary and comparison of the related work.}
\label{table:RelatedWork}
%\vspace{-0.001cm}
\begin{tabular}{
|@{\hspace{0.02cm}}p{2.89cm}
|@{\hspace{0.05cm}}p{2.7cm} 
|@{\hspace{0.05cm}}p{1.5cm} 
|@{\hspace{0.05cm}}p{2.9cm} 
%|@{\hspace{0.05cm}}p{1.9cm} 
|@{\hspace{0.05cm}}p{2.7cm} 
%| @{\hspace{0.05cm}}p{2.0cm} 
|@{\hspace{0.05cm}}p{2.1cm} 
|@{\hspace{0.05cm}}p{2.65cm} |}
\hline
&\textbf{No need for behavioral models or source code for test case selection}&\textbf{Support for PL test case selection}&\textbf{No need to derive upfront all the test cases of PL for test case selection}&\textbf{No need for behavioral models or code for test case prioritization}&\textbf{Support for PL test case/product prioritization}
&\textbf{Support for prioritizing test cases in PL, not products in PL}\\
\hline

\textbf{Our Approach}& $+$ & $+$ & $+$ & $+$ & $+$ & $+$ \\
\hline

Wang et al.~\cite{Wang2016, Wang2017} & $+$ & $+$ & $-$ & $\mathit{NA}$ & $\mathit{NA}$ & $\mathit{NA}$ \\
\hline

Cabral et al.~\cite{Cabral2010} & $+$ & $+$ & $-$ & $\mathit{NA}$ & $\mathit{NA}$ & $\mathit{NA}$ \\
\hline

Knapp et al.~\cite{Knapp2014} & $+$ & $+$ & $-$ & $\mathit{NA}$ & $\mathit{NA}$ & $\mathit{NA}$ \\
\hline

Schurr et al.~\cite{Schurr2010} & $+$ & $+$ & $-$ & $\mathit{NA}$ & $\mathit{NA}$ & $\mathit{NA}$ \\
\hline

Briand et al.~\cite{Briand2009}& $-$ & $-$ & $\mathit{NA}$ & $\mathit{NA}$ & $\mathit{NA}$ & $\mathit{NA}$ \\
\hline
Hemmati et al.~\cite{Hemmati2010}& $-$ & $-$ & $\mathit{NA}$ & $\mathit{NA}$ & $\mathit{NA}$ & $\mathit{NA}$ \\
\hline

Rothermel et al.~\cite{Rothermel2000}& $-$ & $-$ & $\mathit{NA}$ & $\mathit{NA}$ &  $\mathit{NA}$ & $\mathit{NA}$ \\
\hline

Rothermel et al.~\cite{Rothermel1997}& $-$ & $-$ & $\mathit{NA}$ & $\mathit{NA}$ & $\mathit{NA}$ & $\mathit{NA}$ \\
\hline

Binkley~\cite{Binkley1997}& $-$ & $-$ & $\mathit{NA}$ & $\mathit{NA}$ & $\mathit{NA}$ & $\mathit{NA}$ \\
\hline

Harrold et al.~\cite{Harrold2001}& $-$ & $-$ & $\mathit{NA}$ & $\mathit{NA}$ & $\mathit{NA}$ & $\mathit{NA}$ \\
\hline

Qu et al.~\cite{Qu2011}& $-$ & $-$ & $\mathit{NA}$ & $\mathit{NA}$ & $\mathit{NA}$ & $\mathit{NA}$ \\
\hline

Kung et al.~\cite{Kung1995}& $-$ & $-$ & $\mathit{NA}$ & $-$ & $-$ & $\mathit{NA}$ \\
\hline

Muccini et al.~\cite{Muccini2006}& $-$ & $-$ & $\mathit{NA}$ & $\mathit{NA}$ & $\mathit{NA}$ & $\mathit{NA}$ \\
\hline

Vaysburg et al.~\cite{Vaysburg2002}& $-$ & $-$ & $\mathit{NA}$ & $\mathit{NA}$ &  $\mathit{NA}$ & $\mathit{NA}$ \\
\hline

Mirarab et al.~\cite{Mirarab2008}& $-$ & $-$ & $\mathit{NA}$ & $\mathit{NA}$ & $\mathit{NA}$ & $\mathit{NA}$ \\
\hline

Lity et al.~\cite{Lity2016, Lochau2014, Lity2012}& $-$ & $+$ & $+$ & $\mathit{NA}$ & $\mathit{NA}$ & $\mathit{NA}$ \\
\hline

%Riebisch et al.~\cite{riebisch2003uml}& $+$ & $+$ & $+$ &$+$ & $-$ & $-$ & $-$\\
%\hline

Lachmann et al.~\cite{Lachmann2015}& $\mathit{NA}$ & $\mathit{NA}$ & $\mathit{NA}$ & $-$ & $+$ & $+$ \\
\hline

Lachmann et al.~\cite{Lachmann2016}& $\mathit{NA}$ & $\mathit{NA}$ & $\mathit{NA}$ & $-$ & $+$ & $+$ \\
\hline

Lachmann et al.~\cite{Lachmann2017}& $\mathit{NA}$ & $\mathit{NA}$ & $\mathit{NA}$ & $-$ & $+$ & $+$ \\
\hline

%Hartmann et al.~\cite{Hartmann2005}& $ $ & $ $ & $ $ & $ $ & $ $ & $ $ & $ $\\
%\hline

Henard et al.~\cite{Henard2014}& $\mathit{NA}$ & $\mathit{NA}$ & $\mathit{NA}$ & $+$ & $+$ & $-$ \\
\hline

Al-Hajiaji et al.~\cite{Hajiaji2014, Hajjaji2017}& $\mathit{NA}$ & $\mathit{NA}$ & $\mathit{NA}$ & $+$ & $+$ & $-$ \\
\hline

Baller et al.~\cite{Baller2014}& $\mathit{NA}$ & $\mathit{NA}$ & $\mathit{NA}$ & $+$ & $+$ & $-$ \\
\hline

Shrikanth et al.~\cite{Srikanth2005, Srikanth2012}& $\mathit{NA}$ & $\mathit{NA}$ & $\mathit{NA}$ & $+$ & $-$ & $\mathit{NA}$ \\
\hline

Tonella et al.~\cite{Tonella2006}& $\mathit{NA}$ & $\mathit{NA}$ & $\mathit{NA}$ & $+$ & $-$ & $\mathit{NA}$ \\
\hline

Lachmann et al.~\cite{Lachmann2016b}& $\mathit{NA}$ & $\mathit{NA}$ & $\mathit{NA}$ & $+$ & $-$ & $\mathit{NA}$ \\
\hline

Sanchez et al.~\cite{Sanchez2011}& $\mathit{NA}$ & $\mathit{NA}$ & $\mathit{NA}$ & $+$ & $-$ & $\mathit{NA}$ \\
\hline

Hemmati et al.~\cite{Hemmati2017}& $\mathit{NA}$ & $\mathit{NA}$ & $\mathit{NA}$ & $+$ & $-$ & $\mathit{NA}$ \\
\hline

Lachmann et al.~\cite{Lachmann2016b}&$\mathit{NA}$ & $\mathit{NA}$ & $\mathit{NA}$ & $+$ & $-$ & $\mathit{NA}$ \\
\hline

%e.g.,~\cite{Srikanth2014, Srikanth2016, Srikanth2012, , , }

Krishnamoorthi et al.~\cite{Krishnamoorthi2009}&$\mathit{NA}$ & $\mathit{NA}$ & $\mathit{NA}$ & $+$ & $-$ & $\mathit{NA}$ \\
\hline

Arafeen et al.~\cite{Arafeen2013}&$\mathit{NA}$ & $\mathit{NA}$ & $\mathit{NA}$ & $-$ & $-$ & $\mathit{NA}$ \\
\hline

%~\cite{, Kundu2009, , }

Haidry and Miller~\cite{Haidry2013}&$\mathit{NA}$ & $\mathit{NA}$ & $\mathit{NA}$ & $+$ & $-$ & $\mathit{NA}$ \\
\hline

Korel et al.~\cite{Korel2008}&$\mathit{NA}$ & $\mathit{NA}$ & $\mathit{NA}$ & $-$ & $-$ & $\mathit{NA}$ \\
\hline

Kundu et al.~\cite{Kundu2009}&$\mathit{NA}$ & $\mathit{NA}$ & $\mathit{NA}$ & $-$ & $-$ & $\mathit{NA}$ \\
\hline

Tahat et al.~\cite{Tahat2012}&$\mathit{NA}$ & $\mathit{NA}$ & $\mathit{NA}$ & $-$ & $-$ & $\mathit{NA}$ \\
\hline

\end{tabular}
\\
\\

\vspace{-7cm}
%\end{table*}%
\end{sidewaystable}

Lachmann et al.~\cite{Lachmann2015} introduce a test case prioritization technique for product lines using delta-oriented architecture models. %The approach identifies changed parts of the architecture models of products by computing \textit{deltas}, i.e., modifications on products. 
The differences between products are captured in the form of \textit{deltas}~\cite{Clarke2010}, which are modifications between architecture models of products used for integration testing. 
%The proposed approach uses the delta information to rank the test cases for a product to be tested. The basic idea is to rank test cases higher if they cover more changed elements in the architecture model of the product. 
The proposed approach ranks test cases based on the number of changed elements in the architecture.
The approach first identifies the regression deltas specifying the differences between architecture models. For every product, it creates a delta graph, which is later used to compute the degree of changes between the current product under test and the previously tested products. To prioritize test cases, the approach computes the change of each architecture component using the delta graph of the current product under test. The higher the corresponding change, the more likely is the test case to fail.
The approach is later extended using risk factors~\cite{Lachmann2017} and behavioral knowledge of architecture components~\cite{Lachmann2016}. %to achieve a more effective test case prioritization. 
%Fabrizio-25.10: removing grey-box which was not intrioduced earlier
%Lachmann et al. consider grey-box testing in which the approach needs access to product architecture descriptions and to some limited information about component behavior. In contrast, our approach focuses on white-box testing and does not require any design information. It prioritizes system test cases in the context of use case-driven development (\textit{Challenge 2}).
The approach proposed by Lachmann et al. requires access to product architecture descriptions and information about component behavior. In contrast, we do not require any design information but rely on NL requirements specifications, i.e., use case specifications. %(\textit{Challenge 2}).
Our approach does not need to identify any regression delta between requirements of previous and current products. We, instead, rely on logistic regression models that use variability information in PL use case models, the classification of test cases, test execution history and test scenario characteristics. \MREVISION{R4.10}{Compared to the delta-oriented approach, we rely on textual requirements and test case execution history without requiring detailed design models, which rarely exist in industrial settings. For instance, IEE does not produce the detailed design models that the delta-oriented approach requires to prioritize test cases for product lines. Therefore, we expect our approach to be more widely applicable in industrial settings.}

\MREVISION{R4.6, R4.10}{In Table~\ref{table:RelatedWork}, based on a set of features necessary for the selection and prioritization of system test cases in product lines, we summarize the differences between our approach and the closest related work.
For each approach, the symbol '+' indicates that the approach provides a feature, the symbol '-' indicates that it does not do so,
and 'NA' indicates that the feature is out of scope. For instance, the approach by Lachmann et al.~\cite{Lachmann2016b} automatically prioritizes system test cases, but does not classify test cases. Therefore, all the features related to the classification of system test cases are not considered for Lachmann et al.~\cite{Lachmann2016b} in Table~\ref{table:RelatedWork}. 
Some of the existing test case selection approaches do not support product lines~\cite{Binkley1997, Rothermel1997, Rothermel2000, Vaysburg2002, Mirarab2008, Muccini2006}. The approaches that support product lines need either detailed behavioral models~\cite{Lity2016, Lochau2014, Lity2012} or must derive upfront all the test cases of a product line~\cite{Wang2016, Wang2017, Cabral2010, Knapp2014, Schurr2010}.
On the other hand, PL prioritization approaches either need detailed behavioral models~\cite{Lachmann2015, Lachmann2017, Lachmann2016} or prioritize products in a product line, not system test cases~\cite{Henard2014, Hajiaji2014, Hajjaji2017, Baller2014}. Our approach is currently the only approach that automatically classifies and prioritizes system test cases in PL without requiring neither behavioral models nor the upfront provision of all test cases. This is enabled by the capability of automatically analyzing system requirements in NL in the form of use case specifications. }

\MREVISION{R4.6}{There is work using industrial cases studies to evaluate their PL test case classification and prioritisation techniques. However, our work is driven by current practice and its limitations, together with working assumptions, in a specific domain that was not addressed by existing work: use case driven development of embedded, safety-critical systems.}

%
% !TEX root =  Main.tex
\section{Overview of the Approach}
\label{sec:overview}

\begin{wrapfigure}{h}{0.46\linewidth}
\vspace*{-1.75em}
\caption{Overview of the Approach}\label{fig:overview}
%\vspace*{-1em}
\centerline{\includegraphics[width=\linewidth]{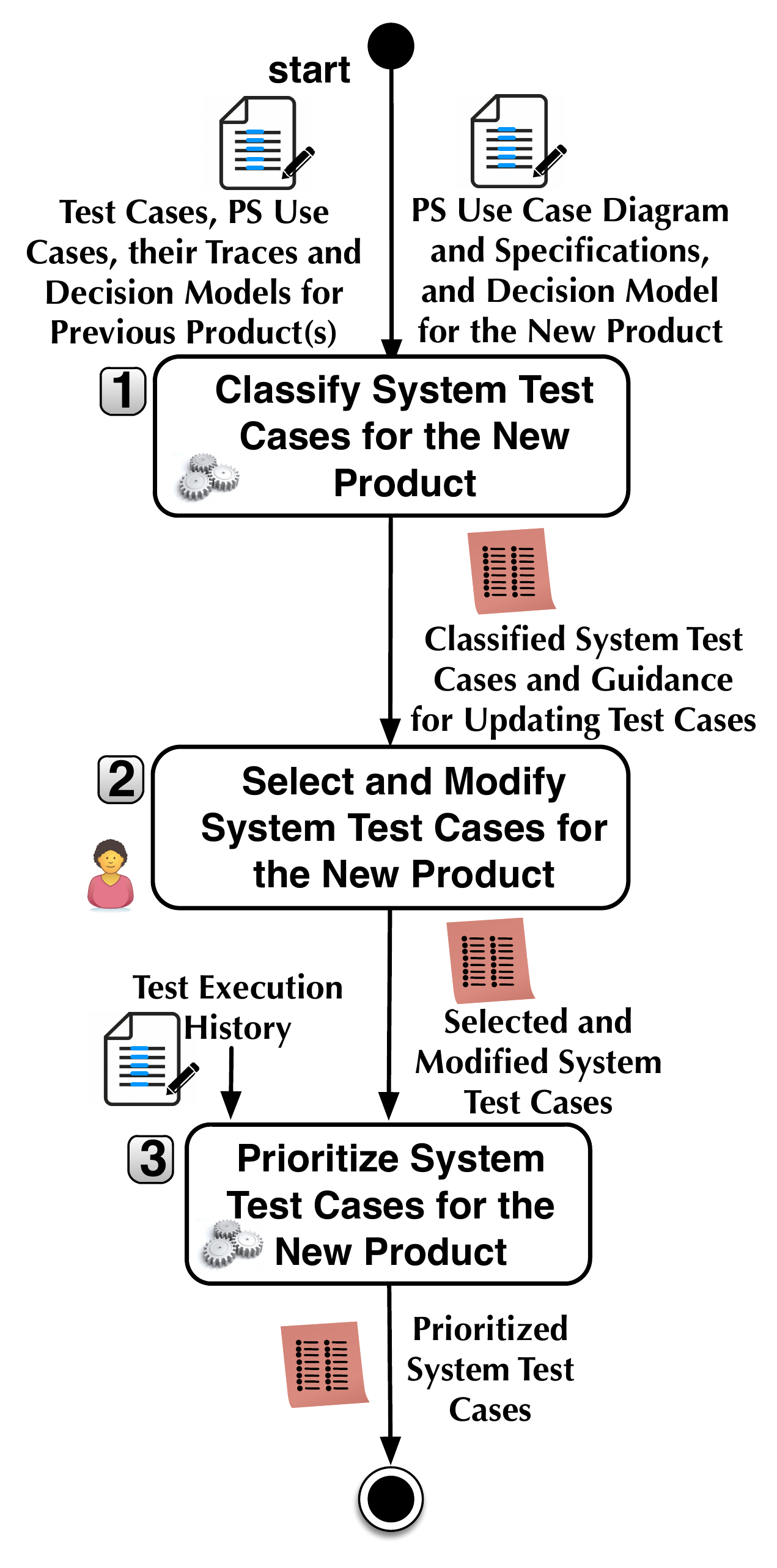}}
\vspace*{-1.7em}
\end{wrapfigure}

The process in Fig.~\ref{fig:overview} presents an overview of our approach. %In Step 1, \textit{Configure product specific use case models for a new product}, the analyst is asked %, for a new product, 
%to input configuration decisions regarding variation points captured in PL use case models. %The PS use case models are automatically generated from the PL use case models and the configuration decisions. 
%Step 1 is semi-automated. It takes as input (1) a \textit{PL use case diagram}, which captures variability, and its constraints and dependencies, and (2) \textit{PL use case specifications}, which detail the variability information captured in the PL use case diagram. The analyst is guided to make configuration decisions on variability information in the PL use case models and the PS use case models are automatically generated from the PL models and the configuration decisions as output. Step 1 relies on our product line modeling and configuration approaches proposed and assessed in our previous work~\cite{Hajri2015}~\cite{Hajri2016c}~\cite{Hajri2016b}. 
In Step 1, \textit{Classify system test cases for the new product}, our approach takes as input (i) system test cases, PS use case models, their traceability links, and configuration decisions for previous product{s} in the product family, and (ii) PS use case models and configuration decisions for the new product, to classify the system test cases for the new product as \textit{obsolete}, \textit{retestable}, and \textit{reusable}, and to provide information on how to modify obsolete system test cases to cover new, untested use case scenarios. %(\textit{Challenge 1}). 

Step 1 is fully-automated. The classification and modification information in output of this step is for the test engineer to decide which test cases to execute for the new product and which modifications to make on the obsolete test cases to cover untested, new use case scenarios. We give the details of this step in Section~\ref{sec:selection}.

In Step 2, \textit{Select and modify system test cases for the new product}, by using the classification information and modification guidelines automatically provided by our approach, the engineer decides which test cases to run for the new product and modifies obsolete test cases to cover untested, new use case scenarios. %The test case classification our approach provides in Step 2 is based on the impact of requirements changes on system test cases. 
The activity is not automated because, %for the final decision of 
for the selection of system test cases, the engineer may also need to consider implementation and hardware changes (e.g., code refactoring and replacing some hardware with less expensive technology) in addition to the classification information provided in Step 2, which is purely based on changes in functional requirements. %, which are simply design decisions not necessarily driven by requirements changes. 
For instance, a reusable test case might need to be rerun because part of the source code verified by the test case is refactored. %In addition, in order to ensure the safety for highly safety-critical systems, the test execution strategy the company follows may enforce all the reusable test cases to be rerun. 
 
%For the final decision of the selection of the system test cases, the engineer may also need to consider implementation and hardware changes, which are simply design decisions not necessarily driven by requirements changes.} 

%In Step 4, \textit{Prioritize system test cases for the new product}, the system test cases are automatically prioritized for the execution of system test cases in the new product using multiple factors given in the test execution history as input, such as fault proneness of requirements, requirements volatility in the product line, and customer priority (\textit{Challenge 2}). % the distribution of test case failures, the number of products in which the test case is executed, and the number of products in which the test case failed (\textit{Challenge 3}). 

In Step 3, \textit{Prioritize system test cases for the new product}, selected test cases are automatically prioritized based on %historical test execution data 
risk factors including fault proneness of requirements, and requirements volatility. %(\textit{Challenge 2}). % the distribution of test case failures, the number of products in which the test case is executed, and the number of products in which the test case failed (\textit{Challenge 3}). 
We discuss this step in Section~\ref{sec:prioritization}. %Steps 2 and 4 are the main focus of this paper.  

%
% !TEX root =  Main.tex
\section{Classification of System Test Cases} %in a Product Family}
\begin{figure}[t]
%\vspace*{-0.5em}
\centerline{\includegraphics[width=1.0\linewidth]{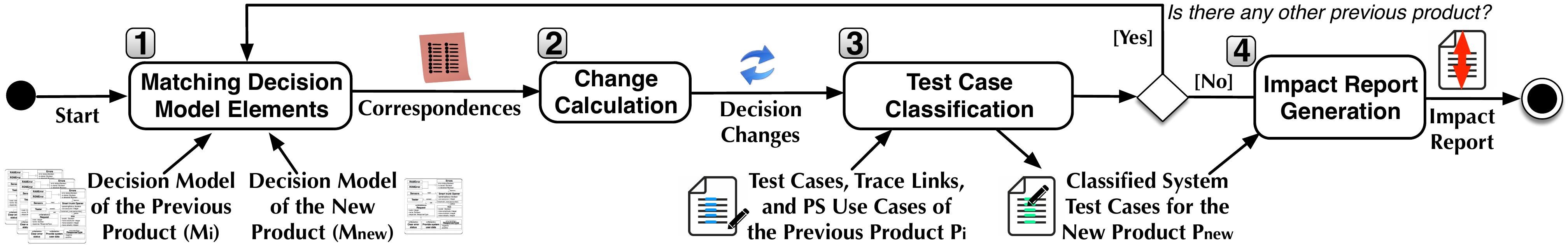}}  
%\vspace*{-0.8em}
%\caption{\protect\scalebox{.95}{Overview of the Model Differencing and Test Case Classification Pipeline}}\label{fig:Pipeline}
\caption{Overview of the Model Differencing and Test Case Classification Pipeline}\label{fig:Pipeline}
%\textcolor{red}{IT WOULD BE BETTER TO HAVE M1.. Mn inputs to the pipeline, and then iterate over Mi and Mn}
%\vspace*{-0.8em}
\end{figure}

\label{sec:selection}
%After PS use cases are configured for a new product, system test cases from the existing test suite(s) of the prior product(s) in the product line are classified to be selected for the new product. 
The test case classification is implemented as a pipeline (see Fig.~\ref{fig:Pipeline}), which takes as input the configuration decisions made for the previous products, the configuration decisions made for the new product, and the previous product's system test cases, traceability links, and PS use case models. The pipeline produces an impact report with the list of existing test cases classified.

\begin{figure*}[!t]
\centerline{\includegraphics[width=0.980\linewidth]{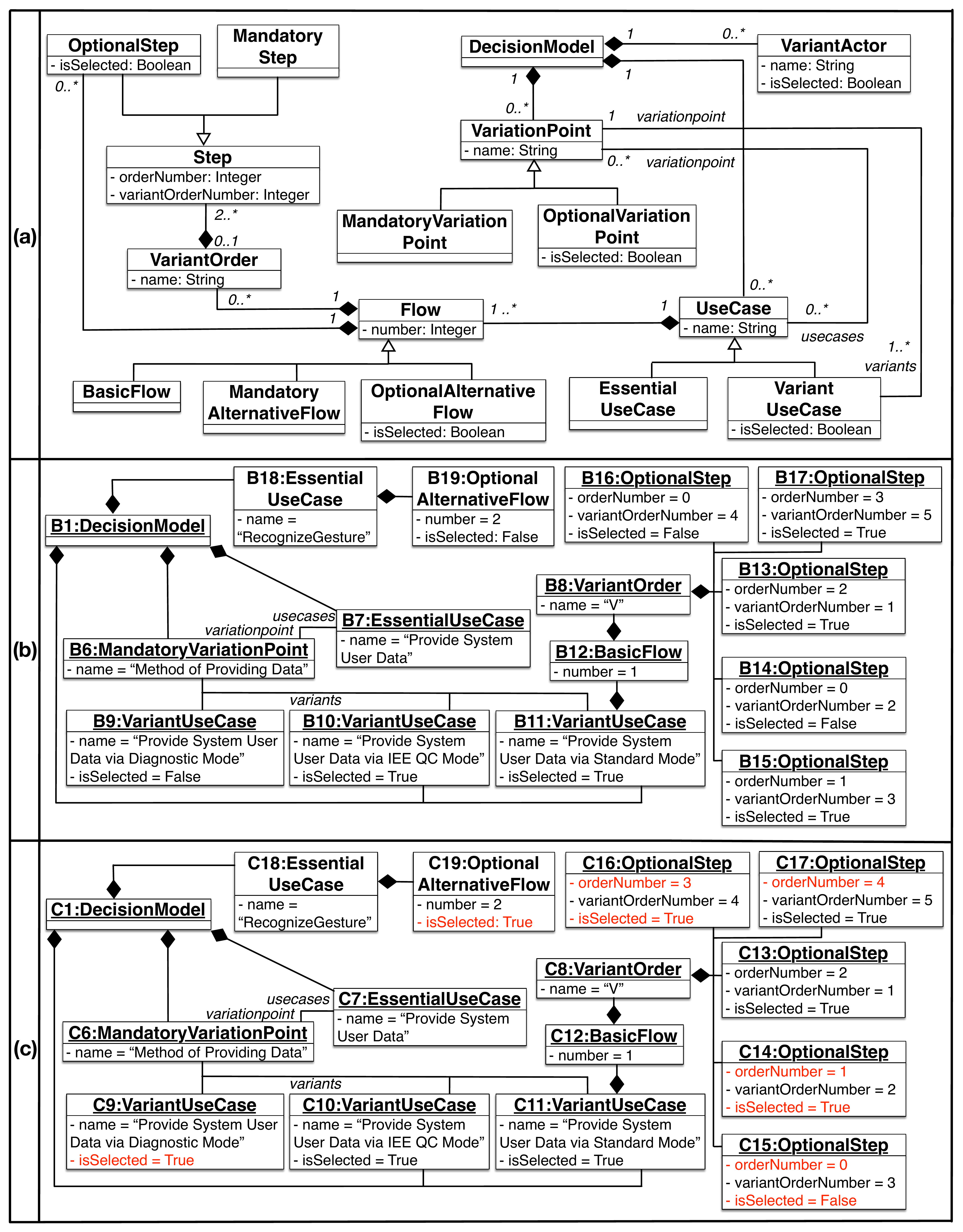}}
%\vspace*{-0.9em}
\caption{(a) Decision Metamodel, (b) Part of the Example Decision Model of the Previous Product ($\mathit{M_{i}}$), and (c) Part of the Example Decision Model of the New Product ($\mathit{M_{new}}$)}\label{fig:decisionModels}
%\vspace*{-2.0em}
\end{figure*}

Configuration decisions are captured in a decision model that is automatically generated by PUMConf during the configuration process. 
The decision model conforms to a decision metamodel described in our previous work~\cite{Hajri2016c}. The metamodel includes the main use case elements for which the user makes decisions (i.e., variation points, optional steps, optional alternative flows, and variant orders). PUMConf keeps a decision model for each configuration in the product line.  
Fig.~\ref{fig:decisionModels} provides the decision metamodel and two decision models for the PL use case models in Fig.~\ref{fig:productlineDiagram} and Table~\ref{tab:useCaseRUCM}.

%The pipeline takes the decision models (\textit{M1} and \textit{M2} in Fig.~\ref{fig:Pipeline}), and the prior product's system test cases, trace links and PS use case models as input. 
%Fabrizio-25.10: I have difficulties understanding the following sentence, is enough what I have added at the beginning of teh section?
%The system test cases are classified to be selected, as output, together with an impact report, i.e., a list of the impacts of the use case changes on the system test cases. %(\textit{Challenge 1}). 
The pipeline has four steps (see Fig.~\ref{fig:Pipeline}). The first three steps are executed for each of the $\mathit{n}$ previous products in the product line, where each one has a decision model $\mathit{M_{i}}$ with $\mathit{i = 1..n}$. Note that we also employ the first two steps of the pipeline in our previous work~\cite{Hajri2018, Hajri2017a}. %to incrementally reconfigure PS use case models for evolving configuration decisions.
In Step 1, \textit{Matching decision model elements}, our approach automatically executes the structural differencing of $\mathit{M_{i}}$ and $\mathit{M_{new}}$ by looking for corresponding model elements representing decisions for the same variations (see Section~\ref{subsec:matching_decisions}).

%In Step 1, \textit{Matching decision model elements}, our approach automatically does the structural differencing of \textit{M1} and \textit{M2} by looking for the correspondences in \textit{M1} and \textit{M2}. %To that end, we devise an algorithm that identifies the matching model elements in \textit{M1} and \textit{M2}. 
%The output of Step 1 is a set of pairs of corresponding model elements, representing decisions for the same variations, in \textit{M1} and \textit{M2} (Section~\ref{subsec:matching_decisions}).

 %In a variation point, the user selects variant use cases to be included for the product. For PL use case specifications, the user selects optional steps and optional alternative flows to be included and determines the order of steps (variant order). %Therefore, the matching elements in Step 1 are the pairs of variation points and use cases including the variation points, the pairs of use cases and optional alternative flows in the use cases, and the triples of use cases, flows in the use cases, and optional steps in the flows.   

\begin{wraptable}{r}{4.00cm}
\vspace*{-1.80em}
\centering
%\footnotesize
\scriptsize
\caption{Change Types for Configuration Decisions}
%\vspace*{-0.9em}
\label{tab:changeTypes}
\scalebox{1.15}{
\begin{tabular}{|p{2.90cm}|}
\hline
\multicolumn{1}{|c|}{\textbf{Change Types}}                                                                                                  \\ \hline
\textbf{.} Add a decision                                                                                                                               \\ \hline
\textbf{.} Delete a decision                                                                                                                            \\ \hline
\textbf{.} Update a decision                                                                                                                            \\ \hline
\begin{tabular}[c]{@{}l@{}}\,\ - Select some unselected \\   \,\  variant element(s)\end{tabular}                                                       \\ \hline
\begin{tabular}[c]{@{}l@{}}\,\ - Unselect some selected \\   \,\ variant element(s)\end{tabular}                                                       \\ \hline
%Fabrizio-26.10: This is not good because you have a type of change that its simply the merge of two changes. It does not make any sense. If the reason why you have this is because some variant element present a multiple choiche, then call it change a multiple choice option.
\begin{tabular}[c]{@{}l@{}}\,\ - Unselect some selected\\  \,\ variant element(s) and \\  \,\  select some unselected\\  \,\ variant element(s)\end{tabular} \\ \hline
\begin{tabular}[c]{@{}l@{}}\,\ - Change order number\\   \,\  of variant step order(s)\end{tabular}                                                       \\ \hline
\end{tabular}
}
\vspace*{-1.05em}
\end{wraptable}

%Fabrizio-25.10: you should try to minimize the number of keywords introduced. "decision-level changes" does not mean much to me, when I read it I start thinking that  there are multiple level of changes considered in the approach. What is a "corresponding element of a change" ? It's not clear.
In Step 2, \textit{Change calculation}, the approach determines %how the two product configurations differ. More specifically, it determines 
how configuration decisions of the two products differ. %and characterizes the difference between the two.}
%Fabrizio-25.10: Can't we replace 'change type' with 'diff type' ? Because after all we are comparing models that are not necessary derived one from the other. I  might understand that 'change types' sounds more interesting than 'diff type'.
%No we cannot because we already take the difference in Step 1. What we do in Step 2 is interpreting those differences in terms of changes in decisions.
Table~\ref{tab:changeTypes} lists the types of decision changes. %In Table~\ref{tab:changeTypes}, 
%We use the term \emphx{decision} for the resolution of variations (i.e., \emphx{Variant Use Case}, \emphx{Optional Alternative Flow}, and \emphx{Optional Step}). 
A decision is represented by means of a n-tuple of model elements in a decision model, e.g., $<$variation point VP, use case UC including VP$>$. A change is of type ``Add a decision'' when a tuple representing a decision in $\mathit{M_{new}}$ has no matching tuple in $\mathit{M_{i}}$.  A change is of type ``Delete a decision'' when a tuple representing a decision in $\mathit{M_{i}}$ has no matching tuple in $\mathit{M_{new}}$. A change is of type ``Update a Decision'' when a tuple representing a decision in $\mathit{M_{i}}$ has a matching tuple in $\mathit{M_{new}}$ with non-identical attribute values (see the red-colored attributes in Fig.~\ref{fig:decisionModels}(c)).

%A decision is about selecting, for the product, variant use cases in the variation point.

%In Step 2, \textit{Change calculation}, decision-level changes are identified from the corresponding model elements (see Section~\ref{subsec:matching_decisions}). Table~\ref{tab:changeTypes} lists the types of decision-level changes.  
%A set of elements in \textit{M1} which does not have a corresponding set of elements in \textit{M2} is considered to be a deleted decision, which we refer to as ``Delete a Decision''. 
%Analogously, a set of model elements in \textit{M2} which does not have a corresponding set of elements in \textit{M1} is considered to be an added decision (``Add a Decision''). A set of corresponding model elements with non-identical attribute values (see the red-colored attributes in Fig.~\ref{fig:decisionModels}(c)) is considered to be a change of the type ``Update a Decision''. 

%In Step 3, \textit{Test case classification}, system test cases of the previous product are classified to be selected for the new product (see Section~\ref{subsec:classification}). The trace links between the system test cases and the PS use cases and the configuration decision changes obtained from Step 2 are used to classify the system test cases for the new product. %For a given set of system test cases verifying a use case, each test case executes a scenario in the use case.  
In Step 3, \textit{Test case classification}, the system test cases of the previous products are classified for the new product by using the decision changes obtained from Step 2 and the traceability links between the system test cases and the PS use case specifications (see Section~\ref{subsec:classification}).
%Fabrizio-25.10: 'path' is not good
%A use case has multiple `paths' that can be taken by an actor at any one time. 
%A use case scenario is a single path through the use case. A system test case exercises a use case scenario. 
A use case can describe multiple use case scenarios (i.e., sequences of use case steps from the start to the termination of the use case) because of the presence of conditional steps. Each system test case is expected to exercise one use case scenario.
%For instance, there are two use case scenarios in the use case \textit{Provide System User Data} in Table~\ref{tab:ExamplePSUseCase}. 
%%Fabrizio-25.10: This section is supposed to provide an overview, we cannot give these details
%Please note that the number of basic and alternative flows in a use case is not necessarily the same with the number of scenarios since a scenario may include multiple flows. 
%%Fabrizio-25.10: something wrong here, first you need to determine which is the corresponding test case to classify it, no? Please FIX IT
%%Fabrizio-25.10: Also, the following is somehow misleading. We classify every test case, even if they are not impacted by changes. Those are simple reusable.
%Arda: We do not say we do not classify those test cases. We just say that if there is any change we calculate the impact of the change on those test cases. By defaut, it is clear that if there is no change, there is no impact.
For each use case of the new product, we identify the impact of the decision change(s) on the use case scenarios, i.e., any change in the execution sequence of the use case steps in the scenario.

%For instance, ???. 
%To this end, we automatically extract, from a use case specification, a use case scenario model that makes the implicit control flow in the specification explicit. We take a use case scenario model and test cases of the product to be classified, and identify the tested use case scenarios in the model. %For each tested scenario, 
%We check whether a decision change modifies the execution sequence of the use case steps in each tested scenario. %To derive use case scenarios of the new product that have not been tested in the previous product, we apply the decision changes on the impacted scenarios of the previous product.  

%Based on the impact of the configuration decision changes on the use case scenarios, we classify the system test cases in the test suite of the previous product. 
A system test case is classified in one of three categories: \textit{obsolete}, \textit{retestable} and \textit{reusable}. A test case is obsolete if it exercises an invalid execution sequence of use case steps in the new product. %of the corresponding use case scenario for the new product. 
%A retestable test case is a test case which remains valid in the new version of the corresponding use case scenario in terms of the sequence of the use case steps except the output steps, i.e., use case steps describing the interaction from the system to the actors. 
A test case is retestable if it exercises an execution sequence of use case steps that has remained valid in the new product, except for internal steps representing internal system operations (e.g., reset of counters). %output steps, i.e., use case steps describing an interaction from the system to actors. 
A test case is reusable if it exercises an execution sequence of use case steps that has remained valid in the new product. %in the new version of the use case scenario in the new product. 
The test case categories are mutually exclusive. %Therefore, any system test case in the test suite of the previous product is either obsolete, retestable or reusable. 
Use case scenarios of the new product that have not been tested for the previous product are reported as new use case scenarios. %Some of the obsolete and retestable test cases can be modified to cover those scenarios. Our approach automatically derives changes for obsolete and retestable test cases from configuration changes. %for use case specifications. 
%To do so, we map configuration decision changes to test case changes. 

In Step 4, \textit{Impact report generation}, we automatically generate an impact report from the classified test cases of each previous product to enable engineers to select test cases from more than one test suite (see Section~\ref{subsec:impact_report_generation}). Steps 1, 2 and 3 are the pairwise comparison of each previous product with the new product. 
%%Fabrizio-25.10: SHouldn't we say that we also produce a merged report?
If there are multiple previous products ($\mathit{n > 1}$ in Fig.~\ref{fig:Pipeline}), test cases of each product are classified separately in $\mathit{n}$ reports in Step 3. The generated impact report compares these $\mathit{n}$ separate reports and lists sets of new scenarios and reusable and retestable test cases for the $\mathit{n}$ previous products.

\subsection{Steps 1 and 2: Model Matching and Change Calculation}
\label{subsec:matching_decisions}

For the first two pipeline steps %, \textit{Matching decision model elements} and \textit{Change calculation}, 
in Fig.~\ref{fig:Pipeline}, 
we rely on a model matching and change calculation algorithm we devised in our prior work~\cite{Hajri2017a, Hajri2018}. In this section, we provide a brief overview of the two steps and their output for the example decision models in Fig.~\ref{fig:decisionModels}(b) and (c). %For the algorithm, the reader is referred to our prior work~\cite{Hajri2018}~\cite{Hajri2017a}.

%\begin{wraptable}{r}{7.10cm}
\begin{table}
%\vspace*{-1.10em}
\centering
%\footnotesize
\scriptsize
\caption{Matching Decisions in $\mathit{M_{i}}$ and $\mathit{M_{new}}$ in Fig.~\ref{fig:decisionModels}}
%\vspace*{-1.2em}
\label{tab:Step1Output}
\scalebox{1.28}{
\begin{tabular}{|c|c|}
\hline
\textbf{Decisions in $\mathit{M_{i}}$}               & \textbf{Decisions in $\mathit{M_{new}}$}             \\ \hline
\textless B6, B7 \textgreater{}        & \textless C6, C7 \textgreater{}        \\ \hline
\textless B18, B19 \textgreater{}      & \textless C18, C19 \textgreater{}      \\ \hline
\textless B11, B12, B13 \textgreater{} & \textless C11, C12, C13 \textgreater{} \\ \hline
\textless B11, B12, B14 \textgreater{} & \textless C11, C12, C14 \textgreater{} \\ \hline
\textless B11, B12, B15 \textgreater{} & \textless C11, C12, C15 \textgreater{} \\ \hline
\textless B11, B12, B16 \textgreater{} & \textless C11, C12, C16 \textgreater{} \\ \hline
\textless B11, B12, B17 \textgreater{} & \textless C11, C12, C17 \textgreater{} \\ \hline
\end{tabular}
}
%\vspace*{-1.25em}
\end{table}

%Fabrizio-27.10: There is a problem with the previous definition based on sets. Sets are not sorted. Check if you agree with presenting them as follows. Maybe is not consistent with the previous paper, but at least is easier to understand.} 
%Fabrizio-27.10: Another thing, I tried to clarify why we capture model elements with tuples because is a piece of information to provide at this stage. I assume we rely on it in later stages of the paper. 
%Fabrizio-27.10: I do not know what you did for the other paper. But conceptually is somehow wrong. You should have said that first you identify model elements that correspond to variations, then you identify added and removed elements, finally for the remaining elements you identify the corresponding ones and classify them. I leave as is because based on previous work but complicates the understanding.
In Step 1, we identify %the model elements that correspond to decisions. Also, we build 
pairs of decisions in $\mathit{M_{i}}$ and $\mathit{M_{new}}$ that are made for the same variants. %In such pairs, one element belongs to the decision model for the previous product, the other to the decision model for the new product. 
 %Each model element is represented by means of a n-tuple of identifiers appearing in the decision model that univocally identify the decision. 
 %Variation points and optional alternative flows are captured by pairs of identifiers, one referring to the variation point itself (or to the optional alternative flow), and one to the use case including the variation point. 
The decision metamodel in Fig.~\ref{fig:decisionModels}(a) includes the main use case elements for which the user makes decisions (i.e., variation point, optional step, optional alternative flow, and variant order). In a variation point included by a use case\footnote{In PL use case diagrams, use cases are connected to variation points with an include dependency.}, the user selects variant
use cases to be included for the product. For PL use case specifications, the user selects optional steps and alternative
flows to be included and determines the order of steps (variant order). Therefore, the matching decisions in Step 1 are (i) the pairs of variation points and use cases including the variation points, (ii) the pairs of use cases and optional alternative flows in the use cases, and (iii) the triples of use cases, flows in the use cases, and optional steps in the flows.
Table~\ref{tab:Step1Output} shows some decisions in Fig.~\ref{fig:decisionModels}(b) and (c). For example, the pairs $\langle B6, \, B7\rangle$ and $\langle C6, \, C7\rangle$ represent two decisions for the variation point \emphx{Method of Providing Data} included in the use case \emphx{Provide System User Data}. The triples $\langle B11,\, B12, \, B13 \rangle$ and $\langle C11,\, C12, \, C13 \rangle$ represent two decisions for an optional step in the basic flow of the use case \emphx{Provide System User Data via Standard Mode} (i.e., for V2 in Line 40 in Table~\ref{tab:useCaseRUCM}).

In Step 2, \textit{Change Calculation}, we first identify deleted and added configuration decisions by checking tuples of model elements in one input decision model ($\mathit{M_{i}}$) which do not have any matching tuples of model elements in another input decision model ($\mathit{M_{new}}$).  %To identify updated decisions, we then look for tuples of model elements in $\mathit{M_{i}}$ that match tuples of model elements in $\mathit{M_{new}}$ with non-identical attribute values. %(i.e., corresponding model elements that present non identical attribute values). %and classify the differences between them.}
To identify updated decisions, we check tuples of model elements in $\mathit{M_{i}}$ that have matching tuples of model elements in $\mathit{M_{new}}$ with non-identical attribute values. The matching pairs of variation points and their including use cases represent decisions for the same variation point
(e.g., $\langle B6, \, B7\rangle$ and $\langle C6, \, C7\rangle$ in Table~\ref{tab:Step1Output}). If the selected variant use cases for the same variation point are not the
same in $\mathit{M_{i}}$ and $\mathit{M_{new}}$, the decision in $\mathit{M_{i}}$ is considered as updated in $\mathit{M_{new}}$. We have similar checks for optional steps, optional alternative flows and variant order of steps.
For instance, an optional step is selected in the decision represented by the triple $\langle B11,\, B12, \, B15 \rangle$ in  $\mathit{M_{i}}$, while the same optional step is unselected in the decision represented by the matching triple $\langle C11,\, C12, \, C15 \rangle$ in  $\mathit{M_{new}}$. For the decision models in Fig.~\ref{fig:decisionModels}, the decisions represented by $\langle B6, \, B7\rangle$, $\langle B18, \, B19 \rangle$, $\langle B11, \, B12, \, B14 \rangle$, $\langle B11, \, B12 \, B15 \rangle$, $\langle B11,\, B12, \, B16 \rangle$, and $\langle B11, \, B12, \, B17 \rangle$ are identified as \textit{updated}. There are no deleted or added decisions for the models in Fig.~\ref{fig:decisionModels}.

\subsection{Step 3: Test Case Classification}
\label{subsec:classification}

System test cases of the previous product are automatically classified based on the identified changes (Step 3 in Fig.~\ref{fig:Pipeline}). 
To this end, we devise an algorithm (see Fig.~\ref{algo:classification})
which takes as input a set of use cases (\emphx{UC}), the test suite of the previous product (\emphx{ts}), and a triple of the sets of configuration changes (\emphx{dc}) detected in Step 2. It classifies the test cases and reports use case scenarios of the new product that are not present in the previous product. %that are not present in the previous product but can be identified from the use case specifications of the new product. 

%\begin{wrapfigure}{h}{0.610\linewidth}
\begin{figure}[t]
%\vspace*{-0.2cm}
\begin{center}
\scalebox{1.02}{\parbox{\linewidth}{\footnotesize
%\hbox{{\bf Algorithm}\;{\sc Diff} } 
%\vspace*{-.36cm}
\begin{tabbing}
Output:x \= $\mathit{Block\_Slice}_r$:abit \=\kill
{\bf Input:} Set of use case specifications of the previous product $\mathit{UC}$, \\ \,\,\,\,\,\,\,\,\,\,\,\,\,\,\,\,\,\,\,\ Test suite of the previous product $\mathit{ts}$,\\ \,\,\,\,\,\,\,\,\,\,\,\,\,\,\,\,\,\,\,\ Triple of sets of decision-level changes $\mathit{dc}$ \\ \,\,\,\,\,\,\,\,\,\,\,\,\,\,\,\,\,\,\,\ (ADD, DELETE, UPDATE) \\
{\bf Output:} Quadruple of sets of classified test cases $\mathit{classified}$ \\ % \\\>(OBSOLETE, REUSE, RETEST, NEW) \\
\> \\
\end{tabbing}
\vspace*{-1.1cm}
\begin{tabbing}
100.\= if \= if \= if \= if \= if \= if \kill
\> \\
1.\> Let $\mathit{OBSOLETE}$ be the empty set for obsolete test cases\\
2.\> Let $\mathit{REUSE}$ be the empty set for reusable test cases\\
3.\> Let $\mathit{RETEST}$ be the empty set for retestable test cases\\
4.\> Let $\mathit{NEW}$ be the empty set for new use case scenarios\\
5.\> Let $\mathit{classified}$ be the quadruple (OBSOLETE, REUSE, RETEST, NEW)\\
6. \> \textbf{foreach} {$(\mathit{u} \in \mathit{UC})$} \textbf{do} \\
7. \>\> \textbf{if} {(there is a change in $\mathit{dc}$ for $\mathit{u}$)} \textbf{then}\\
8. \>\>\> $\mathit{model} \leftarrow$ \textbf{generateUseCaseModel}($\mathit{u}$)\;\\
%9.\>\>\> Let $\mathit{inst}$ be the \textit{UseCaseStart} instance in $\mathit{model}$\\
%10.\>\>\> Let $\mathit{sc}$ be the empty list for tested scenario to be identified in $\mathit{model}$\\
9.\>\>\> Let $\mathit{u_{new}}$ be a new version of $\mathit{u}$ after the changes in $\mathit{dc}$\\
10. \>\>\> $\mathit{Scenarios} \leftarrow$\textbf{identifyTestedScenarios}($\mathit{model}$, $\mathit{ts}$)\\
11. \>\>\> \textbf{foreach} {$(s \in \mathit{Scenarios})$} \textbf{do} \\
12. \>\>\>\> $\mathit{T} \leftarrow$ \textbf{retrieveTestCases}($\mathit{s}$, $\mathit{ts}$, $\mathit{Scenarios}$) \\
%15. \>\>\>\> \textbf{if} {($T \neq \emptyset$)}  \textbf{then} \\
13. \>\>\>\> $\mathit{classified} \leftarrow \mathit{classified} \,\, \cup \,\, $\textbf{analyzeImpact}($\mathit{s}$, $\mathit{T}$, $\mathit{u_{new}}$, $\mathit{dc}$) \\
%17.  \>\>\>\> \textbf{end if}\\
14. \>\>\> \textbf{end foreach}\\
15. \>\> \textbf{else}\\
16. \>\>\> $\mathit{REUSE} \leftarrow$ $\mathit{REUSE} \,\, \cup$ \textbf{retrieveTestCases}($\mathit{u}$, $\mathit{ts}$) \\
17. \>\> \textbf{end if}\\
18. \> \textbf{end foreach}\\
19. \> $\mathit{NEW} \leftarrow$ \textbf{filterNewScenarios}($\mathit{NEW}$)\;\\
20. \> $\textbf{return} \, \mathit{classified}$\\

\end{tabbing}}}
\end{center}
\vspace*{-0.75cm}
\caption{Test Case Classification Algorithm\label{algo:classification}}
%\vspace*{-0.6cm}
\end{figure}

%Fabrizio-27.10: I couldn'yt follow the previous version
For each use case in the previous product, we check whether it is impacted by some configuration changes (Lines 6-7 in Fig.~\ref{algo:classification}). If there is no impact, all the system test cases of the use case are classified as \textit{reusable} (Lines 15-17); otherwise, we rely on the function \textit{generateUseCaseModel} (Line 8) to generate a \emphx{use case model}, i.e., a model that captures the control flow in the use case. This model is used to identify scenarios that have been tested by one or more test cases (\textit{identifyTestedScenarios} in Line 10). 
For each scenario verified by a test case (\textit{retrieveTestCases} in Line 12), we rely on the function \textit{analyzeImpact} (Line 13) to determine how decision changes affect the behaviour of the scenario.

In Sections~\ref{subsubsec:usecase_model_generation},~\ref{subsubsec:scenario_generation},~\ref{subsubsec:matching_scenarios_testcases} and~\ref{subsubsec:impact_identification}, we give the details of the functions \textit{generateUseCaseModel}, \textit{identifyTestedScenarios}, \textit{retrieveTestCases} and \textit{analyzeImpact}, respectively.

\subsubsection{Use Case Model Generation}
\label{subsubsec:usecase_model_generation}

\begin{figure}[t]
%\begin{wrapfigure}{h}{0.490\linewidth}
%\vspace*{-1.05em}
        \centerline{\includegraphics[width=0.780\linewidth]{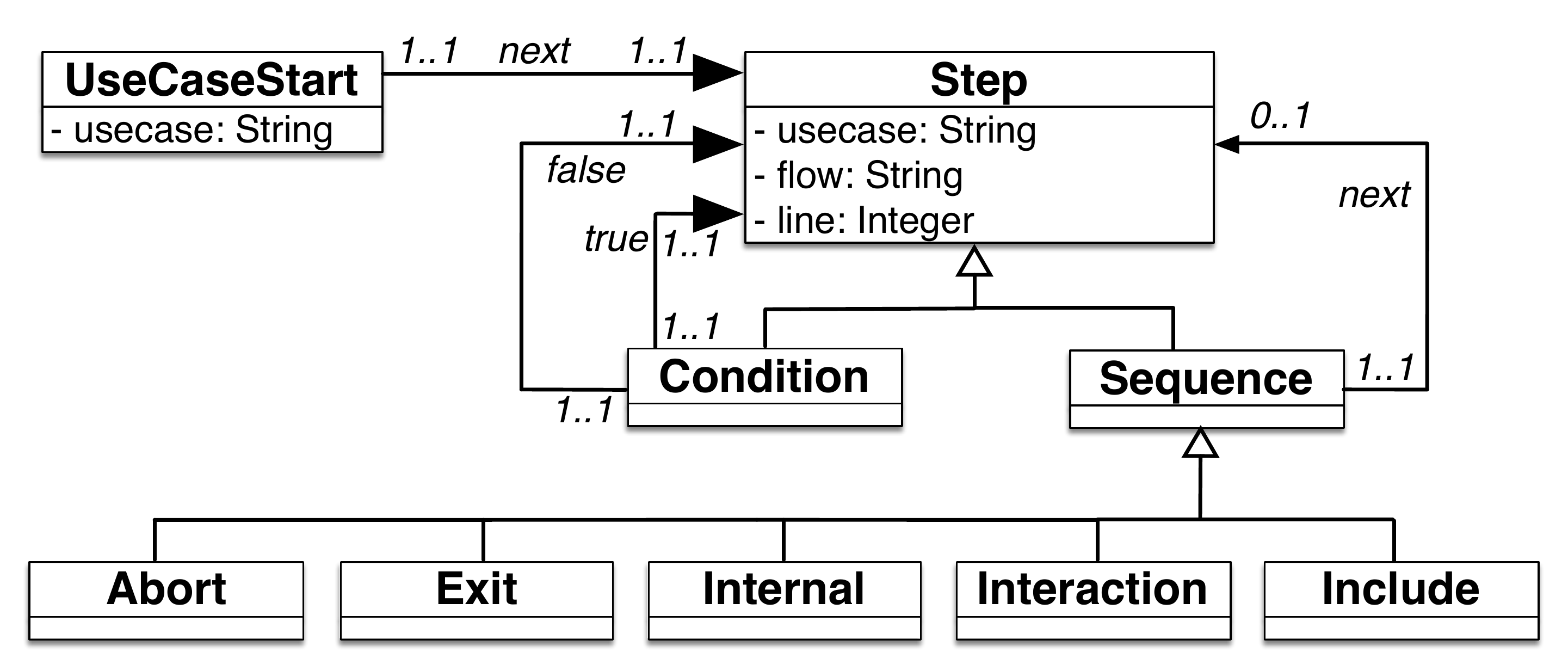}}
        % \centerline{\includegraphics[width=1.0\linewidth]{images/productlineDiagram}}
%\vspace*{-0.8em}
      \caption{Metamodel for Use Case Scenario Models}
      \label{fig:scenarioMetamodel}
%\vspace*{-1.1em}
\end{figure}

To generate a use case scenario model from a PS use case specification, we rely on a Natural Language Processing (NLP) solution proposed by Wang et al.~\cite{Wang2015a}.
It relies on the RUCM keywords and part-of-speech tagging to extract information required to build a use case model.
In this section, we briefly describe the metamodel for use case scenario models, shown in Fig.~\ref{fig:scenarioMetamodel}, and provide an overview of the model generation process. 
%Fig.~\ref{fig:scenarioMetamodel} gives the metamodel for use case scenario models. 
\emphx{UseCaseStart} represents the beginning of a use case with a precondition and is linked to the first \emphx{Step} (i.e., \emphx{next} in Fig.~\ref{fig:scenarioMetamodel}). There are two \emphx{Step} subtypes, i.e., \emphx{Sequence} and \emphx{Condition}. 
%\FAB{\emphx{Sequence} captures a sequence of steps in a use case flow, \emphx{Condition} captures a condition step that branches into an alternative use case flow.}
\emphx{Sequence} has a single successor, while \emphx{Condition} has two successors (i.e., \emphx{true} and \emphx{false} in Fig.~\ref{fig:scenarioMetamodel}).  

%%\vspace*{-1.0em}
%%\vspace*{-0.9em}

%\begin{figure*}[!t]
%\begin{wrapfigure}{h}{0.473\linewidth}
\begin{figure}[h]
%\vspace*{-1.05em}
%\centerline{\includegraphics[width=0.682\linewidth]{images/usecaseModels}}
\centerline{\includegraphics[width=0.71\linewidth]{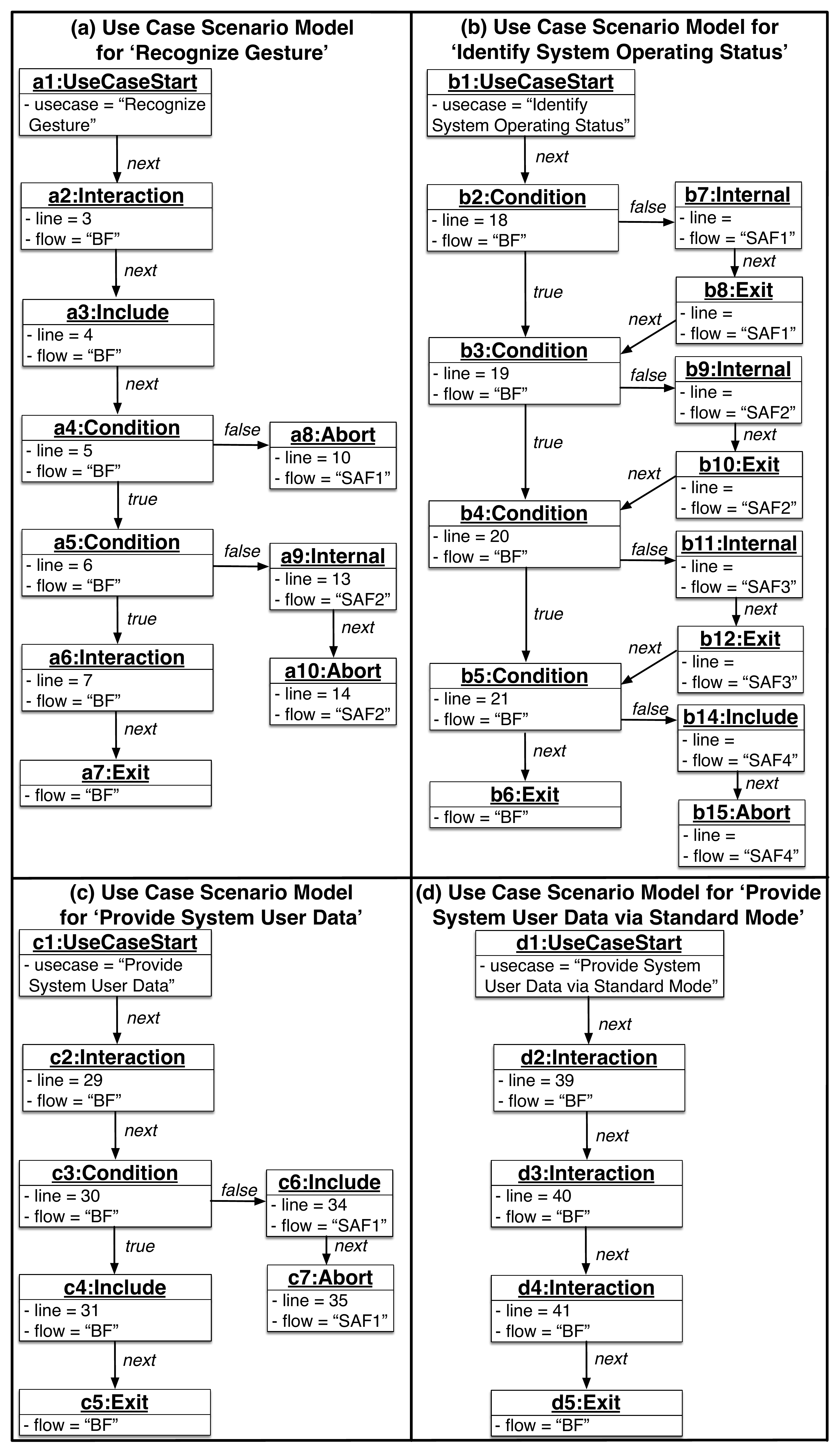}}
%\vspace*{-0.40em}
\caption{Use Case Scenario Models for the Use Case Specifications in Table~\ref{tab:ExamplePSUseCase}}\label{fig:usecaseModels}
%\vspace*{-0.4em}
%\end{figure*}
\end{figure}

\emphx{Interaction} indicates the invocation of an input/output operation between the system and an actor. 
%The \emphx{Interrupt} indicates the presence of inputs that enable the evaluation of a condition in a global or bounded alternative flow. Therefore, it is linked to the \emphx{Condition} (i.e., \emphx{enabling} in Fig.~\ref{fig:scenarioMetamodel}). 
\emphx{Internal} indicates that the system alters its internal state. \emphx{Exit} represents the end of a use case flow, while \emphx{Abort} represents the termination of an anomalous execution flow. 
%We generate a use case scenario model for each use case specification. 
Fig.~\ref{fig:usecaseModels} shows the models generated from the use cases in Table~\ref{tab:ExamplePSUseCase}. %We employ Natural Language Processing (NLP) to process use case specifications and identify use case steps (see Section~\ref{sec:tool}). 
For each \emphx{Interaction}, \emphx{Include}, \emphx{Internal}, \emphx{Condition} and \textit{Exit} step, a \emphx{Step} instance is generated and linked to the previous \emphx{Step} instance.

%Fabrizio-28.10.2018: I tried to rephrase in a simpler way
%For each specific alternative flow, a \emphx{Condition} instance is created and linked to the \emphx{Step} instance of the first step of the alternative flow (e.g., \emphx{a4} and \emphx{a5} in Fig.~\ref{fig:usecaseModels}(a)).
%Global and bounded alternative flows begin with a guard condition in use case specifications. %, i.e., a \emphx{Condition} instance in the scenario model. 
%For each reference flow step a global or bounded alternative flow is referring to, a \emphx{Condition} instance is created and linked to the \emphx{Step} instance of the reference flow step.  
%
For each alternative flow, a \emphx{Condition} instance is created and linked to the \emphx{Step} instance of the first step of the alternative flow (e.g., \emphx{a4} and \emphx{a5} in Fig.~\ref{fig:usecaseModels}(a)).
For multiple alternative flows on the same condition, \emphx{Condition} instances are linked to each other in the order they follow in the specification. For alternative flows that return back to the reference flow, an \emphx{Exit} instance is linked to the \emphx{Step} instance that represents the reference flow step (e.g., \emphx{next} between \emphx{b8} and \emphx{b3} in Fig.~\ref{fig:usecaseModels}(b)). 
%Fabrizio-28.10.2018: not sure is really needed the following
%Please note that we do not give all the specific alternative flows for the use case \textit{Identify System Operating Status} in Table~\ref{tab:ExamplePSUseCase}.  

For alternative flows that abort, an \emphx{Abort} instance is created and linked to the \emphx{Step} instance of the previous step (e.g., \textit{a8}, \emphx{a10}, \emphx{b15} and \emphx{c7} in Fig.~\ref{fig:usecaseModels}). For the end of the basic flow, there is always an \emphx{Exit} instance (e.g., \emphx{a7}, \emphx{b6}, \emphx{c5} and \emphx{d5} in Fig.~\ref{fig:usecaseModels}).

\subsubsection{Identification of Tested Use Case Scenarios}
\label{subsubsec:scenario_generation}

%ARDA: I know it is not the best way to write it but I think we aggree that it will be a seperate section because we have a dedicated function.
%\FIXME{I cannot understand why the identificaton of tested scenarios (this section) cannot go together with the identification of test cases covering the scenario (next section)}

We automatically identify tested use case scenarios in a use case specification. 
A scenario is a sequence of steps that begins with a \emphx{UseCaseStart} instance and ends with an \emphx{Exit} instance in the use case model. 
%It is represented as a linked list of instances of the model. 
Each use case scenario captures a set of interactions that should be exercised during the execution of a test case.
The function \emphx{identifyTestedScenarios} (see Line 12 in Fig.~\ref{algo:classification}) %It takes a use case scenario model (\emphx{sm}), an instance in the use case scenario model (\emphx{inst}), a linked list of instances (i.e., a scenario) in the use case scenario model (\emphx{sc}) and the test suite of the previous product (\emphx{ts}) as input. It provides as output all tested scenarios in the scenario model. % except the scenarios including cycles.  
implements a depth-first traversal of use case models to identify tested scenarios. It visits alternative flows which are tested together with previously visited alternative flows by the same test case. %Before calling \emphx{identifyTestedScenarios}, the input use case scenario model is merged with the use case scenario models of the included use cases in a way similar to the generation of interprocedural control flow graphs~\cite{Harrold1998}. %The \emphx{Include} instances in $sm$ are replaced with the \emphx{UseCaseStart} instances of the included use cases; the \emphx{Exit} instances of the basic flows of the included use cases are linked to the \emphx{Step} instances following the \emphx{Include} instances in $sm$. For instance, the model in Fig.~\ref{fig:usecaseModels}(b) is added to the model in Fig.~\ref{fig:usecaseModels}(a), while the model in Fig.~\ref{fig:usecaseModels}(d) is added to the model in Fig.~\ref{fig:usecaseModels}(c). 

%Before calling the function \emphx{generateScenarios}, the use case scenario models of the included use cases are automatically added to the use case scenario model $sm$ in a way similar to the generation of interprocedural control flow graphs~\cite{Harrold1998}. 

%\begin{figure*}[!t]
%\begin{wrapfigure}{h}{0.470\linewidth}
%\vspace*{-1.25em}
\begin{figure}
%\centerline{\includegraphics[width=0.70\linewidth]{images/exampleScenarios}}
\centerline{\includegraphics[width=0.700\linewidth]{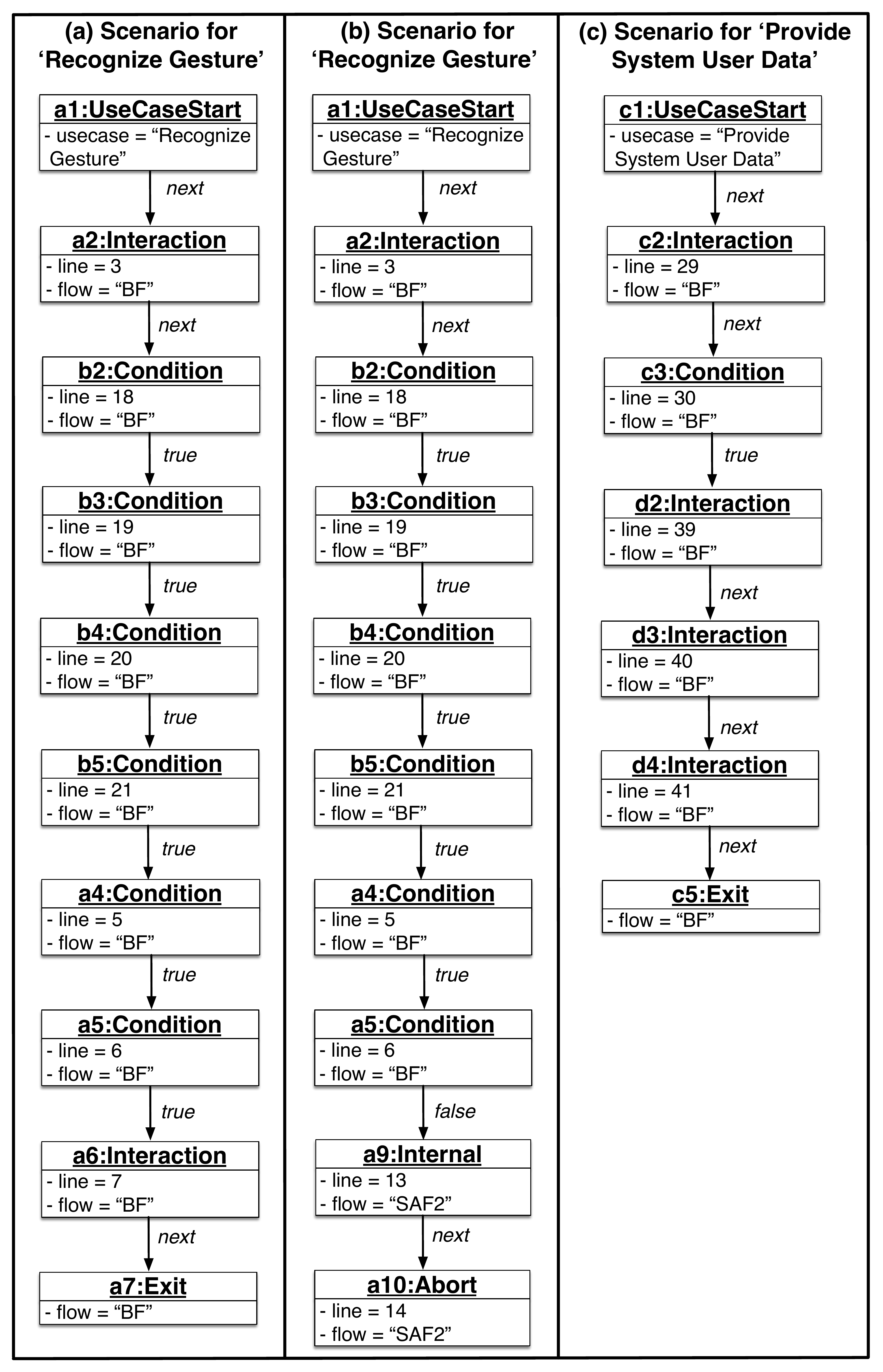}}
%\vspace*{-0.4em}
\caption{Some Tested Use Case Scenarios} %Extracted from the Use Case Scenario Models in Fig.~\ref{fig:usecaseModels}}
\label{fig:exampleScenarios}
%\vspace*{-0.4em}
%\end{figure*}   
\end{figure}

%We call \emphx{identifyTestedScenarios} where \emphx{inst} is a \emphx{UseCaseStart} instance in \emphx{sm}, and \emphx{sc} is initially an empty linked list for model instances (Line 12 in Fig.~\ref{algo:classification}). 

%?? scenarios are generated from those models in total, while ?? scenarios are for the use case \textit{Recognize Gesture}, and ?? scenarios are for the use case \textit{Provide System User Data}.

%For a \emphx{Condition} instance, we first add the \emphx{Condition} instance to the scenario (Line 7 in Fig.~\ref{algo:generateScenarios}) and copy the scenario for true and false branches (Lines 8 and 9). To avoid the infinite number of scenarios, we introduced the threshold attribute for each \emphx{Condition} instance in the scenario. If the threshold is not exceeded for the \emphx{Condition} instance in the scenario, we first visit the true branch and then the false branch (Lines 11 and 12). If the threshold is exceeded, we only take the next instance in the basic flow to proceed in the scenario (Lines 14 to 20). If the \emphx{Condition} instance is for a bounded or alternative flow, the next instance in the basic flow is the one taken when the condition is false (see \emphx{inst.false} in Line 16).

Fig.~\ref{fig:exampleScenarios} shows three tested scenarios extracted from the scenario models in Fig.~\ref{fig:usecaseModels}(a) and (c). The scenario in Fig.~\ref{fig:exampleScenarios}(a) executes the true branch of the \emphx{Condition} instance \emphx{a5} in Fig.~\ref{fig:usecaseModels}(a), while the scenario in Fig.~\ref{fig:exampleScenarios}(b) executes the false branch of the same instance. The scenario in Fig.~\ref{fig:exampleScenarios}(c) executes the basic flows in Fig.~\ref{fig:usecaseModels}(c) and (d).

%%\vspace*{-1.0em}
%\begin{figure}[h]
%\begin{wrapfigure}{h}{0.42\linewidth}
%\vspace*{-1.15em}
%        \centerline{\includegraphics[width=\linewidth]{images/testCaseMappings}}
%        % \centerline{\includegraphics[width=1.0\linewidth]{images/productlineDiagram}}
%\vspace*{-0.8em}
%      \caption{Test Case Mappings to Use Case Scenarios for Deterministic Software}
%      \label{fig:testCaseMappings}
%\vspace*{-1.45em}
%\end{wrapfigure}
%%\vspace*{-0.9em}

\subsubsection{Identification of Test Cases for Use Case Scenarios}
\label{subsubsec:matching_scenarios_testcases}

\begin{figure}[h]
%\begin{wrapfigure}{h}{0.560\linewidth}
%\vspace*{-0.80em}
        %\centerline{\includegraphics[width=0.70\linewidth]{images/TraceabilityMetamodel}}
         \centerline{\includegraphics[width=0.75\linewidth]{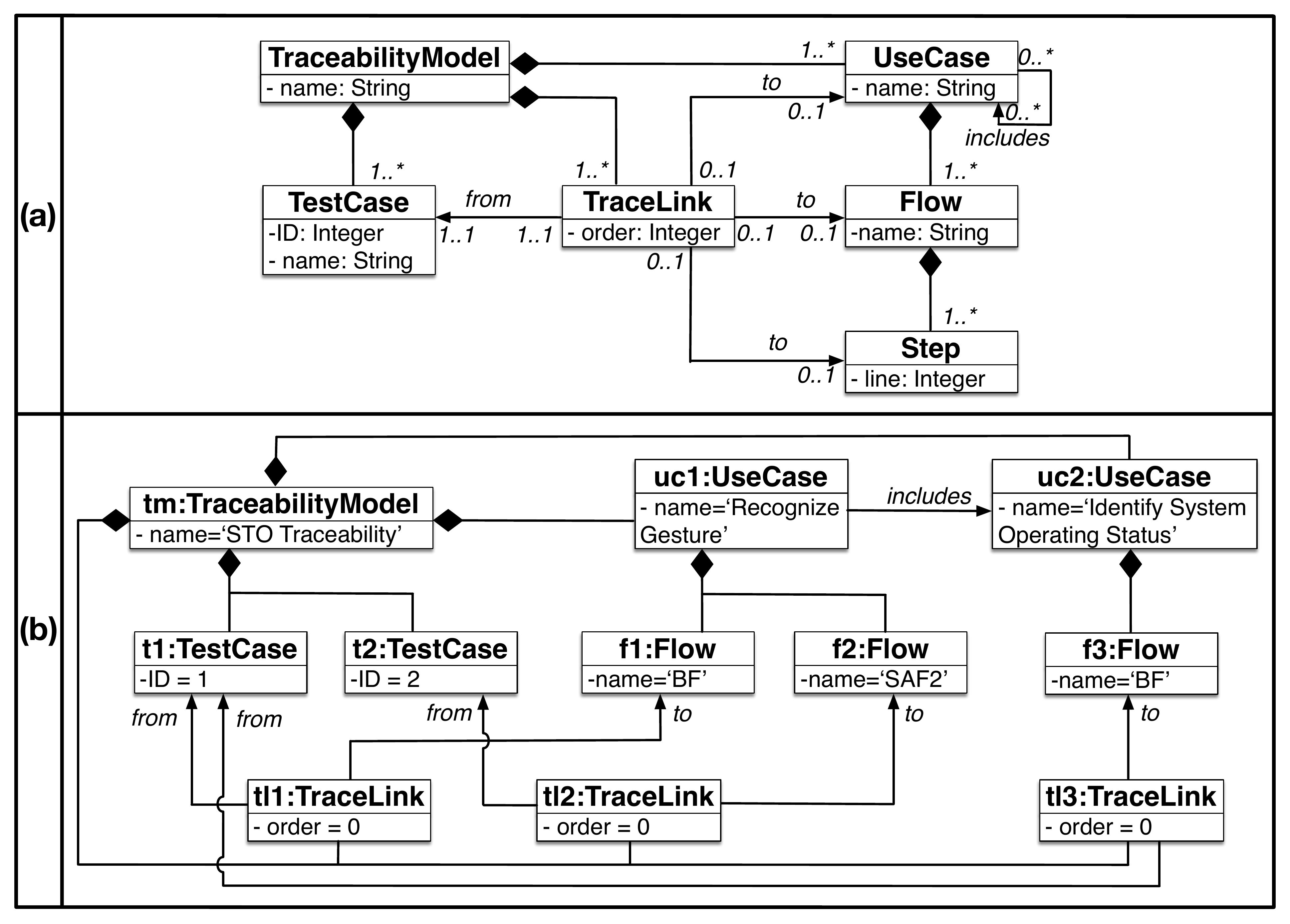}}
%\vspace*{-0.9em}
      \caption{(a)Traceability Metamodel and (b) Example Model} %for Test Cases and Use Cases}
      \label{fig:traceabilityMetamodel}
%\vspace*{-0.65em}
\end{figure}

%After all the tested scenarios have been identified, system test cases of each tested scenario are automatically retrieved from the test suite to be classified. %by analyzing the impact of configuration decision changes on the scenario. 
%We assume that, in our context, the system under test is deterministic. Therefore, there is always a many-to-one relationship between system test cases and use case scenarios~\cite{Ammann2008}. Each test case in the test suite verifies exactly one use case scenario in the use case scenario model (see Fig.~\ref{fig:testCaseMappings}). %Fig.~\ref{fig:testCaseMappings} illustrates a many-to-one relationship between system test cases and use case scenarios. 
We use traceability links between test cases and use case specifications to retrieve test cases for a given scenario. The accuracy of test case retrieval depends on the granularity of traceability links. 
Companies may follow various traceability strategies~\cite{Ramesh:2001aa}, %~\cite{Egyed2009}~\cite{Paige2011}, 
and generate links in a broad range of granularity (e.g., to use cases, to use case flows or to use case steps). We implement a traceability metamodel which enables the user to generate traceability links at different levels of granularity (see Fig.~\ref{fig:traceabilityMetamodel}(a)). % between system test cases and use case specifications at different levels of granularity (see Fig.~\ref{fig:traceabilityMetamodel}).

%%\vspace*{-1.0em}
%%\vspace*{-0.9em}

%Our tool supports both Excel spreadsheets and IBM DOORS link database to specify trace links between use case specifications and system test cases (see Section~\ref{sec:tool}). 
%The user can trace system test cases to only use cases, to basic and alternative flows, or even to use case steps in the flows. If there is more than one alternative flow that a test case verifies, she can also indicate the execution order of the flows (see the attribute \textit{order} of \textit{TraceLink} in Fig.~\ref{fig:traceabilityMetamodel}). 
Fig.~\ref{fig:traceabilityMetamodel} (b) gives part of the traceability model for traceability links, assigned by engineers, between two test cases and the use cases \emphx{Recognize Gesture} and \emphx{Identify System Operating Status} in Table~\ref{tab:ExamplePSUseCase}.
Test case \emphx{t1} is traced to the basic flows of \emphx{Recognize Gesture} and \emphx{Identify System Operating Status} (i.e., ($\emphx{t1}$ $\xrightarrow{\emphx{tl1}}$ $\emphx{f1}$) and ($\emphx{t1}$ $\xrightarrow{\emphx{tl3}}$ $\emphx{f3}$)), while \emphx{t2} is traced to the specific alternative flow \emphx{SAF2} of \emphx{Recognize Gesture} (i.e., ($\emphx{t2}$ $\xrightarrow{\emphx{tl2}}$ $\emphx{f2}$)).

We retrieve, using the traceability links in Fig.~\ref{fig:traceabilityMetamodel}(b), \emphx{t1} for the scenario in Fig.~\ref{fig:exampleScenarios}(a) since it is the only scenario executing the basic flows of \emphx{Recognize Gesture} and \emphx{Identify System Operating Status}. The scenario in Fig.~\ref{fig:exampleScenarios}(b) executes the alternative flow \emphx{SAF2} of \emphx{Recognize Gesture} (see \emphx{a9} and \emphx{a10} in Fig.~\ref{fig:exampleScenarios}(b)). Therefore, \emphx{t2} is retrieved for %the scenario in 
Fig.~\ref{fig:exampleScenarios}(b).

%In the context of our case study, as it is often the case, trace links are assigned from test cases to only basic and alternative flows without indicating the execution order of the flows. There are cases where we need finer-grained trace links to retrieve test cases. %to retrieve the right test case for a given scenario.

Our approach often requires traceability links from test cases to only basic and alternative flows without indicating the execution order of the flows. There are few cases where finer-grained traceability links are needed to retrieve test cases. First, %we need finer-grained trace links 
when multiple scenarios take the same alternative flows with different orders, %Assume there are two scenarios $\emphx{s1}$ and $\emphx{s2}$ taking the specific alternative flow $\emphx{saf}$ and the bounded alternative flow $\emphx{baf}$ with two different orders ($\mathit{saf}$, $\mathit{baf}$) and ($\mathit{baf}$, $\mathit{saf}$), respectively. For the test cases $\emphx{tc1}$ and $\emphx{tc2}$ exercising $\emphx{s1}$ and $\emphx{s2}$, we have the trace links \emphx{tr1}, \emphx{tr2}, \emphx{tr3} and \emphx{tr4}: ($\emphx{tc1}$ $\xrightarrow{\emphx{tr1}}$ $\emphx{saf}$), ($\emphx{tc1}$ $\xrightarrow{\emphx{tr2}}$ $\emphx{baf}$), ($\emphx{tc2}$ $\xrightarrow{\emphx{tr3}}$ $\emphx{saf}$) and ($\emphx{tc2}$ $\xrightarrow{\emphx{tr4}}$ $\emphx{baf}$). %It is not possible to match %, without knowing the execution order of the alternative flows in the test cases, 
%the scenarios and the system test cases. 
%In such a case, 
these orders are needed to match test cases and scenarios (the attribute \textit{order} in Fig.~\ref{fig:traceabilityMetamodel}(a)).

Second, we need finer-grained traceability links when there are more than one scenario taking the same bounded or global alternative flow. Those alternative flows refer to more than one step in a reference flow. Hence, a scenario can take a bounded or global alternative flow from different reference flow steps; we need traceability links indicating the reference flow step in which the flow is taken (see ``\emphx{to}'' from \emphx{TraceLink} to \emphx{Step} in Fig.~\ref{fig:traceabilityMetamodel}(a)). %To match test cases and scenarios for bounded/global alternative flows, we need trace links indicating the reference flow step in which the flow is taken (see the ``\emphx{to}'' from \emphx{TraceLink} to \emphx{Step} in Fig.~\ref{fig:traceabilityMetamodel}). 
If we do not have traceability at the right level of granularity \MREVISION{R4.1, R4.8, R4.12}{in these cases}, we ask the user to match use case scenarios and test cases. %resolve the matching problem of scenarios and test cases.    

\MREVISION{R4.1, R4.8, R4.12}{The two cases above, which represent the most expensive cases of traceability, are expected to happen very rarely. For instance, in our case study (see Section~\ref{sec:evaluation}), we did not encounter them at all and there was no need to manually match use case scenarios and system test cases. Overall, the additional effort entailed by our approach depends on the traceability practice in place. For instance, companies in safety-critical domains must follow guidelines enforced by the international safety standards regarding traceability and introducing our approach would not entail any additional overhead.}

\subsubsection{Impact Identification}
\label{subsubsec:impact_identification}

We analyze the impact of configuration changes on use case scenarios to identify new scenarios and classify retrieved test cases as \emphx{obsolete}, \emphx{retestable} and \emphx{reusable}.
%According to related work~\cite{Yoo2012}, reusable test cases only execute the parts of the product under test that remain unchanged with respect to previous products, while retestable test cases execute parts of the product that have been changed with respect to previous products. Obsolete test cases either no longer test the feature they were designed for or include sequences of inputs and expected outputs that are not correct due to changes in specifications (i.e., they need to be modified in order to cover the same use case scenarios).}
%\CHANGED{To classify test cases, 
To this end, we devise an algorithm (see Fig.~\ref{algo:impact}) 
which takes as input a use case scenario to be analyzed ($\mathit{s_{old}}$), a set of test cases verifying the scenario ($T$), a use case specification for the new product ($u$), and a triple of the sets of configuration changes (\emphx{dc}) produced in Step 2.
If there is no change impacting the scenario, test cases verifying the scenario are classified as \emphx{reusable} (Line 8 in Fig.~\ref{algo:impact}). For any change in the scenario (e.g., %updating the order of use case steps), and 
removing a use case step), 
the test cases are classified as either \emphx{retestable} or \emphx{obsolete} (Line 5) as shown in Table~\ref{tab:classificationTestCases}, 
which describes how test cases are classified based on the types of changes affecting the variant elements covered by a scenario.
A test case is classified as retestable when it does not need to be modified to cover the corresponding scenario. Changes impacting a use case scenario may lead to modifications in source code. Since modifications in the source code may introduce faults, retestable test cases are expected to be re-executed to ensure that the system behaves correctly. \MREVISION{R3.11}{A reusable test case exercises use case steps that remains valid in the new product, while this is not the case for internal steps exercised by retestable test cases.} A test case is classified as obsolete if its the sequence of inputs may no longer enable the execution of the corresponding scenario or when the oracles are no longer correct. Obsolete test cases cannot be reused as is to retest the system but need to be modified.

%Fabrizio-28.10: algorithms and tables should stay in their own files otherwise is difficult to follow anything
% !TEX root =  Main.tex

%Fabrizio 28.10: Why aren't we using an algorithm environment?

%\begin{wrapfigure}{h}{0.60\linewidth}
\begin{figure}[t]
%\vspace*{-0.2cm}
\begin{center}
\scalebox{1.02}{\parbox{\linewidth}{\footnotesize
%\hbox{{\bf Algorithm}\;{\sc Diff} } 
%\vspace*{-.36cm}
\begin{tabbing}
Output:x \= $\mathit{Block\_Slice}_r$:abit \=\kill
{\bf Inputs:} 
%Fabrizio 28.10: why "linked" list? Anyway it seems that we provide a list of scenarios
%Linked list of scenario model instances $\mathit{s_{old}}$, \\ 
Old use case scenario $\mathit{s_{old}}$,  Set of test cases $\mathit{T}$, \\
\ \ \ \ \ \ \ \ \ \ \ \  New use case specification $\mathit{u}$, Decision changes $\mathit{dc},$\\

%Fabrizio 28.10: it's not straightforward to understand what is a triple of sets
%\ \ \ \ \ \ \ \ \ \ \ \ Triple of sets of decision-level changes $\mathit{dc}$, 
%\ \ \ \ \ \ \ \ \ \ \ \ Decision changes $\mathit{dc}$,\\
%\FIXME{Why the following is an input?}
%\\ \ \ \ \ \ \ \ \ \ \ \ \ Quadruple of sets of classified test cases $\mathit{classified}$ \\ \ \ \ \ \ \ \ \ \ \ \ \ (OBSOLETE, REUSE, RETEST, NEW) \\
{\bf Output:} Quadruple of sets of classified test cases $\mathit{classified}$   \\\>(OBSOLETE, REUSE, RETEST, NEW) \\
\> \\
\end{tabbing}
\vspace*{-1.1cm}
\begin{tabbing}
100.\= if \= if \= if \= if \= if \= if \kill
\> \\
1. \>$\mathit{model} \leftarrow$ \textbf{generateScenarioModel}($\mathit{u}$)\;\\
2.\> Let $\mathit{inst}$ be the \textit{UseCaseStart} instance in $\mathit{model}$\\
%\FIXME{What is this $\mathit{s_{new}}$ ? Why we generate such a list? TO DISCUSS} \\
3.\>Let $\mathit{s_{new}}$ be an \textit{empty} scenario\\
4. \> \textbf{if} {(there is at least one change in $\mathit{dc}$ for $\mathit{s_{old}}$)} \textbf{then}\\
5. \>\> $\mathit{classified} \leftarrow$ \textbf{analyzeChangesOnScenario}($\mathit{s_{old}}$, $\mathit{dc}$, $\mathit{T}$)\;\\
%4. \>\> $\mathit{NEW} \leftarrow$  $\mathit{NEW} \,\cup$ \textbf{generateNewScenarios}($\mathit{s}$, $\mathit{um}$)\;\\
6. \>\> $\mathit{NEW} \leftarrow$  \textbf{identifyNewScenarios}($\mathit{model}$, $\mathit{s_{old}}$,  $\mathit{s_{new}}$, $\mathit{inst}$)\;\\
%7. \>\> $\mathit{NEW} \leftarrow$  $\mathit{NEW} \,\cup$ $\mathit{NS}$\;\\ %\textbf{filterNewScenarios}($\mathit{NS}$)\;\\
7. \> \textbf{else}\\
%8. \>\> $\mathit{REUSE} \leftarrow$ $\mathit{REUSE} \,\cup \mathit{T}$  \\
8. \>\> $\mathit{REUSE} \leftarrow \mathit{T}$  \\
9. \> \textbf{end if}\\
%11. \> $\mathit{NEW} \leftarrow$ \textbf{filterNewScenarios}($\mathit{NEW}$)\;\\
10. \> $\textbf{return} \, \mathit{classified}$\\

\end{tabbing}}}
\end{center}
\vspace*{-0.8cm}
\caption{Algorithm for \textit{analyzeImpact}\label{algo:impact}}
\vspace*{-0.5cm}
\end{figure}

\begin{table*}[h!]
%\tiny
\scriptsize
%\FIXME{Why this table was formatted in a strange manner? Why did we have tabulars inside tabulars? I noticed the same formatting of tables (tabular inside other tabular) other times. Can you please fix them?}

\centering
%\vspace*{-0.60em}
\caption{Changes in Use Case Scenarios and Classification of System Test Cases }
\label{tab:classificationTestCases}
%\vspace*{-1.00em}

\begin{tabular}{|p{0.3cm}|p{1.30cm}|p{1.000cm}|p{7.5cm}|}
\hline
\textbf{Rule ID}  
&\textbf{Change in the Scenario}  
& \textbf{Test Case Classification} 
& \textbf{Rationale}                                                                                                                                                                                                                                                                                                                                                                                                                                                                                                                                                                                                                                                                                                                                                                                                                                                                                                                                                                                                                                                                                                                                                                                                                                                                               \\ \hline
R1                                                                                                 
&Add or remove an internal step                                                                                                 
& Retestable                                                                                       
& Internal use case steps represent internal system operations (e.g., reset of counters) and do not directly affect system-actor interactions. Therefore, a test case does not need to be modified to exercise a scenario including added or deleted internal steps (e.g., a new internal step does not imply an additional test input or an update in the test oracle). The test case can be executed against the new product without any change; however, %the changes (e.g., source code changes) applied to the implemented system may introduce faults 
the system may not behave as expected (e.g., because
of a faulty implementation of a new internal use case step) and thus the test case is classified as retestable.
%& Internal use case steps represent internal system operations (e.g., reset of counters) and do not directly affect system-actor interactions. Therefore, a test case is not impacted by adding or deleting an internal step (e.g., a new internal step does not imply an additional test input or an update in the test oracle). The test case can be executed against the new product without any change; however, the system may not behave as expected (e.g., because of a faulty implementation of a new internal use case step) and thus the test case should be retested.                                                                                                                                                                                                                                                                                                                                                                                                                                                                                                                                                                                                     
\\ \hline
R2                                                                                                 
&Update the order of an internal step                                                                                           
& Retestable                                                                                       
&Since internal use case steps do not directly affect system-actor interactions, a test case does not need to be modified in the presence of a change in the order of internal steps (i.e., a different sequence of internal steps does not imply an update in test inputs or oracles). However, the system may not behave as expected (e.g., because of a faulty implementation of the new order of an internal step) and thus the test case is classified as retestable.
\\ \hline
R3                                                                                                 
&Add or remove a condition step where the condition refers exclusively to state variables 
& Retestable                                                                                       
&Condition steps are used to verify properties of input entities and/or state variables. 
A condition step, in practice, restricts the execution of a use case scenario to a subset of the values assigned to the input entities and/or state variables verified by the condition. State variables are used to model the system state, while input entities describe system inputs provided by actors. 
The addition and removal of condition steps that verify the properties of state variables reflect changes in the internal behaviour of the system but not in the system-actor interactions. Therefore, a test case is not modified in the presence of added/removed condition steps that only verify the properties of state variables (e.g., such a new condition step does not imply an update in test inputs
and oracle).
However, the system may not behave as expected (e.g., because of a faulty implementation of the changed state variables) and thus the test case is classified as retestable.
\\ \hline
%Fabrizio-28.10.2018: for removed conditons probably we could have considered the test case as retestable, the condition is removed thus all the inputs go through, no?
%ARDA: but we may remove or add constraint on the inputs which requires the test case be modified
R4                                                                                                 
&Add or remove a condition step where the condition refers to an input entity
& Obsolete 
&%A condition step verifying the properties of an input entity mentions system inputs. 
Adding or removing a condition step referring to input entities may imply an update in the test inputs if the test input values do not satisfy the changed condition. Since we do not inspect executable test cases in our analysis, it is not possible to determine if the test cases of the previous product
%exercising the old scenario 
already provide the values that fulfill the changed condition. To be conservative, we consider test cases of scenarios impacted by such changes as obsolete thus forcing engineers to verify if the test input values exercise the scenario. 
%&When old and new scenarios differ for a condition step verifying an input variable, some of the input values processed by one of the two scenarios are not processed by the other. Since in our analysis we rely on requirements specifications only and we do not inspect the executable test cases, it is not possible for us to determine if the test cases developed to exercise the old scenario for the previous product provide the values that belong to the tested scenario or not. For this reason, to be conservative, we consider the test cases that cover old scenarios affected by this type of change obsolete thus requiring engineers to verify if the provided values are the ones that exercise the scenario.
                                                                                                                                                                                                                                                                                                                                                                                                                                                                                                                                                                                                                                                                                                                                                                                                                                                                                                                                                                                                                                                                                                                                                                                                                                                                                                                         \\ \hline
%Fabrizio-28.10: I left like this, honestly a simple data flow analysis that checks if the variable checked in the condition was defined 
R5                                                                                                 
&Update the order of a condition step
& Obsolete
& When old and new scenarios differ regarding the order in which condition steps appear, then the behaviour triggered by the test case of the previous product might not be the same in the new product (e.g., if the steps that define the variables verified by the condition are between the condition steps that have been changed).
%are not the same in the two scenarios). 
Therefore, we consider a test case that exercises an old scenario affected by such changes as obsolete.
\\ \hline
R6                                                                                                 
&Add or remove an input/output step
& Obsolete                                                                                         
& Input and output use case steps represent system-actor interactions. Therefore, 
the implementation of the test case needs to be modified to exercise the targeted scenario
%test case is impacted by adding or removing an input/output step 
when input and output steps are added or removed 
(e.g., a new input step implies an additional test input in the test case).
%& Changes in the input and output steps directly affects the way a test case provides inputs to the system (e.g., the number and type of input values) or the way it verifies the results produced (e.g., by inspecting the outputs specified in the scenarios).                                                                                                                                                                                                                                                                                                                                                                                                                                                                                                                                                                                                                                                                                                                                                                                                                                                                                                                                                                                                                                                                                                                                                                                                                                                                                                                           
\\ \hline
R7                                                                                                 
&Update the order of an input/output step
& Obsolete  
& Since input and output use case steps represent system-actor interactions, 
%a test case is impacted by updating the order of an input/output step 
the implementation of the test case needs to be modified to exercise the targeted scenario when the order of input and output steps is updated
(e.g., a new order of input steps implies an update in the sequence of test inputs). %in the test case).                                                                                                                                                                                                                                                                                                                                                                                                                                                                                                                                                                                                                                                                                                                                                                                                                                                                                                                                                                                                                                                                                                            
%& The order of input and output steps in use case specifications reflects the order in which inputs and outputs are received or produced by the implemented system. For this reason, a change in the order of input and output steps makes the test case obsolete.                                                                                                                                                                                                                                                                                                                                                                                                                                                                                                                                                                                                                                                                                                                                                                                                                                                                                                                                                                                                                                                                                                            
\\ \hline
R8                                                                                                 
&Remove an alternative flow
&Obsolete                                                                                         
&Alternative flows capture sequences of interactions taking place under certain execution conditions. 
If a use case scenario of the previous product covers an alternative flow that does not exist in the new product, the corresponding test case should be considered as obsolete because the interactions verified by the test case cannot take place with the new product.
\\ \hline
R9                                                                                                 
&Multiple changes in the use case scenario
& 
Obsolete or Retestable
& 
A test case is classified as \emphx{obsolete} if there is at least one change in the scenario that makes the test case obsolete. A test case is classified as \emphx{retestable} if there are no changes in the scenario that make the test case obsolete and if there is at least one change in the scenario that makes the test case retestable.                                                                                                                                                                                                                                                                                                                                                                                                                                                                                                                                                                                                                                                                                                                                                                                                                                                                                                                                                                                                \\ \hline
\end{tabular}
%\vspace*{-2.00em}
\end{table*}

% !TEX root =  Main.tex

%\begin{wrapfigure}{h}{0.72\linewidth}
\begin{figure}[h]
%\vspace*{-0.5cm}
%\begin{center}
\scalebox{1.02}{\parbox{\linewidth}{\footnotesize
%\hbox{{\bf Algorithm}\;{\sc Diff} } 
%\vspace*{-.36cm}
\begin{tabbing}
Output:x \= $\mathit{Block\_Slice}_r$:abit \=\kill
{\bf Input:} New scenario model $\mathit{sm}$, old scenario $\mathit{s_{old}}$, new scenario $\mathit{s_{new}}$, \\model instance $\mathit{inst}$,\\%\\  \,\,\,\,\,\,\,\,\,\,\,\,\,\,\,\,\,\,\,\ new scenario $\mathit{s_{new}}$,\\
%Linked list of scenario model instances of the previous product $\mathit{s_{old}}$,\\  \,\,\,\,\,\,\,\,\,\,\,\,\,\,\,\,\,\,\,\ Linked list of scenario model instances of the new product $\mathit{s_{new}}$,\\  %\,\,\,\,\,\,\,\,\,\,\,\,\,\,\,\,\,\,\,\ Scenario model instance $\mathit{inst}$\\  %\,\,\,\,\,\,\,\,\,\,\,\,\,\,\,\,\,\,\,\ Triple of sets of decision-level changes $\mathit{dc}$ \\
%{\bf Output:} Set of linked lists of model instances $\mathit{S}$ \\ % \\\>(OBSOLETE, REUSE, RETEST, NEW) \\
{\bf Output:} Set of triples of new scenario, old scenario and guidance $\mathit{S}$ \\ % \\\>(OBSOLETE, REUSE, RETEST, NEW) \\
\> \\
\end{tabbing}
\vspace*{-1.1cm}
\begin{tabbing}
100.\= if \= if \= if \= if \= if \= if \kill
\> \\
1. \> Let $S$ be the empty set for triples of old scenario, new scenario and \\ \,\,\,\,\,\,\,\,\,\,guidance \\
%\CHANGED{1. \> Let $S$ be an empty set.}\\
2. \>\textbf{if} {($inst$ is a $\mathit{UseCaseStart}$, $\mathit{Interaction}$ or $\mathit{Internal}$ instance)} \textbf{then}\\
3. \>\> \textbf{addToScenario}($\mathit{inst}$, $\mathit{s_{new}}$) \\
4. \>\> $\mathit{S} \leftarrow$ $\mathit{S} \,\cup$ \textbf{identifyNewScenarios}($\mathit{sm}$, $\mathit{s_{old}}$, $\mathit{s_{new}}$, $\mathit{inst.next}$) \\
5. \> \textbf{end if}\\
6. \>\textbf{if} {($\mathit{inst}$ is a $\mathit{Condition}$ instance)} \textbf{then}\\
7. \>\> \textbf{addToScenario}($\mathit{inst}$, $\mathit{s_{new}}$) \\
8. \>\>\textbf{if} {($\mathit{inst}$ exist in $\mathit{s_{old}}$)} \textbf{then}\\
9. \>\>\> Let $\mathit{t}$ be the instance after $\mathit{inst}$ in the branch taken in $\mathit{s_{old}}$ \\
10. \>\>\> Let $\mathit{t_{new}}$ be the instance corresponding to $t$ in $\mathit{sm}$ \\
11. \>\>\>$\mathit{S} \leftarrow$ $\mathit{S} \,\cup$ \textbf{identifyNewScenarios}($\mathit{sm}$, $\mathit{s_{old}}$, $\mathit{s_{new}}$, $\mathit{t_{new}}$) \\
12. \>\> \textbf{else}\\
13. \>\>\>\textbf{if} {($\mathit{inst}$ represents a condition leading to a specific alternative flow)} \textbf{then}\\
14. \>\>\>\>\textbf{if} {($\mathit{inst}$ and $\mathit{inst.false}$ exist together in $\mathit{s_{new}}$)} \textbf{then}\\
%\CHANGED{15a. \>\>\>\>\textbf{if} {($\mathit{inst}$ belongs to basic flow)} \textbf{then}\\}
%9. \>\>\>\> \textbf{addToScenario}($\mathit{inst}$, $\mathit{s_{new}}$) \\
%8. \>\>\> $sc_{T} \leftarrow sc$ \\
%10. \>\>\>\> $s_{F} \leftarrow s_{new}$ \\
15. \>\>\>\>\>$\mathit{S} \leftarrow$ $\mathit{S} \,\cup$ \textbf{identifyNewScenarios}($\mathit{sm}$, $\mathit{s_{old}}$, $\mathit{s_{new}}$, $\mathit{inst.true}$) \\
%\CHANGED{15b. \>\>\>\>\textbf{fi}}\\
%10. \>\>\textbf{if} {($inst.threshold$ $<$ $\mathit{2}$)} \textbf{then}\\
16. \>\>\>\> \textbf{else}\\
%13. \>\>\>\> \textbf{addToScenario}($\mathit{inst}$, $\mathit{s_{new}}$) \\
%13. \>\>\>\> $s_{F} \leftarrow s_{new}$ \\
17. \>\>\>\>\> $s_{cpy} \leftarrow \textbf{clone}(s_{new})$ \\
18. \>\>\>\>\>$\mathit{S} \leftarrow$ $\mathit{S} \,\cup$ \textbf{identifyNewScenarios}($\mathit{sm}$, $\mathit{s_{old}}$, $\mathit{s_{new}}$, $\mathit{inst.true}$) \\
19. \>\>\>\>\>$\mathit{S} \leftarrow$ $\mathit{S} \,\cup$ \textbf{identifyNewScenarios}($\mathit{sm}$, $\mathit{s_{old}}$, $\mathit{s_{cpy}}$, $\mathit{inst.false}$) \\
20. \>\>\>\> \textbf{end if}\\

%13. \>\>\> $\mathit{inst.threshold} \leftarrow$ $\mathit{inst.threshold}  \, + \mathit{1}$\\  
21. \>\>\> \textbf{else}\\
22. \>\>\>\>\textbf{if} {($\mathit{inst}$ and $\mathit{inst.true}$ exist together in $\mathit{s_{new}}$)} \textbf{then}\\
%21. \>\>\>\> \textbf{addToScenario}($\mathit{inst}$, $\mathit{s_{new}}$) \\
%20. \>\>\>\> $s_{T} \leftarrow s_{new}$ \\
%9. \>\>\>\> $sc_{F} \leftarrow sc$ \\
%\CHANGED{23a. \>\>\>\>\textbf{if} {($\mathit{inst}$ belongs to basic flow)} \textbf{then}}\\
23. \>\>\>\>\>$\mathit{S} \leftarrow$ $\mathit{S} \,\cup$ \textbf{identifyNewScenarios}($\mathit{sm}$, $\mathit{s_{old}}$, $\mathit{s_{new}}$, $\mathit{inst.false}$) \\
%\CHANGED{23b. \>\>\>\>\textbf{fi}}\\
24. \>\>\>\> \textbf{else}\\
%25. \>\>\>\> \textbf{addToScenario}($\mathit{inst}$, $\mathit{s_{new}}$) \\
%23. \>\>\>\> $s_{T} \leftarrow s_{new}$ \\
25. \>\>\>\>\> $s_{cpy} \leftarrow \textbf{clone}(s_{new})$ \\
26. \>\>\>\>\>$\mathit{S} \leftarrow$ $\mathit{S} \,\cup$ \textbf{identifyNewScenarios}($\mathit{sm}$, $\mathit{s_{old}}$, $\mathit{s_{new}}$, $\mathit{inst.false}$) \\
27. \>\>\>\>\>$\mathit{S} \leftarrow$ $\mathit{S} \,\cup$ \textbf{identifyNewScenarios}($\mathit{sm}$, $\mathit{s_{old}}$, $\mathit{s_{cpy}}$, $\mathit{inst.true}$) \\
28. \>\>\>\> \textbf{end if}\\
29. \>\>\> \textbf{end if}\\
30. \>\> \textbf{end if}\\
31. \> \textbf{end if}\\
32. \>\textbf{if} {($inst$ is an $\mathit{Exit}$ or $\mathit{Abort}$ instance)} \textbf{then}\\
33. \>\>\textbf{if} {($inst$ is an $\mathit{Exit}$ instance for the included use case)} \textbf{then}\\
34. \>\>\> $\mathit{S} \leftarrow$ $\mathit{S} \,\cup$ \textbf{identifyNewScenarios}($\mathit{sm}$, $\mathit{s_{old}}$, $\mathit{s_{new}}$, $\mathit{inst.next}$) \\
35. \>\> \textbf{else}\\
36. \>\>\> \textbf{addToScenario}($\mathit{inst}$, $\mathit{s_{new}}$) \\
37. \>\>\> $\mathit{G} \leftarrow$ \textbf{generateGuidance}($\mathit{s_{old}}$, $\mathit{s_{new}}$) \\
38. \>\>\> $\mathit{S} \leftarrow$ $\mathit{S}  \,\cup\,\{<s_{new}, s_{old}, G>\}$\\  
39. \>\> \textbf{end if}\\
40. \> \textbf{end if}\\
41. \> $\textbf{return} \,\, \mathit{S}$\\
\end{tabbing}}}
%\end{center}
\vspace*{-0.8cm}
\caption{Algorithm for \textit{identifyNewScenarios}\label{algo:identifyNewScenarios}}
\vspace*{-0.4cm}
\end{figure}

For the example configuration changes identified in Section~\ref{subsec:matching_decisions}, the scenarios in Fig.~\ref{fig:exampleScenarios}(a) and (b) are classified as \emphx{retestable} while the scenario in Fig.~\ref{fig:exampleScenarios}(c) is classified as \emphx{obsolete}. %The tuple $(\mathit{B18}, \mathit{B19})$ represents the updated configuration decision for the use case \emphx{Recognize Gesture} (see Section~\ref{subsec:matching_decisions}). With the configuration change, the unselected optional bounded alternative flow of the use case \emphx{Recognize Gesture} is selected in the new product (see Fig.~\ref{fig:decisionModels}(c)). 
%\FIXME{TO DISCUSS: I cannot understand the relation between the scenarios in Fig 11 and the examples below. It's easier if we discuss by voice.}
%ARDA: I checked the identifiers as we discussed before but it is not possible to come up with the same indentifiers since the size of the models are not the same. If I try to match the identifiers then the order of the identifiers in the scenarios is messed up!
The tuple $\langle\mathit{B18}, \mathit{B19}\rangle$ represents an updated decision; the unselected optional bounded alternative flow of the use case \emphx{Recognize Gesture} is selected in the new product (see Section~\ref{subsec:matching_decisions}).
The selected optional flow contains a condition, i.e., \emphx{``voltage fluctuation is detected''} in Line 10 in Table~\ref{tab:useCaseRUCM}, which does not refer to any entity in the input steps. Since the condition step is added in the scenarios in Fig.~\ref{fig:exampleScenarios}(a) and (b), these two scenarios are classified as \emphx{retestable}. The triples $\langle\mathit{B11}, \mathit{B12}, \mathit{B14}\rangle, \langle\mathit{B11}, \mathit{B12}, \mathit{B15}\rangle$, $\langle\mathit{B11}, \mathit{B12}, \mathit{B16}\rangle$, and $\langle\mathit{B11}, \mathit{B12}, \mathit{B17}\rangle$ in Fig.~\ref{fig:decisionModels} represent updated decisions for the use case \emphx{Provide System User Data} (see Section~\ref{subsec:matching_decisions}). Some of the unselected output steps are selected while one selected output step is unselected and the order of the output steps are updated in the basic flow of \emphx{Provide System User Data} (see Fig.~\ref{fig:decisionModels}). Therefore, the test case verifying the scenario in Fig.~\ref{fig:exampleScenarios}(c) for the basic flow of \emphx{Provide System User Data} is classified as \emphx{obsolete} (see rules R6, R7 and R9 in Table~\ref{tab:classificationTestCases}).

We process scenarios impacted by the configuration changes to identify new scenarios for the new product (Line 6 in Fig.~\ref{algo:impact}). Furthermore, for each new scenario, we provide guidance to support the engineers in the implementation of test case(s). To this end, we devise an algorithm (Fig.~\ref{algo:identifyNewScenarios}) which takes as input a use case model of the new product (\emphx{sm}), 
%Fabrizio 28.10: We should have an algorithm that receive three inputs, then an internal function that process the current step in the s_new. TO DISCUSS.
a use case step in the model (\emphx{inst}), a use case scenario of the previous product ($\mathit{s_{old}}$) that has been exercised by either an obsolete or retestable test case, and a new scenario ($\mathit{s_{new}}$) which is initially empty. 
%Fabrizio 28.10: teh following sentence was somehow misleading to me
%The algorithm generates a set of triples $\langle$ new scenario ($\mathit{s_{new}}$), old scenario ($\mathit{s_{old}}$) and guidance ($\mathit{G}$) $\rangle$. 
The algorithm generates a set of triples $\langle$ $\mathit{s_{new}}$, $\mathit{s_{old}}$, $\mathit{G}$ $\rangle$, where  $\mathit{s_{new}}$ is the new scenario,
$\mathit{s_{old}}$ is the old scenario of the previous product, %which $\mathit{s_{new}}$ is derived from,
and $G$ is the guidance, a list of suggestions indicating how to modify test cases covering $\mathit{s_{old}}$ to generate test cases covering $\mathit{s_{new}}$.

%%\vspace*{-1.0em}
\begin{figure}[h]
%\begin{wrapfigure}{h}{0.361\linewidth}
%\vspace*{-1.15em}
        \centerline{\includegraphics[width=0.600\linewidth]{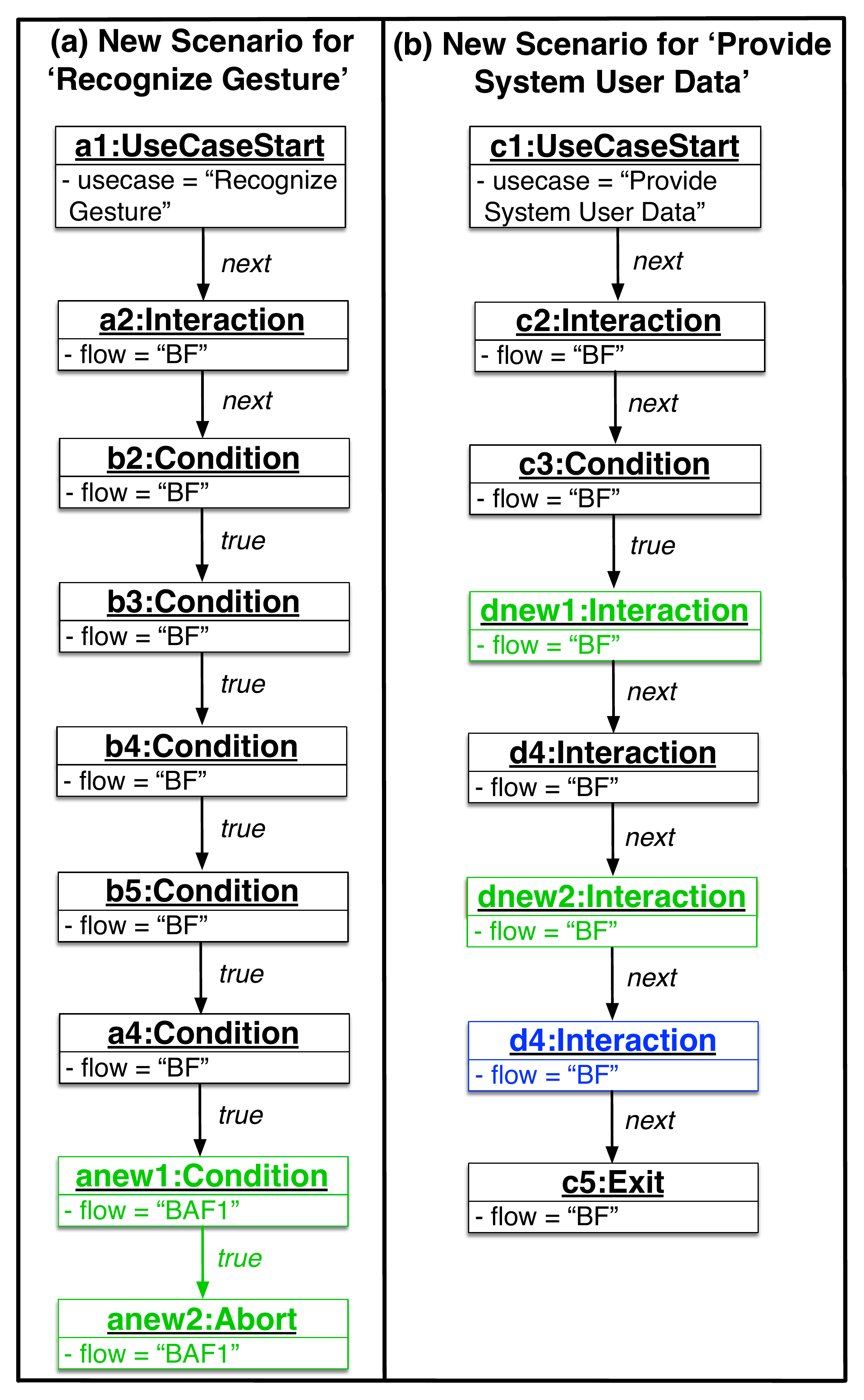}}
        % \centerline{\includegraphics[width=1.0\linewidth]{images/productlineDiagram}}
%\vspace*{-1.05em}
      \caption{Two New Scenarios Derived from the Scenarios in Fig.~\ref{fig:exampleScenarios}}
      \label{fig:newScenarios}
%\vspace*{-1.50em}
\end{figure}
%%\vspace*{-0.9em}  

%Fabrizio 28.10: "guided by" is generating and does not give the idea of how we perform it
%The algorithm implements a depth-first traversal of \emphx{sm} guided by $\mathit{s_{old}}$. 
In Fig.~\ref{algo:identifyNewScenarios}, the algorithm follows a depth-first traversal of \emphx{sm} by following use case steps in \emphx{sm} that have corresponding steps in $\mathit{s_{old}}$. 
%Fabrizio 28.10: the meaning of the following sentence does not mean much, it seems correct first, but in the end what does it mean? The point is not that you take alternative flows, the point is that you follow the branch exiting a condition step according to what you have in the previous product.
%To this end the algorithm takes alternative flows that exist both in the new and previous product if they have already been taken in $\mathit{s_{old}}$ (Lines 8-10 in Fig.~\ref{algo:identifyNewScenarios}). 
To this end, when traversing condition steps, the algorithm follows alternative flows taken in $\mathit{s_{old}}$ (Lines 8-11). 
Whenever a \emphx{Condition} instance is encountered, the algorithm checks if the \emphx{Condition} instance exists also in $\mathit{s_{old}}$ (Line 8). 
If so, the algorithm proceeds with the condition branch taken in $\mathit{s_{old}}$, i.e., the step following the \emphx{Condition} instance in $\mathit{s_{old}}$ (Line 11); otherwise, it takes the condition branch(es) which have not yet been taken in $\mathit{s_{new}}$ (Lines 12-30).

%it creates two new scenarios, one following the true branch and the other following the false branch.  
%takes the condition branches which have not yet been taken in $\mathit{s_{new}}$ %, i.e., the new scenario 
%More precisely, we build new success scenario first. For this reason in the case of specific alternative flows (Lines 25-26) we traverse first the true branch, while in the case of bounded or global alternative flows (Lines 17-18) we traverse first the false branch (i.e., the case in which the condition triggering the alternative flow does not hold).

Alternative flows may lead to execution loops; this happens when alternative flows resume the execution of steps belonging to the originating flows. 
In our current implementation we generate scenarios that cover each loop body once. To this end, when processing condition steps, the algorithm checks if the branches that may lead to cycles have already been traversed (i.e., Lines 14 and 22). If it is the case, the traversal of the scenario is directed towards the branch that brings the scenario out of the cycles (i.e, the true branch for specific alternative flows and the false branch for bounded or global alternative flows as shown in Lines 15 and 23, respectively).

%\begin{figure}[h]
 %\vspace*{-0.7em}
%\begin{wrapfigure}{h}{0.280\linewidth}
\begin{figure}
%\vspace*{-1.30em}
  \centerline{\includegraphics[width=0.50\linewidth]{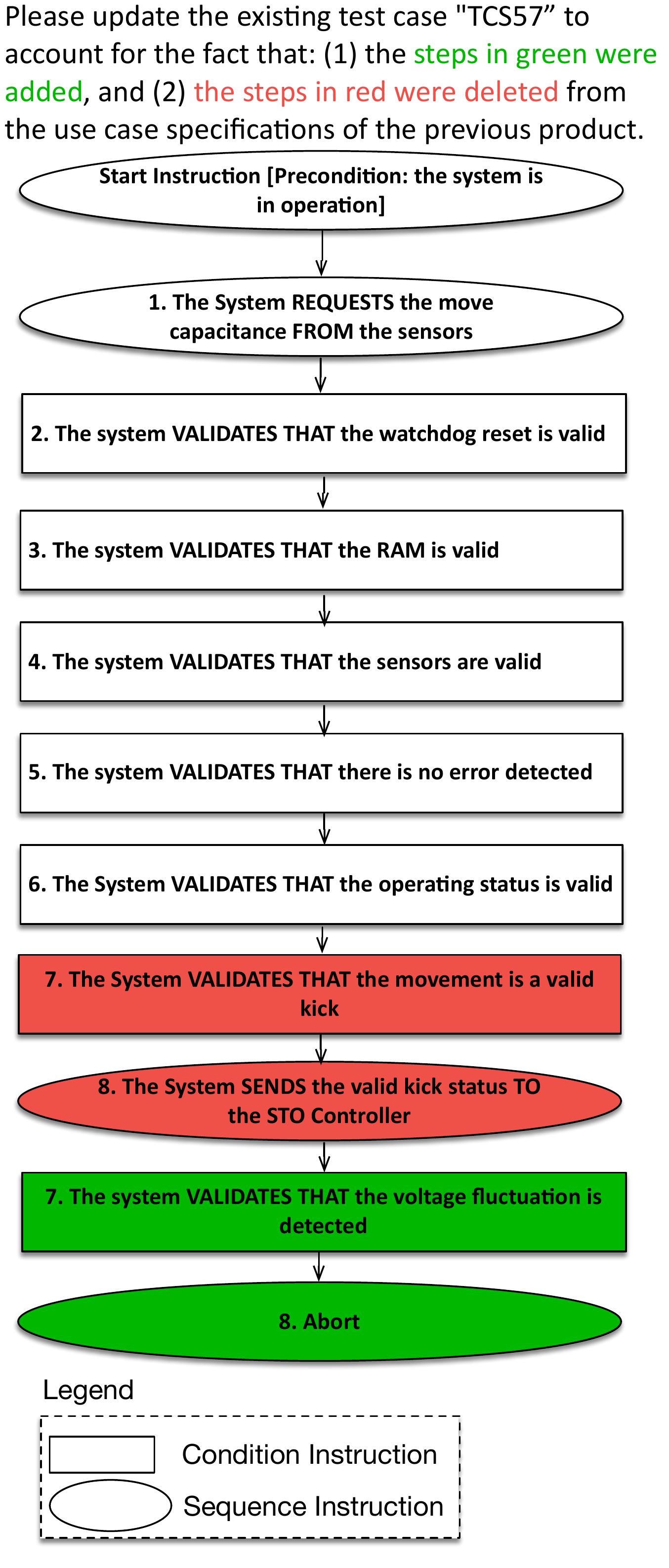}}
% \vspace*{-0.9em}
 \caption{PUMConf's User Interface for Guidance} %to Modify System Test Cases}
 \label{fig:GuidelinesScreenshot}
% \vspace*{-1.4em}
\end{figure}

%Fabrizio: 08.04.2020 Previous version
%\MREVISION{R2.1}{The algorithm in Fig.~\ref{algo:identifyNewScenarios} always terminates since } the generation of $\mathit{s_{new}}$ terminates when an \emphx{Exit} or \emphx{Abort} step is reached (Line 32). 

\MREVISION{R2.1}{The algorithm in Fig.~\ref{algo:identifyNewScenarios} always terminates since (1) the same alternative flow is covered only once and (2) the recursive traversal of the scenario model $sm$ stops when an \emphx{Exit} or \emphx{Abort} step is reached (Line 32). 
The only exception is that of \emphx{Exit} steps of included use cases, which lead to %the traversal of the node
the step that follows the \emphx{Include} step (Line 33).
This \emphx{Exit} step belongs to the use case containing the \emphx{Include} step; therefore, the traversal will eventually reach an \emphx{Exit} or \emphx{Abort} step terminating the recursion.
%Fabrizio: 08.04.2020 Previous version
%Before adding  $\mathit{s_{new}}$ to the result tuple, we automatically compare $\mathit{s_{old}}$ and $\mathit{s_{new}}$, and determine their differences to generate guidance for new test cases ($\mathit{G}$ in Line 37). We provide a set of suggestions for adding, removing and updating test case steps corresponding to added, removed and updated use case steps in  $\mathit{s_{old}}$ and  $\mathit{s_{new}}$.
Before stopping the recursive traversal, the algorithm automatically compares $\mathit{s_{old}}$ and $\mathit{s_{new}}$, and determines 
their differences to generate guidance for new test cases ($\mathit{G}$ in Line 37). We provide a set of suggestions for adding, removing and updating test case steps corresponding to added, removed and updated use case steps in  $\mathit{s_{old}}$ and  $\mathit{s_{new}}$.
Finally, the algorithm adds $\mathit{s_{new}}$, $\mathit{s_{old}}$ and $G$ to the result tuple (Line~38).}
%Using the guidance, the engineer modifies test cases verifying $\mathit{s_{old}}$ to derive new test cases verifying $\mathit{s_{new}}$.   

Fig.~\ref{fig:newScenarios} gives two new scenarios derived from the scenarios in Fig.~\ref{fig:exampleScenarios}. Fig.~\ref{fig:newScenarios}(a) is derived from Fig.~\ref{fig:exampleScenarios}(a) and (b); Fig.~\ref{fig:newScenarios}(b) is derived from Fig.~\ref{fig:exampleScenarios}(c). 
The new scenario in Fig.~\ref{fig:newScenarios}(a) executes the new selected optional bounded alternative flow in which the use case \emphx{Recognize Gesture} aborts due to the voltage fluctuation (see Lines 8-12 in Table~\ref{tab:useCaseRUCM}). While traversing \emphx{sm} for $\mathit{s_{old}}$ in Fig.~\ref{fig:exampleScenarios}(a), the new \emphx{Condition} instance \emphx{anew1} and the new \emphx{Abort} instance \emphx{anew2} (green-colored in Fig.~\ref{fig:newScenarios}(a)) are added in $\mathit{s_{new}}$ to execute the bounded alternative flow. $\mathit{s_{new}}$ in Fig.~\ref{fig:newScenarios}(b) executes the basic flow of the use case \emphx{Provide System User Data} of the new product where the order of one step is updated (blue-colored in Fig.~\ref{fig:newScenarios}(b)) and some new steps are introduced (green-colored) while some others are removed.

Fig.~\ref{fig:GuidelinesScreenshot} shows the generated guideline to modify the test case verifying the retestable scenario in Fig.~\ref{fig:exampleScenarios}(a) for the new scenario in Fig.~\ref{fig:newScenarios}(a). The red and green colors, with a legend, on the scenario explains impacted parts of the corresponding test case. The red steps are deleted while the green ones are added to the scenario. Using this information, the engineer adds and deletes test case steps to cover the new scenario.

\begin{figure}[h]
%\begin{wrapfigure}{h}{0.620\linewidth}
%\vspace*{-1.2em}
    %\centerline{\includegraphics[width=0.82\linewidth]{images/NewTestCase}}
    \centerline{\includegraphics[width=0.99\linewidth]{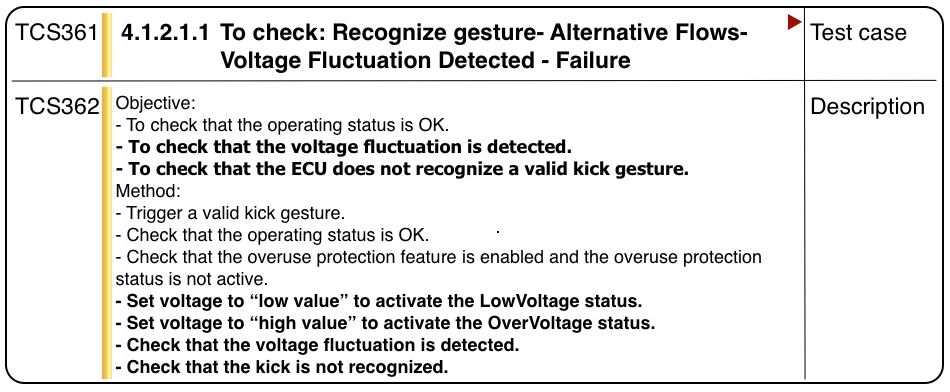}}
%\vspace*{-0.7em}
      \caption{System Test Case derived from the System Test Case in Fig.~\ref{fig:TestCase}}
%\vspace*{-1.5em}
      \label{fig:NewTestCase}

\end{figure}

\begin{figure}[h]
%\begin{wrapfigure}{h}{0.620\linewidth}
%\vspace*{-1.2em}
    %\centerline{\includegraphics[width=0.82\linewidth]{images/NewTestCase}}
    \centerline{\includegraphics[width=0.99\linewidth]{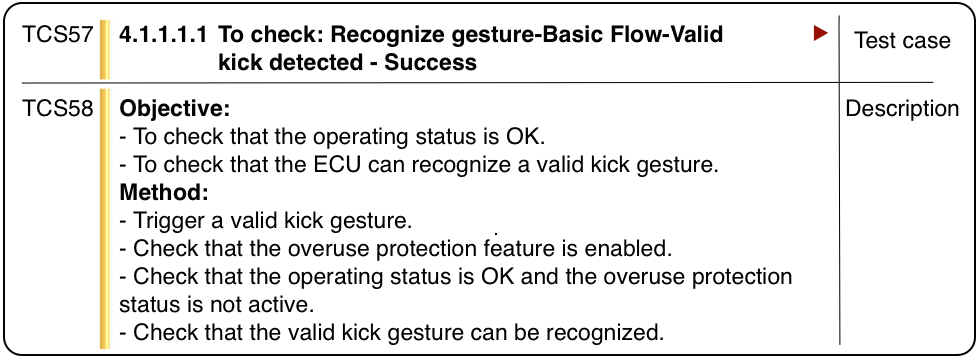}}
%\vspace*{-0.7em}
      \caption{System Test Case for the Scenario in Fig.~\ref{fig:exampleScenarios}(a)}
%\vspace*{-1.5em}
      \label{fig:TestCase}

\end{figure}

\MREVISION{R2.20}{Fig.~\ref{fig:NewTestCase}} shows the header of the test case verifying the new scenario in Fig.~\ref{fig:newScenarios}(a) with the description of the functions under test. For simplification, we omit the implementations of the executable test case. We use the guidance to derive the new test case from the test case in Fig.~\ref{fig:TestCase} verifying the scenario in Fig.~\ref{fig:exampleScenarios}(a). 
The bold lines in Fig.~\ref{fig:NewTestCase} are the new objectives and methods of the test case that correspond to the new use case steps in Fig.~\ref{fig:newScenarios}(a) (i.e., \emphx{anew1} and \emphx{anew2}).

%\FIXME{With the following there are two problems. (1) It does not describe the approach from a general perspective but from a single case. (2) It describes a case handled by merging the results generated from multiple previous scenarios in a section that describes the generation of new scenarios from single previous scenarios. Should we move to the next subsection as part of the report?}
%ARDA: In the following paragraph, I addressed the problems mentioned above. 
%A new scenario might be derived from different old scenarios. For instance, 
A new scenario might be derived separately from multiple old scenarios.
After all the new scenarios are identified for the new product, we automatically detect such new scenarios and provide guidance for only the test cases of the old scenarios from which the engineer generates the new test cases with the least possible changes (Line 19 in Fig.~\ref{algo:classification}). We rank those old scenarios according to the number of changes. If the number of changes are the same, we give priority to scenarios with more changes removing test case steps. We assume that %, while generating a new test case from a old test case, 
removing test case steps is more convenient than adding new steps. 
For instance, the new scenario in Fig.~\ref{fig:newScenarios}(a) is derived from two scenarios in Fig.~\ref{fig:exampleScenarios}(a) and (b). To generate a test case verifying the new scenario in Fig.~\ref{fig:newScenarios}(a), the engineer can modify one of the test cases verifying the scenarios in Fig.~\ref{fig:exampleScenarios}(a) and (b). %If a new scenario can be derived from multiple old scenarios seperately, 
In Fig.~\ref{fig:newScenarios}(a), our approach provides guidance for both scenarios because the number of changes and the number of removed and added test case steps are the same for the two scenarios.       

%\vspace{-2mm}

%\vspace{-3mm}
%\vspace{-2mm}

\subsection{Step 4: Impact Report Generation}
\label{subsec:impact_report_generation}

%We automatically generate an impact report from the classified system test cases of each previous product (Step 4 in Fig.~\ref{fig:Pipeline}). We identify two new scenarios in Fig.~\ref{fig:newScenarios}(a) and (b) when we compare the decision model of the new product in Fig.~\ref{fig:decisionModels}(c) with the decision model of the previous product in Fig.~\ref{fig:decisionModels}(b). However, these two scenarios might have been tested in other previous product(s) in the same product line. Therefore, we also compare with each other new scenarios we identify per previous product. 

We automatically generate an impact report from the classified test cases of each previous product in a product line (Step 4 in Fig.~\ref{fig:Pipeline}). %The report enables engineers to identify a test suite of an existing product to consider for a new product. In addition, 
To enable engineers to select test cases from more than one test suite and thus maximize the number of test cases that can be inherited from previous products, we compare all the test suites in the product line. We then identify sets of new scenarios and reusable and retestable test cases for the product line. 
Assume that there are $\mathit{N}$ previous products in a product line. $\mathit{S_{new1}}$, $\mathit{S_{new2}}$, ..., and $\mathit{S_{newN}}$ are the sets of new scenarios we identify when we compare a new product with each previous product. %$\mathit{T_{reus1}}$, $\mathit{T_{reus2}}$, ..., and $\mathit{T_{reusN}}$ are the sets of reusable test cases; $\mathit{T_{ret1}}$, $\mathit{T_{ret2}}$, ..., and $\mathit{T_{retN}}$ are the sets of retestable test cases we identify when we compare a new product with each previous product. 
%To minimize the number of new scenarios to be covered with new test cases, 
To minimize the number of new test cases the engineer needs to generate, we compute the intersection of the sets of new scenarios ($\mathit{S_{new}}$ = $\mathit{S_{new1}} \,\, \cap \,\, \mathit{S_{new2}} \,\, \cap \,\, ... \,\, \cap \,\ \mathit{S_{newN}}$). 
%Fabrizio: my suggestion
\MREVISION{R2.21}{In other words, a scenario is considered as new only if has not been exercised in any of the previous products. Indeed, the scenarios which are not in the intersection of the sets are covered by at least one reusable or retestable test case in one of the previous products.}
%Fabrizio: we had this previously
%\MREVISION{R2.21}{For instance, the scenario in Fig.~\ref{fig:newScenarios}(b) is added to $\mathit{S_{new}}$ and considered as new scenario for the new product if it is derived as new from each previous product in the product line.}
%The scenarios which are not in the intersection of the sets 
%are covered by at least one reusable or retestable test case in one of the previous products.  
%Fabrizio: my suggestion
%ARDA: The following is a repetition: 'All the remaining scenarios  can thus be exercised by relying on test cases implemented for previous products.'
\MREVISION{R2.21}{If a scenario is exercised only by reusable test cases, 
%Fabrizio: recent change
we select the test case belonging to the most recent product, based on the date of creation of the product. Our rationale is that recent test case implementations are more likely to be reusable (e.g., they do not require updated setup instructions).
%Fabrizio: this is the old sentence 
%we randomly select one of them (they are supposed to be identical). 
We do the same when a scenario is exercised only by retestable test cases. Instead, if a scenario is exercised by both reusable and retestable test cases, we list the previous products in which the test case is identified as either retestable or reusable. Engineers then should decide from which previous product to take the test case.}
%Fabrizio: previous
%Therefore, we take the union of the sets of reusable test cases ($\mathit{T_{reus}}$ = $\mathit{T_{reus1}} \,\, \cup \,\, \mathit{T_{reus2}} \,\, \cup \,\, ... \,\, \cup \,\ \mathit{T_{reusN}}$) and the union of the sets of retestable test cases ($\mathit{T_{ret}}$ = $\mathit{T_{ret1}} \,\, \cup \,\, \mathit{T_{ret2}} \,\, \cup \,\, ... \,\, \cup \,\ \mathit{T_{retN}}$). 
%If a test case is considered both retestable and reusable (i.e., ($\mathit{T_{reus}}  \,\, \cap \,\, \mathit{T_{ret}}) \, \neq  \, \emptyset$), we list the previous products in which the test case is identified as retestable and reusable. 
%\MREVISION{R2.21}{For instance, the previous product(s) and their test case exercising the scenario in Fig.~\ref{fig:newScenarios}(b) are provided to the engineer if the test case exercising this scenario is identified as retestable in a previous product while, at the same time, it is identified as reusable in another previous product.}

Based on the system under test, engineers decide whether to select test cases from a single test suite or from multiple test suites in the product line. For example, if multiple products include different setup procedures (e.g., due to different \MREVISION{R3.13}{hardware} architecture or library versions being used) that need to be executed at the beginning of each test case, it is more practical to select test cases from a single test suite. 

%The generated impact report lists the classified system test cases per previous product and gives the overall classification in the entire product line. The analyst/test engineer uses the generated impact report to decide to select which system test cases from which test suite.  

%
% !TEX root =  Main.tex
\section{Prioritization of System Test Cases} %in a Product Family}
\label{sec:prioritization}

\begin{figure}[t]
%\vspace*{-0.6em}
\centerline{\includegraphics[width=0.80\linewidth]{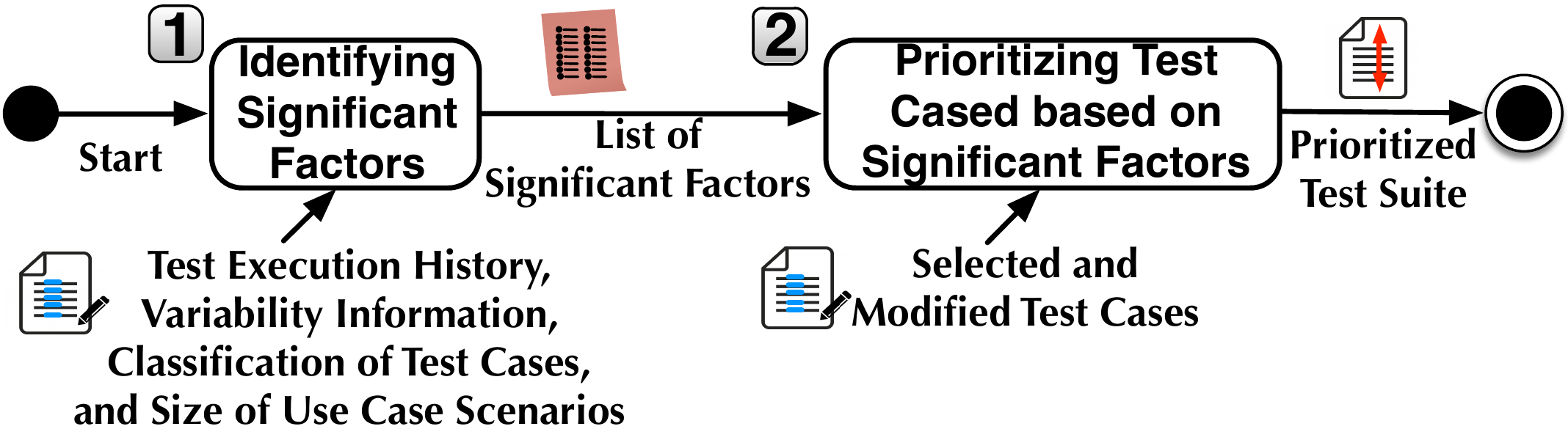}}  
%\vspace*{-0.6em}
\caption{Overview of the Test Case Prioritization Pipeline}\label{fig:PrioritizationPipeline}
%\vspace*{-1.0em}
\end{figure}

Test case prioritization is implemented as a pipeline (see Fig.~\ref{fig:PrioritizationPipeline}). The pipeline takes as input the test suite of the new product, the test execution history of the previous products 
(i.e., the outcome of each test case of the product test suite, for each previous product and version), 
%(e.g., the number of failing products and versions for each test case) 
the size of the use case scenarios exercised by the test cases, the classification of the test cases (i.e., reusable or retestable), and \MREVISION{R3.14}{the PL use case.} A new version of a product is deployed after the previous version has been tested and fixed. Requirements remain identical from version to version of the same product. \MREVISION{R3.15}{When requirements evolve, they are considered to characterize a different, new product in the product line.}
Based on a prediction model using these factors, the test cases of the given test suite are sorted 
%The input test cases are prioritized as output for the new product. More specifically, the test cases are sorted 
to maximize the likelihood of executing failing test cases first.

The prioritization pipeline gives the highest priority to test cases covering new scenarios (i.e., scenarios not available for previous products) since they exercise features that have never been tested before. %and are thus the natural starting point for testing.
% in the test suite of the new product. In our prioritizat
%We were missing a motivation for each factor.
The prioritization of retestable and reusable test cases is instead driven by a set of factors typically correlated with the triggering of failures, according to the relevant literature (e.g.,~\cite{Srikanth2005, Engstrom2011b, Wong1997, Rothermel2001, Li2007}): \textit{the number of previous products in which the test case failed}, \textit{the number of previous products' versions in which the test case failed}, \textit{the size of the scenario exercised by the test case}, 
\textit{the degree of variability in the use case scenario exercised by the test case}, 
and \textit{the classification of the test case (i.e., reusable or retestable)}. 
Note that different versions of a product share the same test suite because functional requirements do not vary across the versions of the same product.
\CHANGED{Since this test suite does not contain obsolete test cases, they are not considered to build the regression model.}  
The number of previous products in which the test case failed  and the number of versions in which the test case failed capture the fault proneness of the test cases, a factor typically considered by other test case prioritization approaches~\cite{Srikanth2005, Engstrom2011b}. 
The size of the use case scenario exercised by a test case is measured in terms of the number of use case steps it contains.
The scenario size captures the complexity of the operations performed by the system during the execution of the test case, under the assumption that longer scenarios require more complex software implementations. Implementation complexity is one of the factors considered in other requirements-based prioritization approaches~\cite{Srikanth2005}. 
The degree of variability in the use case scenario exercised by a test case is measured by counting the number of decision elements included in the scenario.  
In the presence of high variability, it is more likely that some of the system properties verified by the test case are not implemented properly. 
Finally, the classification of a test case as retestable is considered for prioritization since, by definition, the scenario exercised by a retestable test case might be affected by changes in behaviour and thus may trigger a failure.

%The pipeline has three steps (see Fig.~\ref{fig:PrioritizationPipeline}). The first two steps take the variability information of the product line and the test execution history of the previous products to identify the factors correlated with the presence of faults in the previous products. With variables representing the factors and coefficients representing the weights of the factors, we construct a formula to be used to sort test cases in the third step (see \emphx{prioritization formula} in Fig.~\ref{fig:PrioritizationPipeline}).

All these factors mentioned above may have varying importance for test case prioritization in different product lines due to technical and organizational factors. Some factors may even not significantly affect test case prioritization for some product lines. 
To account for the varying importance of risk factors, the pipeline identifies factors significantly correlated with the presence of failures. Test cases are then prioritized based on a prediction model relying on the correlated factors.

The prioritization pipeline includes two steps. In Step 1, \textit{Identifying significant factors}, our approach automatically identifies significant factors for prioritizing the test cases of a new product. To this end, we employ logistic regression~\cite{Hosmer2013}, i.e., a predictive analysis to determine the relationship between one dependent binary variable (i.e., the failure of a test case) and one or more independent variables, which might be either numeric (e.g., the number of the products in which the test case failed in the past) or binary (e.g., the fact that a test case has been classified as retestable). 
%More specifically, logistic regression estimates the parameters of a binary logistic model which has a dependent variable with two possible values, such as pass/fail, which are labeled "0" and "1". Independent variables can each be a binary variable (two classes) or a continuous variable (any real value). 

% !TEX root =  Main.tex
\begin{table*}
\scriptsize
\caption{Excerpt of the Training Data Set used for Logistic Regression}
% \vspace*{-1.3em}
\begin{center}
\begin{tabular}{|p{0.75cm}|p{0.75cm}||p{0.45cm}|p{0.40cm}||p{0.98cm}|p{1.10cm}|p{1.00cm}|p{1.10cm}|p{1.1cm}|}

\hline

\textbf{Product ID}&\textbf{Version ID}&\textbf{Test Case ID}& \textbf{Fails}& \textbf{Retestable (R)}& \textbf{Size of the Use Case Scenario (S)}&\textbf{Degree of Variability of the Scenario (V)}&\textbf{\# of Previous Products in which it Fails (FP)}&\textbf{\# of Previous Versions in which it Fails (FV)}\\

\hline
\hline
\CH&	V1&	TC1	&	1	&	0	&8&	2&0&0\\
\CH&	V1&	TC2	&	0	&	0	&4&	1&0&0\\
\hline
\CH&	V2&	TC1	&	1	&	0	&8&	2&0&1\\
\CH&	V2&	TC2	&	0	&	0	&4&	1&0&0\\
\hline
\CH&	V3&	TC1	&	0	&	0	&8&	2&0&2\\
\CH&	V3&	TC2	&	0	&	0	&4&	1&0&0\\
\hline
\CH &V4&	TC1	&	0	&	0	&8&	2&0&2\\
\CH &V4&	TC2	&	0	&	0	&4&	1&0&0\\
\hline
\hline
%The TC becomes retestable so the size of the scenario may change
\SGM& V1& TC1	&	1	&	1	&9&	3&1&2\\
\SGM& V1& TC2	&	0	&	0	&4&	1&0&0\\
\SGM& V1& TC3	&	0	&	0	&4&	1&0&0\\
\hline
\SGM& V2& TC1	&	0	&	1	&9&	3&1&3\\
\SGM& V2& TC2	&	1	&	0	&4&	1&0&0\\
\SGM& V2& TC3	&	0	&	0	&4&	1&0&0\\
\hline
\SGM& V3& TC1	&	0	&	1	&9&	3&1&3\\
\SGM& V3& TC2	&	0	&	0	&4&	1&0&1\\
\SGM& V3& TC3	&	0	&	0	&4&	1&0&0\\
\hline
\hline
%The TC is again retestable, it's size does not change because they alter the order
\DM& V1& TC1	&	1	&	1	&9&3&2&3\\
\DM& V1& TC2	&	1	&	1	&5&2&1&1\\
\DM& V1& TC3	&	0	&	0	&4&	1&0&0\\
\hline
\DM& V2& TC1	&	1	&	1	&9&3&2&4\\
\DM& V2& TC2	&	0	&	1	&5&2&1&2\\
\DM& V2& TC3	&	0	&	0	&4&	1&0&0\\
\hline

\end{tabular}
\end{center}
\label{table:logisticRegressionInput}
% \vspace*{-1.2em}
\end{table*}

In our context, the logistic regression model estimates the logarithm of the odds that a test case fails. 
The logistic regression model is trained using variability information, the size of the use case scenarios exercised by the test cases, the classification of the test cases, and the execution history of the test cases for previous products. 
The logistic regression model has the following form:

% for one the products used in our empirical evaluation (P5, see Section~\ref{sec:evaluation}) is}
%\begin{math}
$\mathit{ln} \left(  \frac{p(\mathit{TC}_x)}{1-p(\mathit{TC}_x) } \right) = \beta_{0} + \beta_{1} * \mathit{V} + \beta_{2} * \mathit{S}  + \beta_{3} * \mathit{FP}  + \beta_{4} * \mathit{FV} + \beta_{5} * \mathit{R}$
%\end{math}

where $p(\mathit{TC}_x)$ is the probability that test case $\mathit{TC}_x$  fails, $V$ is the degree of variability of the scenario exercised by the test case (i.e., the number of decision elements in the scenario), $S$ is the size of the use case scenario exercised by the test case (i.e., the number of steps), 
$\mathit{FP}$ is the number of failing products, $\mathit{FV}$ is the number of failing versions, and $R$ indicates whether the test case has been classified as retestable.  
$\beta_{0}$ is the intercept, while  $\beta_{1} ... \beta_{5}$ are coefficients which are derived, using the iteratively reweighted least squares approach~\cite{Coleman-SSI-1980}, to estimate the effect size on the failure probability. %\footnote{The probability that the test case fails is computed as $p(\mathit{TC}_x) = \frac{1}{1+e^{-(\beta_{0} + \beta_{1} * \mathit{V} + \beta_{2} * \mathit{S}  + \beta_{3} * \mathit{FP}  + \beta_{4} * \mathit{FV} + \beta_{5} * \mathit{R})}}$}.

%, and they capture how likely each factor impacts on the probability of a failure.}

%In our approach, 
%The logistic regression model is trained using the variability information and execution history of the test cases developed for the previous products. 

We rely on the R environment~\cite{Rproject} to derive the logistic regression model. 
Our toolset automatically generates from the available data the training data set to be processed by the R environment. Table~\ref{table:logisticRegressionInput} shows an excerpt of an example training data set generated by our toolset. 

Table~\ref{table:logisticRegressionInput} includes the failure history of products $\CH$, $\SGM$~and $\DM$ to be used to prioritize the test cases for $\GAC$. Each row in Table~\ref{table:logisticRegressionInput} reports the information belonging to a single test case executed against a version of a product.
The first and second columns represent the product and its version, respectively. 
The third column reports the test case identifier, while the fourth column indicates whether the test case fails (i.e., the dependent variable). 
The rest of the columns in Table~\ref{table:logisticRegressionInput} represent independent variables used to predict failure. 
%The fifth column gives the classification of the test case, i.e., the output of our test case classification technique. 
The fifth column indicates if the test case has been classified as retestable.
The sixth column reports the size of the use case scenario exercised by the test case. 
%The sixth column reports the size of the use case scenario exercised by the test case (i.e., the number of steps). 
The seventh column reports the degree of variability of the scenario exercised by the test case.
%The seventh column reports the degree of variability of the scenario exercised by the test case (i.e., the number of decision elements in the scenario). 
Table~\ref{table:logisticRegressionInput}, for instance, shows that test case $TC1$ executed against $\SGM$ covers nine use case steps while the same test case covers eight use case steps when executed against $\CH$; this is due to the covered use case scenario in $\SGM$ including one additional variant element than the use case scenario covered in $\CH$ (see column \emphx{Degree of Variability}). 
%The test case $T1$ when executed against \SGM. 
Test case $TC3$ has been introduced in $\SGM$~to cover one additional use case scenario not present in $\CH$. 
The eighth and ninth columns report the number of products and the number of versions in which the test case fails, respectively.

To identify the significant factors for test case prioritization, we apply \textit{the p-value method of hypothesis testing} based on Wald test~\cite{Rice2007}.
 The method relies on the failure probability predicted by the regression model to determine whether there is evidence to reject the null hypothesis that 
\emphx{there is no relationship between the two variables}. 
%The alternative hypothesis is \emphx{there exist a relationship between the two variables}.
%which is the initial claim about a population of statistics. 
%In hypothesis testing, we have two hypothesis:
%
%%We have a two-step technique that uses logistic regression to identify significant factors for test case prioritization. As a first step, we apply \textit{the p-value method of hypothesis testing}~\cite{Rice2007} that uses calculated probability to determine whether there is evidence to reject the null hypothesis which is the initial claim about a population of statistics. In hypothesis testing, we have two hypothesis:
%
%\textbf{H0 (the null hypothesis):} There is no relationship between the two variables.
%
%\textbf{H1 (the alternative hypothesis):} There exist a relationship between the two variables.
The p-value indicates the likelihood of observing the data points when the null hypothesis is true. Therefore, if the p-value is smaller than a given threshold (we use 0.05) then it is unlikely that the dataset has been generated by chance and, consequently, the null hypothesis can be rejected (i.e., there is a relationship between the factor and the dependent variable). In the model, we keep the given factors whose p-value is smaller than the threshold.
To automatically determine significant factors, we rely on the p-value computed by the Wald test on the logistic regression model trained by including all the factors.
Finally, we derive a new, multivariate logistic regression model that includes only the significant factors.
For example, the logistic regression model derived for one of the products used in our empirical evaluation (see P4 in Section~\ref{sec:evaluation}) is the following:

\begin{math}
\mathit{ln} \left(  \frac{p(\mathit{TC}_x)}{1-p(\mathit{TC}_x) } \right) = - 1.50 - 0.25 * \mathit{V}  + 0.04 * \mathit{S}  +0.53 * \mathit{FV}  - 1.01  * \mathit{R}
%\mathit{ln} \left(  \frac{p(\mathit{TC}_x)}{1-p(\mathit{TC}_x) } \right) = - 1.17 -1.02 * \mathit{V}  + 0.04 * \mathit{S}  -0.09 * \mathit{FP}  + 0.63 * \mathit{FV} - 0.67 * \mathit{R}
\end{math}

This model, for example, does not include the number of failing products ($\mathit{FP}$) since it is not significant according to the computed p-value.

The generated logistic regression model is a predictive model that returns, based on the significant factors, the probability that a test case fails. %Because of this, 
%The formula above thus computes the logarithm of odds that a test case fails, and, in turn the probability that a test case fails. The coefficients in the formula capture how much each factor impacts on the probability of a failure; in the formula above, the number of failing versions is the factor that impacts more.}
In Step 2, \textit{Prioritize test cases}, we prioritize test cases by relying on the probability calculated by the regression model. The test cases are sorted in descending order of probability and presented to engineers.

\section{\MREVISION{R4.16}{Tool Support}}
\label{sec:tool}
 
We have implemented our test case selection and prioritization approach as an extension of PUMConf. For accessing the tool and some representative models, see:~\url{https://sites.google.com/site/pumconf/}.

\begin{figure}[h]
 %\vspace*{-0.7em}
 \centerline{\includegraphics[width=0.80\linewidth]{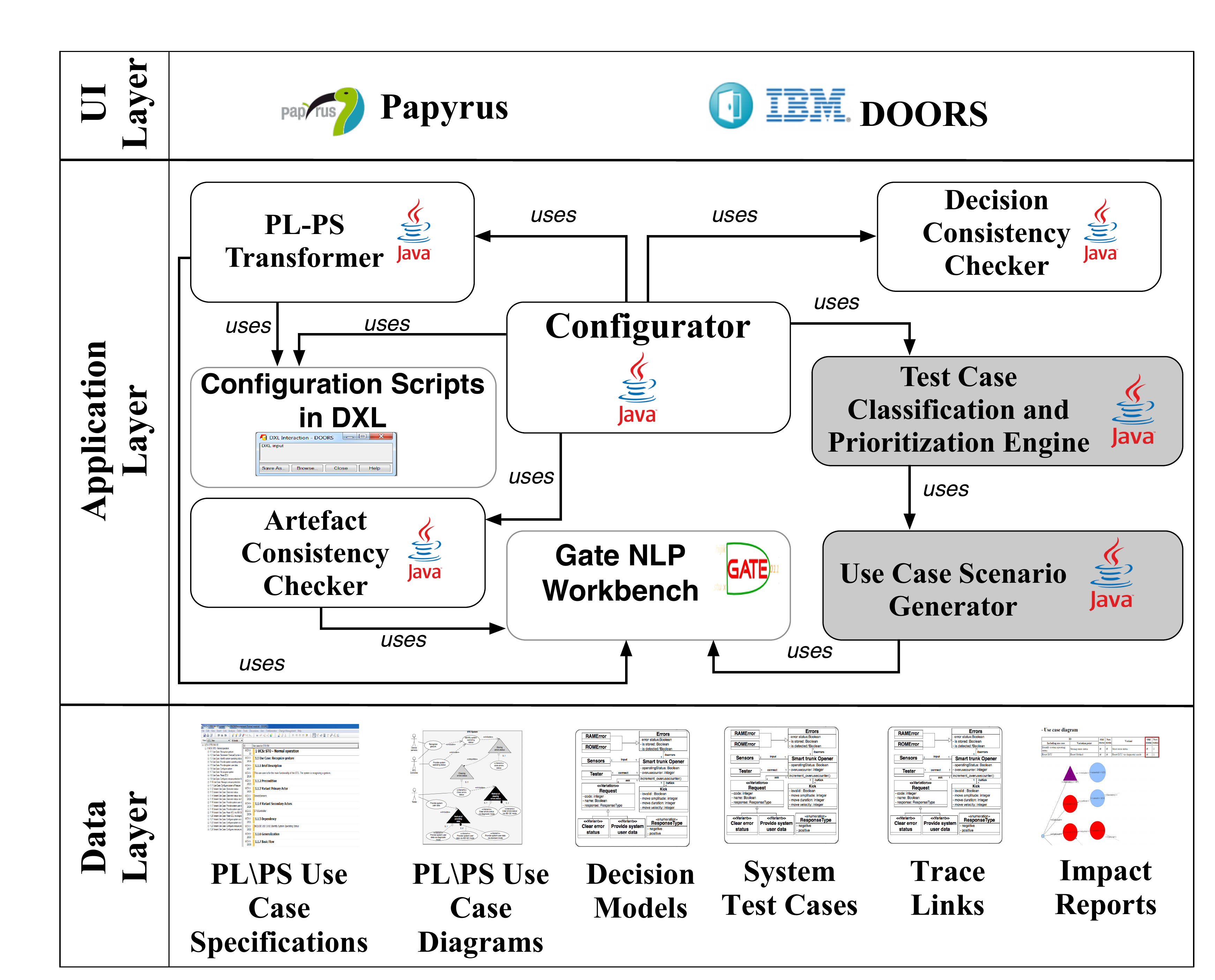}}
% \vspace*{-0.9em}
 \caption{Layered Architecture of PUMConf}
 \label{fig:ToolArchitecture}
% \vspace*{-0.7em}
\end{figure}

Fig.~\ref{fig:ToolArchitecture} shows the layered architecture of our tool PUMConf. It is composed of three layers: (i) the \textit{User Interface (UI) layer}, (ii) the \textit{Application layer}, and (iii) the \textit{Data layer}. We briefly introduce each layer and explain the new components, i.e., the gray boxes in Fig.~\ref{fig:ToolArchitecture}.

\textbf{User Interface (UI) Layer} supports creating and viewing PL and PS use case models (i.e., use case diagrams and specifications) and system test cases, and displaying the generated impact reports. We employ Papyrus (\small \url{https://www.eclipse.org/papyrus/}\normalsize) for use case diagrams and IBM Doors (\small \url{www.ibm.com/software/products/ca/en/ratidoor/}\normalsize) for use case specifications and system test cases. The impact reports are visualized as part of IBM Doors output using JGraph (\small \url{https://www.jgraph.com/}\normalsize), Microsoft Excel (\small \url{https://products.office.com/en/excel/}\normalsize) and html. 

\textbf{Application Layer} supports, with the new components, the main activities of our approach in Fig.~\ref{fig:overview}. %: \textit{classifying system test cases}, \textit{providing guidance to modify system test cases}, and \textit{prioritizing the selected system test cases}.
The \textit{Configurator} component coordinates the other components in the application layer. The \textit{Artifact Consistency Checker} and \textit{Decision Consistency Checker} components were introduced in our previous work~\cite{Hajri2016c, Hajri2018}. The \textit{Artifact Consistency Checker} employs NLP to check the consistency of the PL use case diagram and the PL use case specifications complying with the RUCM template. To perform NLP, our tool employs the GATE workbench (\small\url{http://gate.ac.uk/}\normalsize), an open source NLP framework. The \textit{Decision Consistency Checker} supports inferring decision restrictions and checking their consistency. The \textit{PL-PS Transformer} component annotates the use case specifications using NLP to automatically generate PS use case specifications. It uses scripts written in the Doors eXtension Language (DXL) to automatically (re)configure PS use case specifications. %The DXL scripts are also used to load the (re)configured use case specifications into Doors.

We further implemented some new components: \textit{Test Case Classification and Prioritization Engine} and \textit{Use Case Scenario Generator}. The \textit{Use Case Scenario Generator} also employs the GATE workbench to extract control flow information, i.e., the order of alternative flows and their conditions, from use case specifications. With the extracted control flow information, it identifies the new and already tested use case scenarios. These scenarios are used by the \textit{Test Case Classification and Prioritization Engine} component to classify system test cases and to provide guidance to modify system test cases for new use case scenarios that have not been tested before. To prioritize system test cases, the \textit{Test Case Classification and Prioritization Engine} employs R scripts (\small \url{https://www.rdocumentation.org/}\normalsize) that implement logistic regression.

%Generalized Linear Models (GLM) function as an R script (\small \url{https://www.rdocumentation.org/packages/stats/versions/3.5.1/topics/glm}\normalsize) implementing logistic regression.

\textbf{Data Layer.} The PL and PS use case specifications are stored in the native IBM DOORS format while the PL and PS use case diagrams are stored as UML models. The decision models are saved in Ecore~\cite{EclipseEMF}. 
We generate the impact reports as Microsoft Excel spreadsheets and html pages. Depending on industrial practice, the traceability links between use case specifications and system test cases can be saved in Excel spreadsheets or in IBM DOORS link database. 
%The generated impact reports are stored as Excel spreadsheets and html pages.

%
% !TEX root =  Main.tex

\newcommand{\STOproductsN}{five\xspace}
\newcommand{\STOproductsNv}{5}

\newcommand{\STOVersionsN}{54\xspace}
\newcommand{\STOVersionsNv}{\RED{XX}\xspace}

\section{Evaluation}
\label{sec:evaluation}

%We have conducted an empirical evaluation with the aim to respond to the following research questions:
Our objective is to assess, in an industrial context, %the effectiveness of our approach. 
%the effectiveness of our test case classification and prioritization approach. %in industrial settings. 
%We evaluate whether we could improve reuse and reduce manual effort by automatically classifying and prioritizing system test cases in a product line. 
whether our approach could improve test case reuse and reduce testing effort.
% by automatically classifying and prioritizing system test cases in a product line.
%of previously developed products for each new product in a product line.
%We report on the evaluation of our approach, in terms of (i) system test case classification capability when configuring use cases of a new product of an industrial product line; (ii) capability of prioritizing system test cases selected for a new product of an industrial product line under multiple risk factors.
%We assess the effectiveness of our approach by answering the following research questions:
This empirical evaluation aims to answer the following research questions (RQs):
\begin{itemize}
\item \textit{RQ1. Does the proposed approach provide correct test case classification results?} This research question aims to evaluate the precision and recall of the procedure adopted to classify the test cases developed for previous products.

\item \textit{RQ2. Does the proposed approach accurately identify new scenarios that are relevant for testing a new product?} This research question aims to evaluate the precision and recall of the approach in identifying the new scenarios to be tested for a new product (i.e., new requirements not covered by existing test cases).

%\item \textit{RQ3. Are the provided guidelines for the definition of new system test cases beneficial to software engineers?} This research question aims to determine whether the guidelines provided by the approach ease the implementation of test cases covering new use case scenarios.

%FABRIZIO: I hided the following because we will never be able to answer it.
%\item \textit{RQ3. Does the proposed approach improve the quality of test suites compared to current industrial practice?}
%This research question aims to determine whether the approach does not miss relevant scenarios and identifies new scenarios that are likely to be missed when the classification of test cases is done manually.

\item \textit{RQ3. Does the proposed approach successfully prioritize test cases?} 
This research question aims to determine whether the approach  is able to effectively prioritize system test cases 
that trigger failures and thus can help minimize testing effort while retaining maximum fault detection power. 

%compared to the current practice at IEE.

\item \textit{RQ4. Can the proposed approach significantly reduce testing costs compared to current industrial practice?}
This research question aims to determine to what extent the proposed approach can help significantly reduce the cost of defining and executing system test cases. 

\end{itemize}

%Our approach takes into account multiple test suites.
%\newline \textbf{Objective}: minimizing the effort of testing by considering all the configurations together. 
%\newline To evaluate our approach we will:
%\begin{itemize}
%\item check the correctness of (1) the classification of existing test cases, and (2) the guidance for building test cases covering new scenarios from existing test cases.
%\item check the completeness of the identification of new scenarios (we don't miss any new scenario)
%\item measure the saving effort of testing
%\newline providing minimum number of retestable/obsolete test cases and maximum number of reusable test cases.
%\newline saving in the test case execution and test case generation (discard obsolete test cases)
%\item check the usefulness of our approach: is it really useful?
%\item check the practicality of our approach: the reusable test case are really reusable in practice?
%\end{itemize}

\subsection{Subject of the Study}

The subject of our study is the Smart Trunk Opener (STO) system developed by our industry partner IEE. %, introduced in Section~\ref{sec:background}.
STO has been selected for the assessment of our approach since it is one of the newest IEE products involving multiple customers requiring varying features. %STO has been developed and tested following the PL testing strategy \textit{opportunistic reuse of test assets}. 
%The author is referred to Section~\ref{sec:context} for the current PL engineering practice for STO. 
The development history of the STO product line includes \STOproductsN products delivered to different car manufacturers. 
STO customers include major car manufacturers working in the European, Asian and US markets,  with 2017 sales ranging from 200,000 to 3 million vehicles.
For each product, IEE engineers developed multiple versions, each sharing the same functional requirements but differing with respect to non-functional requirements (e.g., hardware selection or performance optimizations). In total, STO includes \STOVersionsN versions. 

\begin{table*}[h]
% \vspace*{-0.6em}
\scriptsize
\caption{Overview of the STO Product Line Use Cases}
 \vspace*{-0.9em}
\begin{center}
\begin{tabular}{|p{1.455cm}|p{0.85cm}|p{1.0cm}|p{1.00cm}|p{0.80cm}|p{0.5cm}|p{1.4cm}|p{1.39cm}|}

\hline 
   
  &\textbf{\# of Use Cases}
  &\textbf{\# of Variation Points}
  &\textbf{\# of Basic Flows} 
  & \textbf{\# of Altern. Flows}
  & \textbf{\# of Steps}
  &\textbf{\# of Optional Altern. Flows} 
    &\textbf{\# of Optional Steps} \\
  
\hline
\textbf{Essential UCs} & 15 & 5 & 15 & 70 & 269 & 5 & 14 \\
\textbf{Variant UCs} & 14	& 3 & 14 & 132 & 479 &  8 & 13  \\
\hline
\textbf{Total} &	29 & 8 & 29 & 202 & 748 & 13 & 27\\
\hline

\end{tabular}
\end{center}
\label{table:exp:PLmodelsSize}
% \vspace*{-0.8em}
\end{table*}

\MREVISION{R2.2, R4.11}{To develop the STO system, IEE engineers elicited requirements as use cases from an initial customer. 
For each new customer, they cloned the current use cases and identified differences to produce new use cases.  
%With such practice, IEE did not have any documentation of commonalities and variabilities across STO products. 
%IEE engineers followed the strategy \textit{opportunistic reuse of test assets} to test the STO products. 
IEE provided their initial STO documentation, which contained a use case diagram, use case specifications, and supplementary requirements specifications describing nonfunctional requirements and domain concepts. The initial documentation was the output of their current clone-and-own reuse practice. That documentation contains variability information only in the form of some brief textual notes attached to the relevant use case specifications. To model the STO requirements according to our product line use case modeling method, PUM~\cite{Hajri2015, Hajri2016c}, we first examined the initial STO documentation. Since the initial documentation contains almost no structured variability information, we had to work together with two IEE engineers (one software development manager and one embedded software engineer) to build and iteratively refine our models. When we started to study the STO documentation, the STO project was in its initial phase and there was only one prototype implementation to discuss with some potential customers. One may argue that it is not always easy to identify variations in requirements when a new project starts. However, the IEE engineers stated that, most of the time in their domain of applications, requirements and their variability can be identified with the first customer.}

\MREVISION{R2.2, R4.11}{After studying the initial STO documentation and meeting with the IEE engineers, we built the PL use case diagram and specifications for STO.} Table~\ref{table:exp:PLmodelsSize} provides an overview of the STO product line. The data in Table~\ref{table:exp:PLmodelsSize} shows that the system implements 29 use cases, each one being fairly complex since the use cases in total include 202 alternative flows (i.e., alternative cases to be considered when implementing the use case). The STO product line is highly configurable, with 14 variant use cases, 8 variation points, 13 optional alternative flows and 27 optional steps. \MREVISION{R2.2}{Except for the conflict relationship between use cases and the variant order group, we used all the PUM features in the STO PL use case diagram and specifications.} STO has the size and characteristics of typical embedded product line systems managing automotive components.
%To apply the proposed approach, we have considered STO requirements written according to PUM~\cite{Hajri2015}~\cite{Hajri2016c}. %, derived by the first two authors of this paper in collaboration with IEE engineers for our previous work~\cite{Hajri2015}~\cite{Hajri2016c}.

\MREVISION{R2.2, R4.11}{When discussions start with a customer regarding a specific product, IEE engineers need to make decisions on variability aspects documented in PL use case models. At a later stage, when IEE had already developed various STO products for different car manufacturers, we used PUMConf, together with engineers, to configure the PS use case models for five STO products. Configuration decisions were made on the PL use case models using the guidance provided by PUMConf. IEE also provided the test suites of the products. } %and the traceability links between the test cases in the test suites and the use case specifications they had already specified for the products.} 

\MREVISION{R2.2, R4.11}{All the generated PS use case models were confirmed by the IEE engineers to be correct and complete. The PL use case models that we derived from the initial STO documentation were sufficient to make all the configuration decisions needed in PUMConf and to generate the correct and complete PS use case models for the five STO products.}

Table~\ref{table:exp:details} reports information about the STO products including the number of versions for each product. In Table~\ref{table:exp:details}, the products are sorted according to their delivery date, with P1 being the first product of the product line, and P5 being the last.

\begin{table}[h]
 %\vspace*{-1.0em}
\scriptsize
\caption{Details of the Configured Products in the STO Product Line}
 %\vspace*{-1.3em}
\begin{center}
\begin{tabular}{|p{1.3cm}|p{1.4cm}||p{1.6cm}|p{0.95cm}|p{1.6cm}||p{1.6cm}|}

\hline
\textbf{Product ID}& \textbf{\# of Versions}&\multicolumn{3}{c||}{\textbf{\# of Use Case  Elements}}& \textbf{\# of Test Cases}\\
\hline
&&\textbf{Use Cases}&\textbf{Use Case Flows}&\textbf{Use Case Steps}&\\

\hline

\CH& 22	& 28 & 236& 689 & 110	\\
\SGM& 8 &25  &169 & 568 & 86 \\
\DM& 10	& 28 &234&685 & 96	\\
\GAC& 5 & 26 & 212 &618 & 83\\
\VW& 9 & 28 & 238 & 695 & 113 \\
%\BAIC& 1 & &&& \\
%\Brilliance& 1 & &&& \\
%\GWM& 4 & &&& \\

\hline
\end{tabular}
\end{center}
\label{table:exp:details}
 %\vspace*{-1.2em}
\end{table}

%From product to product, the degree of test cases being reused depend also on the testing team, which often vary and may decide to not rely on exiting test cases. 
The different STO products are characterized by different test suites of different sizes while the same test suite is shared by all the versions of the same product since their functional requirements do not vary.  
The test cases have been traced to the use case specifications by IEE engineers. %by relying on the features provided by IBM Doors.
\MREVISION{R2.2}{Note that requirements traceability is a systematic practice at IEE since it is enforced by automotive functional safety standards such as ISO-26262~\cite{ISO26262}.} Column \emphx{\# Test Cases} in Table~\ref{table:exp:details} shows, for every product, the number of test cases belonging to the functional test suite of the product. 
%In our experiments we considered only the functional test suite not the robustness test suite because  

\MREVISION{R1.3}{All the faults considered in our evaluation are real faults previously identified by IEE during testing. Because of the confidentiality agreement we signed with our industry partner, we cannot share fault data.}

\subsection{Experiment Setup}

%This empirical evaluation targets both the test case classification and the test case prioritization strategies proposed in this paper.

Our approach for test case classification can be applied using single-product settings (i.e., to classify and prioritize test cases that belong to a previous product) and whole-line settings (i.e., to classify test cases of multiple previous products).
To evaluate our approach for test case classification and to spot differences in terms of classification results with the two configurations (e.g., number of test cases that can be reused),
%for each STO product except the first one (i.e., P1), %we applied our approach to classify test cases of previous products. %We excluded the first product of the product line (i.e., P1) since no previous product are available in this case.
we applied the approach using both settings.
%single-product settings (i.e., to classify and prioritize test cases that belong to a single previous product) and whole-line settings (i.e., to classify test cases of multiple previous products).
To evaluate test case prioritization, %we applied it to the test suites we generated after we used our approach for test case classification. 
we prioritized test suites developed to test different STO products.
% (we used the original STO test suites since these are the ones for which historical information is available).
We applied test case prioritization to the entire test suite since its execution is required by safety standards for every product being released. 
%To perform this evaluation 
%We considered test suites generated in both the single-product and whole-line settings.

%The proposed approach require that test cases are linked to requirements. To apply our approach to STO, for every STO product, we have manually identified the links between the test cases and the use case specifications by relying on our toolset. More specifically, we have derived traceability links with the granularity of use case flows (i.e., for every test case we indicate the alternative flows it covers).

%
%To perform the experiment we applied the proposed approach
%For every product, we have the PS use case specifications derived from PL

\subsection{Results}
\MREVISION{R2.24}{This section discusses the results of our case study, addressing in turn each of the research questions.}

\subsubsection{RQ1: \MREVISION{R2.23}{Does the proposed approach provide correct test case classification results?}}

%RQ1 aims to determine whether our approach correctly classifies test cases.
%To this end, we have manually inspected the classified test cases with the help of IEE engineers to determine if the test cases were correctly classified as retestable, reusable and obsolete. 
To answer RQ1, we, together with \MREVISION{R3.17}{two IEE engineers}, inspected the classification results produced by the approach. 
%To evaluate the approach we compute the accuracy in terms of portion of test cases being correctly classified.
\MREVISION{R3.17}{We chose these engineers for their complementary roles and extensive experience. One was an embedded software engineer of the STO team  and the other was in charge of managing STO development.}
We evaluated the approach in terms of the average precision and recall we computed over the three different classes according to standard formulas~\cite{Sokolova:2009:SAP}. 
%We computed precision and recall for each of the three classes. 
In our context, a true positive is a test case correctly classified according to the expected class (e.g., a reusable test case classified as reusable). A false positive is a test case incorrectly classified as being part of a given class (e.g., a retestable test case classified as reusable). A false negative is a test case that belongs to a given class but has not been classified as such (e.g., a reusable test case not classified as reusable).

% !TEX root =  Main.tex
\begin{table}
\scriptsize
\caption{Test Case Classification Results for Single-Product Settings}
% \vspace*{-1.3em}
\begin{center}
\begin{tabular}{|p{1.7cm}|p{1.7cm}||p{1.05cm}|p{1.0cm}|p{1.3cm}||p{0.8cm}|p{0.6cm}|}

\hline

\textbf{Classified Test Suite}&\textbf{Product to be Tested}& \textbf{\# of Reusable} & \textbf{\# of Retestable}& \textbf{\# of Obsolete}& \textbf{Precision}&\textbf{Recal}l\\
\hline
\CH&	\SGM&		94	&	2	&	14	&1.0&1.0\\
\CH&	\DM&		105	&	2	&	3	&1.0&1.0\\
\CH&	\GAC&	   102	&	2	&	6	&1.0&1.0\\
\CH&	\VW&		84	&	22	&	4	&1.0&1.0\\
\SGM&	\DM&		85	&	0	&	1	&1.0&1.0\\
\SGM&	\GAC&		83	&	0	&	3	&1.0&1.0\\
\SGM&	\VW&	   67	&	16	&	3	&1.0&1.0\\
\DM&	\GAC&		91	&	0	&	5	&1.0&1.0\\
\DM&	\VW&		77	&	17	&	2	&1.0&1.0\\
\GAC&	\VW&	77	&	5	&	1	&1.0&1.0\\
\hline
\end{tabular}
\end{center}
\label{table:exp:RQ1single}
 %\vspace*{-0.2em}
\end{table}

% !TEX root =  Main.tex
\begin{table}[t]
\scriptsize
\caption{Test Case Classification Results for Whole-Line Settings}
% \vspace*{-1.3em}
\begin{center}
\begin{tabular}{|p{1.4cm}|p{1.40cm}||p{0.85cm}|p{0.95cm}|p{0.85cm}||p{0.83cm}|p{0.6cm}|}

\hline

\textbf{Classified Test Suites}&\textbf{Product to be Tested}& \textbf{Reusable} & \textbf{Retestable}& \textbf{Obsolete}& \textbf{Precision}&\textbf{Recall}\\

\hline
\CH&	\SGM&			94	&	2	&	14	&1.0&1.0\\
\CH, \SGM&	\DM&			107	&	0	&	2	&1.0&1.0\\
\CH, \SGM, \DM&	\GAC&	102	&	0	&	12	&1.0&1.0\\
\CH, \SGM, \DM, \GAC&\VW&	93	&	15	&	1	&1.0&1.0\\
\hline

\end{tabular}
\end{center}
\label{table:exp:RQ1whole}
% \vspace*{-1.2em}
\end{table}

Tables~\ref{table:exp:RQ1single} and~\ref{table:exp:RQ1whole} provide the results for the single-product and whole-line settings, respectively.
The first two columns %of Tables~\ref{table:exp:RQ1single}~and~\ref{table:exp:RQ1whole} 
report the ID of the product(s) whose test suite(s) have been considered for classification and the ID of the product being tested, respectively.
The next three columns %of Tables~\ref{table:exp:RQ1single}~and~\ref{table:exp:RQ1whole}   
provide the number of test cases belonging to the three classes.
The last two columns indicate precision and recall. 
%The number of retestable, reusable and obsolete test cases vary for different products, thus reflecting the fact that different products differ for their configuration.
 %and, as a consequence, some of the test cases belonging to the test suites of previous products become obsolete while other need to be re-executed to ensure that the product complies with its specifications. 
We observe that the approach has perfect precision and recall. This is the result of meticulous requirements modeling and \MREVISION{R4.2}{system testing practices} in place at IEE where functional requirements are documented by means of use cases together with proper traceability to test cases. These practices enable a precise identification of impacted scenarios and consequently the correct classification of test cases. It is typical practice for companies developing embedded, safety-critical systems, since requirements need to be traced and tested to comply with international safety standards (e.g., ISO 26262~\cite{ISO26262} and DO178C~\cite{DO178C}). \MREVISION{R4.2}{At IEE, functional requirements are elicited in the form of use case specifications, and system test cases are manually derived from the use case specifications. IEE engineers manually analyze use case specifications and write system test cases for use case scenarios to be tested. In our case study, each system test case exercises a use case scenario described in the use case specifications. Therefore, our approach is perfectly accurate, i.e., perfect precision and recall for RQ1. In general, our approach is expected to work well in industrial contexts where there is traceability, at the appropriate level of granularity, between requirements and system test cases.}

 \MREVISION{R3.19, R4.1, R4.12}{As we stated in Section~\ref{subsubsec:matching_scenarios_testcases}, there are few cases (i.e., multiple scenarios taking the same alternative flows with different orders and more than one scenario taking the same bounded or global alternative flow) where finer-grained traceability links are needed to retrieve test cases. However, in our case study, we did not encounter the two cases we mentioned above, which are expected to be rare.}

%We did not need to ask IEE engineers to provide additional traceability links to match scenarios with test cases since all the traceability links were at the right level of granularity required by our approach. 

\subsubsection{RQ2: Does the proposed approach accurately identify new scenarios that are relevant for testing a new product?}
\label{sec:rq2}

%RQ2 aims to determine whether new scenarios identified by our approach are relevant for testing a new product under development. To this end, we inspected the new scenarios identified by our approach in the case study. 
To answer RQ2, we checked if the new scenarios were exercised by the test cases in the manually implemented test suites of the new products. If so, we considered those new scenarios relevant. 
In addition, we, together with \MREVISION{R3.17}{IEE engineers}, checked whether the new scenarios that were not exercised were relevant for testing these new products. \MREVISION{R3.20}{Based their domain knowledge, engineers described irrelevant scenarios as scenarios having various errors which are unlikely to happen at the same time during system execution (e.g., having temperature and voltage errors at the same time). These scenarios were not considered worth testing.}

%Based on this information, 
We classified the new scenarios as true positive (i.e., a scenario identified by our approach and relevant for testing), false positive (i.e., a scenario identified by our approach but not relevant for testing), and false negative (i.e., a scenario tested by IEE but not identified by our approach). We computed precision and recall accordingly.

% !TEX root =  Main.tex
\begin{table}[h]
\scriptsize
\caption{Relevance of Scenarios Identified using Single-Product Settings}
% \vspace*{-1.3em}
\begin{center}
\begin{tabular}{|p{1.00cm}|p{0.80cm}||p{0.8cm}|p{0.80cm}|p{0.60cm}|p{1.05cm}|p{1.10cm}||p{0.95cm}|p{0.70cm}|}

\hline

\textbf{Classified}&\textbf{Product}& \multicolumn{4}{c|}{\textbf{New Scenarios Identified}} & \textbf{New Scenarios} & \textbf{Precision}&\textbf{Recall}\\
\textbf{Test Suite}&\textbf{to be Tested}& \textbf{Relevant (TP)} & \textbf{Tested by Engineers} & \textbf{Not Tested} & \textbf{Not Relevant (FP)} & \textbf{Not Identified (FN)}& &\\
\hline
\CH&	\SGM&		3	&	3	& 0&		0&0& 1.0&1.0\\
\CH&	\DM&	3	&	3	& 0&		0&0& 1.0&1.0\\
\CH&	\GAC&		2	&	1	& 1&	0&0& 1.0&1.0\\
\CH&	\VW&	27	&	23	&4&		0&0& 1.0&1.0\\
\SGM&	\DM&		1	&	1	&0&	0&0& 1.0&1.0\\
\SGM&	\GAC&	 1&	1	&0&		0&0& 1.0&1.0\\
\SGM&	\VW&	22	&	22	&0&		0&0& 1.0&1.0\\
\DM&	\GAC&	1	&	1	&0&		0&0& 1.0&1.0\\
\DM&	\VW&	26	&	23	&3&		0&0& 1.0&1.0\\
\GAC&	\VW&	10	&	10	&0&		0&0& 1.0&1.0\\

\hline
\end{tabular}
\end{center}
\label{table:exp:RQ2single}
 %\vspace*{-0.2em}
\end{table}

% !TEX root =  Main.tex
\begin{table}[h]
\scriptsize
\caption{Relevance of Scenarios Identified using Whole-Line Settings}
% \vspace*{-1.3em}
\begin{center}
\begin{tabular}{|p{1.4cm}|p{0.70cm}||p{0.80cm}|p{0.80cm}|p{0.70cm}|p{0.650cm}|p{1.00cm}||p{0.9cm}|p{0.75cm}|}

\hline

\textbf{Classified} &\textbf{Product} & \multicolumn{4}{c|}{\textbf{New Scenarios Identified}}&{\textbf{New Scenarios}}& \textbf{Precision}&\textbf{Recall}\\
\textbf{Test Suites}&\textbf{to be Tested}& \textbf{Relevant (TP)} & \textbf{Tested by Engineers}& \textbf{Not Tested}& \textbf{Not Relevant (FP)}& \textbf{Not Identified (FN)}&\textbf{}&\\
\hline
\CH&	\SGM&			3	&	3	& 0	&0&0 &1.0&1.0\\
\CH, \SGM&	\DM&	1	&	1	&0	&0&0 &1.0&1.0\\
\CH, \SGM, \DM&	\GAC&	0	&	0	&0	&0&0 &1.0&1.0\\
\CH, \SGM, \DM, \GAC&\VW&	14	&	9	&5	&0&0 &1.0&1.0\\
\hline

\end{tabular}
\end{center}
\label{table:exp:RQ2whole}
% \vspace*{-1.2em}
\end{table}

Tables~\ref{table:exp:RQ2single} and~\ref{table:exp:RQ2whole} report the results obtained using the single-product and whole-line settings, respectively. The third, fourth, and fifth columns %of Tables~\ref{table:exp:RQ2single} and~\ref{table:exp:RQ2whole} 
provide the number of relevant scenarios identified by our approach, and, among these, the number of scenarios tested and not tested by IEE engineers. The sixth column (\emphx{Not Relevant}) indicates the number of irrelevant scenarios. The columns named \emphx{New Scenarios Not Identified} provide the number of scenarios tested by IEE engineers but not identified by our approach. The last two columns report precision and recall. 
%Tables~\ref{table:exp:RQ2single} and ~\ref{table:exp:RQ2whole} show that 
All the new scenarios identified by our approach are relevant; %(i.e., column \emphx{Not Relevant} contains only zero) 
%and that 
they are covered by the test cases produced by IEE engineers. %(i.e., column \emphx{New Scenarios Not Identified} contains only zero). 
Consequently, the approach has perfect precision and recall.

In addition, we observe from Table~\ref{table:exp:RQ2whole} that %one benefit of the whole-line settings is that 
the availability of additional products in the whole-line settings enables the identification of additional new scenarios, and consequently more accurate testing. This is what happens for product P5, in which the whole-line settings lead to the identification of 14 new scenarios. Five of these new scenarios have not been tested by engineers in any of the existing products. More precisely, the test suites of P1 and P3 enable the identification of four and three scenarios not tested in the test suite of P5, respectively; only two of these scenarios are tested by both for a total of five new scenarios identified.
This difference between existing test suites is explained by the fact that certain test teams have defined more complete test suites (i.e., the test team for P1 and P3).
Since new scenarios are identified based on existing test cases (see Section~\ref{subsubsec:impact_identification}), for products with more complete test suites, the availability of more test cases may lead to the identification of additional new scenarios.

\subsubsection{RQ3: Does the proposed approach successfully prioritize test cases?}

To answer RQ3 in a realistic fashion, we applied our test case prioritization approach %presented in Section~\ref{sec:prioritization} 
to sort test cases in the test suites of four STO products (i.e., P2, P3, P4 and P5).
In total, we built four logistic regression models, one for each STO product.
To evaluate the quality of the predictions, we relied on historical data. 
%Using test execution history, we verified that higher failure probabilities were accurately predicted for test cases that have failed in the past.
To maximize the realism of results, we prioritized the test cases that belonged to the test suites originally developed by IEE engineers and ignored the new test scenarios we had identified (Section~\ref{sec:selection}). %instead of prioritizing the test cases of the test suite generated by using our test case classification technique. The difference between the two test suites is that our approach identifies new scenarios that have not been tested by IEE engineers. %(i.e., the test suite developed by IEE engineers consists of test cases classified as reusable and retestable by our approach). 
This did not introduce bias in the evaluation since test cases exercising new scenarios are always on top of the prioritized test suite and their execution is always necessary independently from their predicted likelihood to trigger failures.
In the following, we discuss our results including the identification of significant factors and the effectiveness of prioritizing failing test cases for each product. 
\MREVISION{R1.2}{Finally, we evaluate to what extent the availability of additional historical data positively affects test cases prioritization.}

% !TEX root =  Main.tex
\begin{table}
\scriptsize
\caption{Analysis of Significant Factors identified by Logistic Regression}
 %\vspace*{-1.1em}
\begin{center}

%\begin{tabular}{|p{1.4cm}|p{1.2cm}||p{1.6cm}|p{2.5cm}|p{1cm}|p{0.8cm}||p{2.70cm}|}
%\hline
%\textbf{Classified} &\textbf{Product to} & \textbf{Relevant} & \textbf{Odd Ratios} & \multicolumn{2}{c||}{\textbf{Fit of the model}} & \textbf{Prioritization AUC ratio}\\
%\textbf{test suite}&\textbf{be tested}& \textbf{factors}& & \textbf{Precision} & \textbf{Recall} & \textbf{(Observed/Optimal)}\\
%\hline	
%\CH&	\SGM&	V, S, FV&	0.35; 1.08; 2.09 & 0.91	& 0.9 & 0.97 (69.05/70.99) \\		
%\CH, \SGM&	\DM& V, S, FV& 0.35, 1.06; 1.85 & 0.79	& 0.95 & 0.98 (85.48/86.48)\\
%\CH, \SGM, \DM&	\GAC& V, S, FV, R &	0.78; 1.04; 1.71; 2.76 & 0.72 & 0.89 & 0.91 (80.31/87.48) \\
%\CH, \SGM, \DM, \GAC&\VW&	V, S, FP, FV, R & 0.36; 1.04; 0.92; 1.87; 1.95 & 0.30	&1 &0.92 (93/100.97)\\
%\hline

%\begin{tabular}{|p{1.4cm}|p{1.50cm}||p{1.6cm}|p{2.5cm}|p{2.1cm}||p{2.25cm}|}
%\hline
%\textbf{Classified} &\textbf{Product to} & \textbf{Significant} & \textbf{Odd Ratios} & \textbf{Fit of the } & \textbf{Prioritization }\\
%\textbf{test suite}&\textbf{be tested}& \textbf{factors}& & \textbf{model (Recall)} & \textbf{AUC ratio (Observed/Optimal)}\\
%\hline	
%\CH&	\SGM&	V, S, FV&	0.35; 1.08; 2.09 & 0.91					& 0.97 (69.05/70.99) \\		
%\CH, \SGM&	\DM& V, S, FV& 0.35, 1.06; 1.85 & 0.79					& 0.98 (85.48/86.48)\\
%\CH, \SGM, \DM&	\GAC& V, S, FV, R &	0.78; 1.04; 1.71; 2.76 & 0.72 		& 0.91 (80.31/87.48) \\
%\CH, \SGM, \DM, \GAC&\VW&	V, S, FP, FV, R & 0.36; 1.04; 0.92; 1.87; 1.95 	&1 &0.92 (93/100.97)\\
%\hline

%\end{tabular}

\begin{tabular}{|p{1.4cm}|p{1.2cm}||p{1.6cm}|p{2.68cm}|}
\hline
\textbf{Classified Test Suites} &\textbf{Product to be Tested} & \textbf{Significant Factors} & \textbf{Odds Ratio} \\
\hline	
\CH&	\SGM&	V; S; FV&	0.35; 1.08; 2.09  \\		
\CH, \SGM&	\DM& V; S; FV& 0.35; 1.06; 1.85  \\
\CH, \SGM, \DM&	\GAC& V; S; FV; R &	0.78; 1.04; 1.71; 0.36\\
\CH, \SGM, \DM, \GAC&\VW&	V; S; FP; FV; R & 0.36; 1.04; 0.92; 1.87; 1.95 	 \\
\hline

\end{tabular}

 \vspace*{0.7em}

Legend: V=Degree of Variability, S=Size, FP=Failing Products, FV=Failing Versions, R=Retestable.
\end{center}
\label{table:exp:significantFactors}
 %\vspace*{-1.7em}
\end{table}

\begin{table*}
\scriptsize
\caption{Test Case Prioritization Results}
% \vspace*{-1.4em}
\begin{center}
\begin{tabular}{|p{1.50cm}|p{0.80cm}||p{2.1cm}|p{1.60cm}|p{1.20cm}|p{1.7cm}|}
\hline 
\textbf{Classified Test} &\textbf{Product to be} &\textbf{AUC Ratio} & \multicolumn{2}{c|}{\textbf{\%Test Cases Executed to Cover}}  & \textbf{\%\CHANGED{Failing Test Cases Covered}}\\
\textbf{Suites}&\textbf{Tested} &\textbf{(Observed/Ideal)} & \textbf{\CHANGED{All the Failing Test Cases}} & \textbf{80\% of the \CHANGED{Failing Test Cases}} & \textbf{with 50\% of the Test Cases}\\
\hline	
\CH&	\SGM&	0.98 (65.46/66.48) &
72.09\% 
 & 38.37\% 
 & 97.43\% \\

\CH, \SGM&	\DM& 0.99 (82/82.48) 
& 41.66\% 
& 22.91\% 
& 100\% \\

\CH, \SGM, \DM& \GAC&  0.97 (71.02/72.97) & 
51.80\% & 
22.89\% & 
95\% \\

\CH, \SGM, \DM, \GAC&\VW&	0.95 (101.32/105.97) & 
26.54\% & 
18.58\% & 
100\% \\
\hline

\end{tabular}

\end{center}
\label{table:exp:prioritizationResults}
% \vspace*{-1.7em}

%\begin{tabular}{|p{1.3cm}|p{1.15cm}||p{1.5cm}|p{2.5cm}||p{2.05cm}|p{1.4cm}|p{1.4cm}|}
%\hline
%\textbf{Classified} &\textbf{Product to} & \textbf{Significant} & \textbf{Odd Ratios} & \textbf{Prioritization}&
% \textbf{\% TC}& 
% \textbf{\% TC}\\
%\textbf{test suites}&\textbf{be tested}& \textbf{factors}& &\textbf{AUC ratio}&
%\textbf{executed}&\textbf{executed}\\
%& &  & &\textbf{(Observed/Optimal)}&
%\textbf{to cover}&\textbf{to cover}\\
%& &  & &&
%\textbf{all failures}&\textbf{80\% of failures}\\
%\hline	
%\CH&	\SGM&	V; S; FV&	0.35; 1.08; 2.09 &  0.97 (69.05/70.99)& 96\% & 42\% \\		
%\CH, \SGM&	\DM& V; S; FV& 0.35; 1.06; 1.85 &  0.98 (85.48/86.48)& 57\% & 30\% \\
%\CH, \SGM, \DM&	\GAC& V; S; FV; R &	0.78; 1.04; 1.71; 2.76 & 0.91 (80.31/87.48)& 58\% & 30\% \\
%\CH, \SGM, \DM, \GAC&\VW&	V; S; FP; FV; R & 0.36; 1.04; 0.92; 1.87; 1.95 	&0.92 (93/100.97)& 33\% & 25\% \\
%\hline
%
%\end{tabular}
%
%
%Legend: V=Variation, S=Size, FP=Failing Products, FV=Failing Versions, R=Retestable.
%\end{center}
%\label{table:exp:prioritizationResults}
% \vspace*{-1.0em}
\end{table*}

%Fabrizio-08.04.20: Previous version
%Table~\ref{table:exp:significantFactors} provides detailed information about statistically significant factors identified according to logistic regression results (Wald test).
Table~\ref{table:exp:significantFactors} provides detailed information about the statistically significant factors identified by our approach based on logistic regression results \MREVISION{R4.13}{(see Section~\ref{sec:prioritization}).}

Column \emphx{Significant factors} lists the significant factors identified for each product. \MREVISION{R1.2}{In our analysis, we emulate past development by following the chronology of the products, in order to obtain realistic results that would have been obtained in practice.}  As expected, when more historical information is available, more factors tend to significantly correlate with observed failures.
% this depends on the fact that for certain factors, for the initial products of the product line, the data points are scattered and logistic regression cannot determine the regression coefficients. 
For example, we observe that the classification of a test case as retestable becomes significant after three products are included in the development history of the product line. This can be explained by the fact that updated configuration decisions impact a limited number of scenarios (i.e., the number of retestable test cases is usually low) and thus, this factor only becomes significant when enough examples of retestable test cases have occurred in previous products. %it is not possible to derive regression coefficients that lead to statistically significant predictions when historical data is available only for two products. 
As expected, the number of failing products also becomes significant after a sufficient number of products in the product line.

Column \emphx{Odds Ratio} presents the odds ratio of each significant factor. The odds ratio captures the effect size of the factor on the outcome of the regression model (i.e., the probability of observing a failing test case). A value above one indicates that the factor positively contributes to the outcome; otherwise it negatively contributes to the outcome. Note, however, that a statistical interaction with another factor may affect the odds ratio. The results show that the  number of failing versions is the factor that impacts most positively the probability of failure. It is highly likely that a test case that failed in the past will fail again, which is in line with previous research results. The odds ratio for the number of failing versions varies between 1.71 and 2.09. We observe that, as expected, the number of failing products statistically interacts with the number of failing versions. This has been determined by running logistic regression on each factor separately. Certain factors show a positive regression parameters when considered alone but become negative when interacting with other factors in the multivariate regression model. In this case, this is probably due to the two factors being correlated.

%Fabrizio-08.04.20: Previous version
%To evaluate the effectiveness of test case prioritization, we measured the percentage of test cases to be executed to trigger all the failures, and compared our approach with the ideal case that executes all the failing test cases first. Table~\ref{table:exp:prioritizationResults} summarizes our findings.

\begin{figure}
\subfigure[P2]{\includegraphics[width=5.50cm]{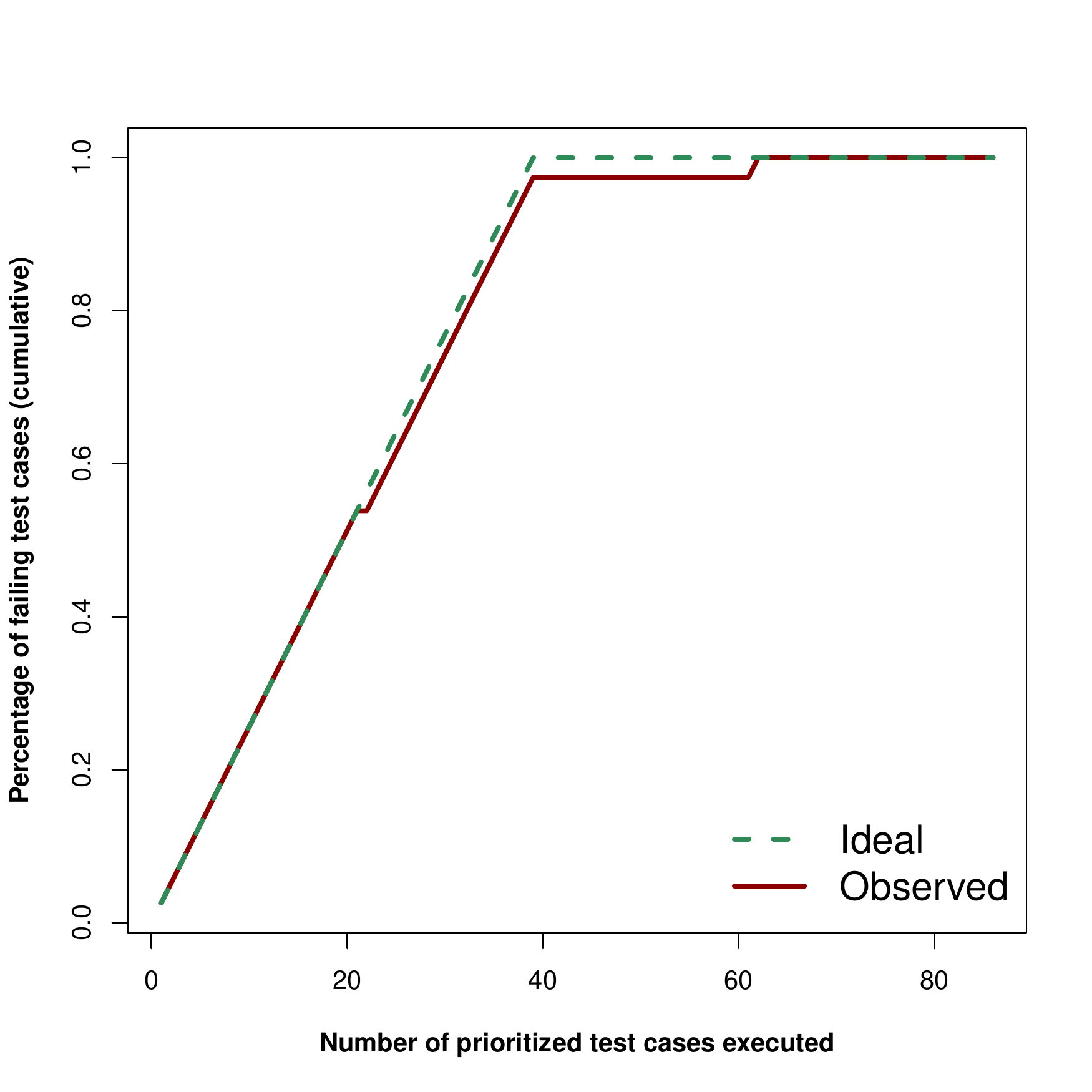}}
\hfill
\subfigure[P3]{\includegraphics[width=5.50cm]{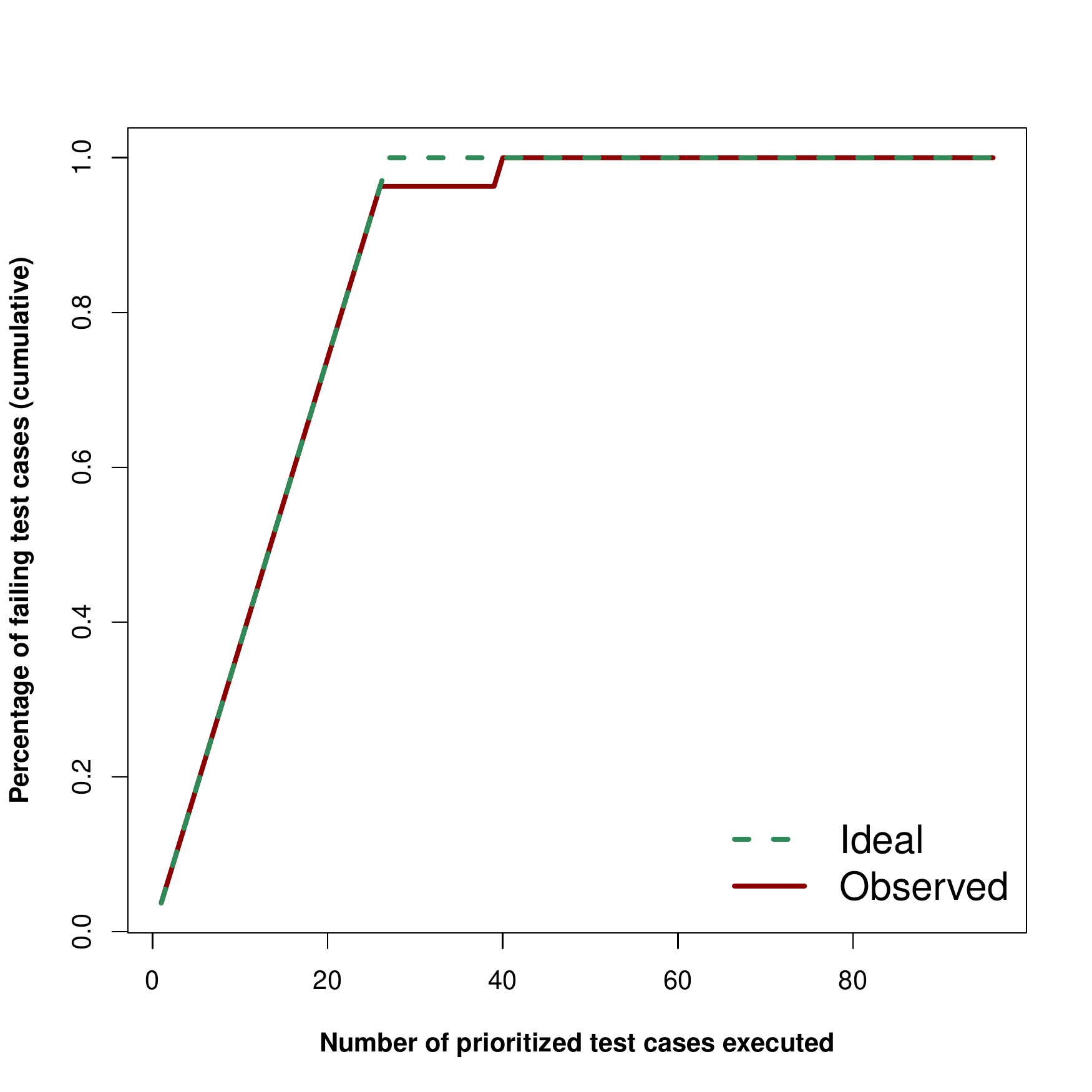}}
\subfigure[P4]{\includegraphics[width=5.50cm]{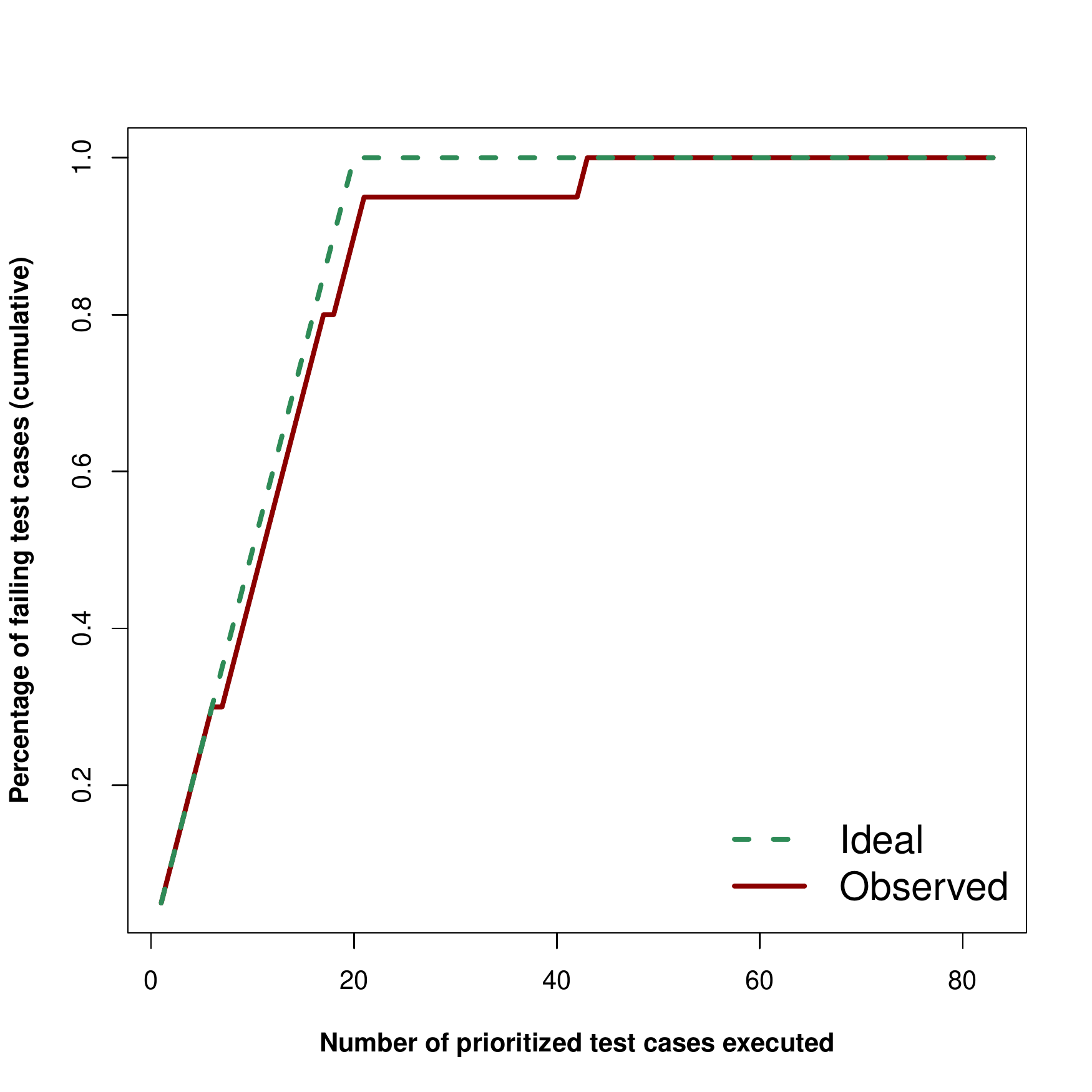}}
\hfill
\subfigure[P5]{\includegraphics[width=5.50cm]{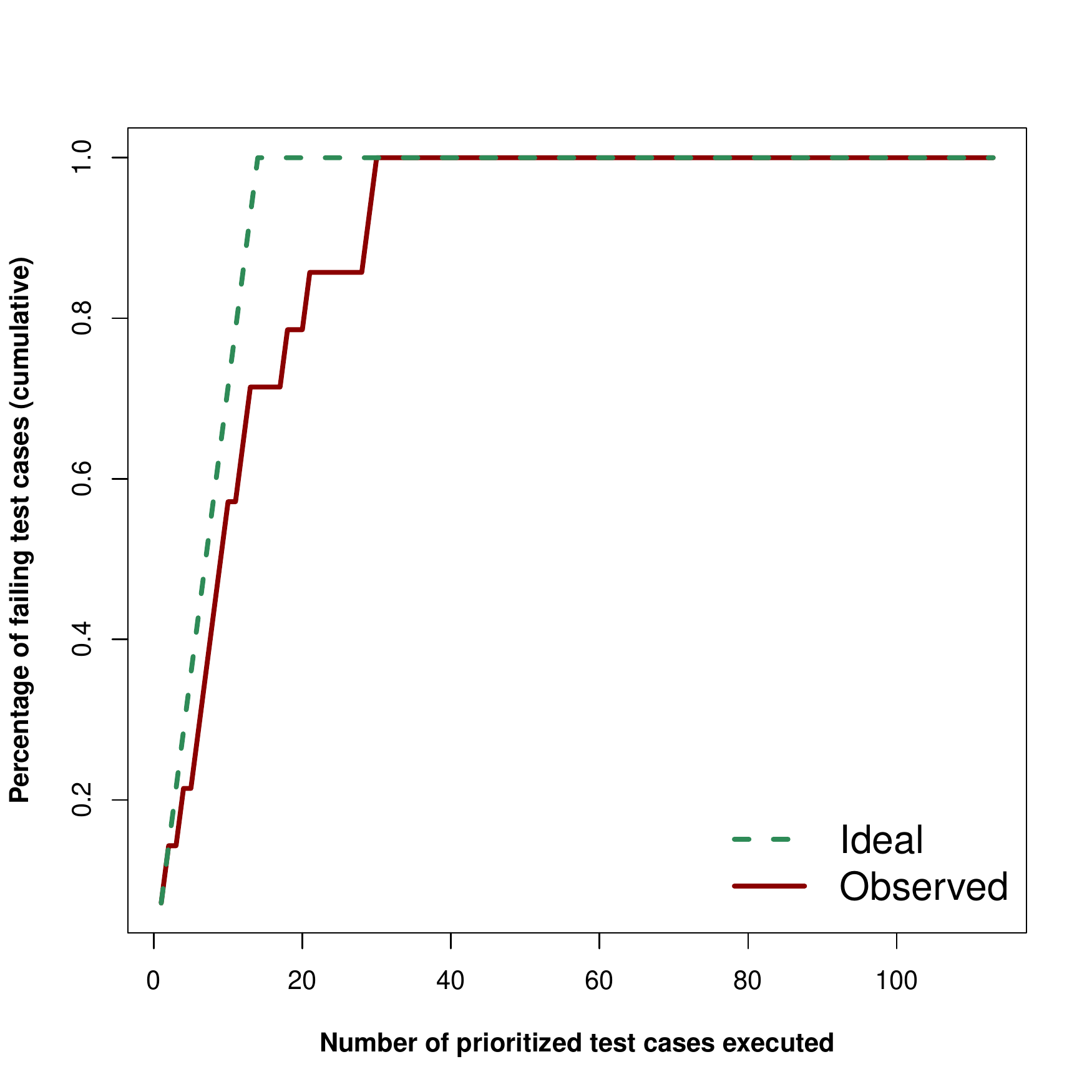}}
%\vspace*{-1.0em}
\caption{Prioritization results: percentage of failing test cases after running $x$ prioritized test cases.}
\label{fig:results:auc}
%\vspace*{-0.35em}
\end{figure}

\MREVISION{R4.13}{To evaluate the effectiveness of test case prioritization, we measured (1) the percentage of failing test cases in the first half of the prioritized test cases, which provides insights regarding the effectiveness of the approach with half of the test budget, (2) the percentage of test cases that must be executed to cover 80\% of the failing test cases, which determines the cost of running most of the failing test cases, and (3) the percentage of test cases that must be executed to exercise all failing test cases, to indicate the cost of achieving optimal fault detection. Finally, we compared our approach with the ideal case that executes all the failing test cases first.
Table~\ref{table:exp:prioritizationResults} and Fig.~\ref{fig:results:auc} summarize our findings.}

Table~\ref{table:exp:prioritizationResults} shows that, for all the products in our evaluation, our approach \CHANGED{covers} more than \CHANGED{80\% of the failing test cases} by executing less than 50\% of the test cases (see Columns \emphx{\%\CHANGED{Failing test cases covered} with 50\% of the test cases} and \emphx{\%Test cases executed to \CHANGED{cover 80\% of the failing test cases}}). 
We notice that the number of test cases required to cover all \CHANGED{the failing test cases} 
drops below 55\% when the test execution history of at least two products becomes available (see Column \emphx{\%TCs executed to \CHANGED{cover all the failing test cases}}).
%decreases when more historical information is available (see Column \emphx{\%TCs executed to identify all the failures}). 
In the case of P5, for example, the execution of 27\% of the test cases is sufficient to \CHANGED{cover all the failing test cases}. This is explained by the fact that newer products are more mature (i.e., they tend to fail less frequently) but is also due to logistic regression models improving over time. Indeed, for newer products, though there are fewer failing test cases, our approach remains accurate at giving higher priority to failing test cases. This capability is particularly relevant for industry since the early identification of failures enables early maintenance activities and, consequently, speeds up the product release.

To compare our approach with the ideal case, we computed the Area Under Curve (AUC) for the cumulative percentage of \CHANGED{failing test cases} in the executed test cases for both the ideal case and our prioritization approach, and computed the AUC ratio of the two. \MREVISION{R1.1}{AUC is similar to Average Percentage of Faults Detected (APFD)~\cite{do2006use}, a standard measure to assess regression test prioritization. The difference is that our y-axis is the percentage of failing test cases, not the percentage of detected faults. Since each test case in our methodology exercises a distinct use case scenario, in a safety context, engineers told us this is more relevant than the number of faults as the number of failing test cases captures more accurately the level of risk.} %detected per executed test case for both approaches, and computed the AUC ratio. 
Fig.~\ref{fig:results:auc} shows the two curves for all products after the initial one.
%cumulative percentage of failures detected by the number of test cases prioritized by our approach and by the ideal approach. 
%\CHANGED{As an additional metric, we also compute the AUC ratio considering the proposed and the ideal approach. In our context, the AUC ratio simplifies the comparison of different execution results in the presence of different test suite size and a different number of failing test cases. Intuitively the AUC ratio captures how good is the proposed approach compared to the optimal approach. 
As for APFD, the best result is achieved when the AUC ratio is equal to one (i.e., the AUC for the observed data matches the ideal AUC). The results show that the proposed approach achieves impressive results since the AUC ratio is always greater than or equal to 0.95.
% In the case of P5 the AUC ratio is lower than in the other three cases, this mainly depends on the presence of a much higher number of retestable test cases in the case of P5. Given that only a few retestable test cases have been observed in previous products, the regression coefficient derived for retestable test cases lead to imprecise results with the effect that, in our experiments, some of the non-failing retestable test cases receive higher priority than complex (i.e., with a big size) non retestable ones.}

\MREVISION{R1.2}{Finally, we checked if the logistic regression models improve over time due to the increase in available historical data. To this end, we compare 
the identification of significant factors and the effectiveness of prioritizing failing test cases when having historical data for a different number of past products. We focused on the prioritization of the test suite for P5, as it is the last product with the most historical data available. We therefore inspected and compared the P5 results
obtained with data from (1) P1, (2) P1+P2, (3) P1+P2+P3, and (4) P1+P2+P3+P4.
We also compared these results for P5 with the results obtained for earlier products (i.e., Tables~\ref{table:exp:significantFactors} and \ref{table:exp:prioritizationResults}).}

\CHANGED{Table~\ref{table:exp:significantFactorsHD} shows the significant factors identified. 
Similarly to Table~\ref{table:exp:significantFactors}, the number of significant factors increases with the number of available product versions. Notably, the number of failing products becomes significant after a sufficient number of products is present in the product line. Different from Table~\ref{table:exp:significantFactors}, in Table~\ref{table:exp:significantFactorsHD} the classification of a test case as retestable is always significant, which is due to the fact that, in P5, updated configuration decisions impact a higher number of scenarios than in the other cases. 
%For example, when testing P5 with the test suite of P1, we have 22 retestable test cases.
The numbers of retestable test cases for the different configurations in Table~\ref{table:exp:significantFactorsHD} are 22, 20, 20, and 15, respectively.
Concerning \emphx{Odds Ratios}, we see values that are close to the ones in Table~\ref{table:exp:significantFactors}. 
%Fabrizio: I do not explain why we say that it statistically interact because was already discussed before
Further, similar to Table~\ref{table:exp:significantFactors}, the number of failing products statistically interacts with the number of failing versions. However, different from Table~\ref{table:exp:significantFactors}, the classification of a test case as retestable has always an odds ratio above 1. This is due to a larger and different set of retestable test cases selected for P5 when compared to P4 (because of obsolete test cases, the set of test cases selected to build the regression model varies, see Section~\ref{sec:prioritization}); P4 is the other product in Table~\ref{table:exp:significantFactors} for which the classification of a test case as retestable is significant.}

\CHANGED{Concerning the test case prioritization results, Table~\ref{table:exp:prioritizationResultsHD} shows that an increasing number of available product versions leads to better results, i.e., a higher number of failing test cases detected for the same subset of the test suite. More precisely, the percentage of test cases executed to cover all the failing test cases decreases from 43.15\%, when only one product is available, to 26.54\%, when all the four products are available. A similar trend can also be observed for the percentage of test cases executed to cover 80\% of the failing test cases. All the failing test cases can be covered by executing half of the test suite.}

\begin{figure}
\center
\includegraphics[width=9.2cm]{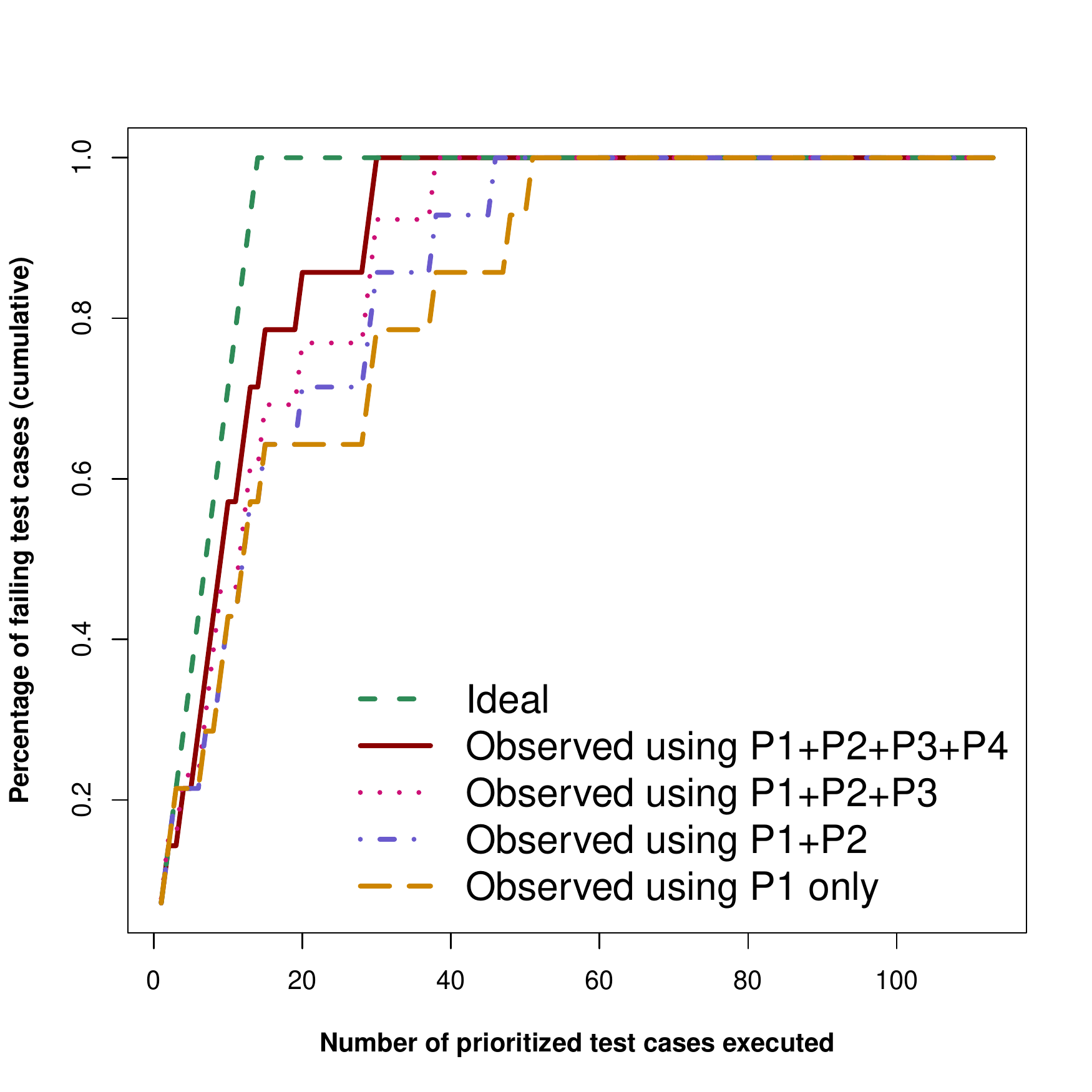}
%\vspace*{-1.0em}
\caption{Prioritization results for P5: percentage of failing test cases observed after running $x$ prioritized test cases, for different sets of product versions used to train the regression model.}
\label{fig:results:auc:histData}
%\vspace*{-0.35em}
\end{figure}

\CHANGED{Finally, in Fig.~\ref{fig:results:auc:histData}, we compared the results of the ideal case for P5 with the results achieved when relying on the different sets of product versions. Unsurprisingly, 
the curves obtained by relying on more products are closer to the ideal one, i.e., better results are achieved when an increasing number of products is available.} 

%detected per executed test case for both approaches, and computed the AUC ratio. 
%Fig.~\ref{fig:results:auc} shows the two curves for all products after the initial one.
%%cumulative percentage of failures detected by the number of test cases prioritized by our approach and by the ideal approach. 
%%\CHANGED{As an additional metric, we also compute the AUC ratio considering the proposed and the ideal approach. In our context, the AUC ratio simplifies the comparison of different execution results in the presence of different test suite size and a different number of failing test cases. Intuitively the AUC ratio captures how good is the proposed approach compared to the optimal approach. 
%As for APFD, the best result is achieved when the AUC ratio is equal to one (i.e., the AUC for the observed data matches the ideal AUC). The results show that the proposed approach achieves impressive results since the AUC ratio is always greater than or equal to 0.95.

% !TEX root =  Main.tex
\begin{table}
\scriptsize
\caption{Impact of Historical Data on Logistic Regression Results}
 %\vspace*{-1.1em}
\begin{center}

\begin{tabular}{|p{1.4cm}|p{1.2cm}||p{1.6cm}|p{2.68cm}|}
\hline
\textbf{Classified Test Suites} &\textbf{Product to be Tested} & \textbf{Significant Factors} & \textbf{Odds Ratio} \\
\hline	
\CH&	\VW&	V; S; FV; R&	0.47; 1.09; 1.97; 2.11  \\		
\CH, \SGM&	\VW& V; S; FV; R& 0.37; 1.06; 1.87; 1.78  \\
\CH, \SGM, \DM&	\VW& V; S; FP; FV; R &	0.38; 1.04; 0.83; 1.95; 2.08\\
\CH, \SGM, \DM, \GAC&\VW&	V; S; FP; FV; R & 0.36; 1.04; 0.92; 1.87; 1.95 	 \\
\hline

\end{tabular}

 \vspace*{0.7em}

Legend: V=Degree of Variability, S=Size, FP=Failing Products, FV=Failing Versions, R=Retestable.
\end{center}
\label{table:exp:significantFactorsHD}
 %\vspace*{-1.7em}
\end{table}

\begin{table*}
\scriptsize
\caption{Impact of Historical Data on Test Case Prioritization Results}
% \vspace*{-1.4em}
\begin{center}
\begin{tabular}{|p{1.50cm}|p{0.80cm}||p{2.1cm}|p{1.60cm}|p{1.20cm}|p{1.7cm}|}
\hline 
\textbf{Prioritized Test} &\textbf{Product to be} &\textbf{AUC Ratio} & \multicolumn{2}{c|}{\textbf{\%Test Cases Executed to Cover}}  & \textbf{\%\CHANGED{Failing Test Cases Covered}}\\
\textbf{Suites}&\textbf{Tested} &\textbf{(Observed/Ideal)} & \textbf{\CHANGED{All the Failing Test Cases}} & \textbf{80\% of the \CHANGED{Failing Test Cases}} & \textbf{with 50\% of the Test Cases}\\
\hline	
\CH&	\VW&	0.89 (94.32/105.97) &
45.13\%	&33.62\%	&100\% \\

\CH, \SGM&	\VW& 0.91 (96.68/105.97) &
40.70\%	&26.54\%	&100\%\\

\CH, \SGM, \DM& \VW&  0.92 (98.46/106.46) & 
33.62\%	&25.66\%	&100\%\\

\CH, \SGM, \DM, \GAC&\VW&	0.95 (101.32/105.97) & 
26.54\% & 
18.58\% & 
100\% \\
\hline

\end{tabular}

\end{center}
\label{table:exp:prioritizationResultsHD}
\end{table*}

\subsubsection{RQ4: Can the proposed approach significantly reduce testing costs compared to current industrial practice?}

\begin{table}
\scriptsize
\caption{Test Development Costs Savings}
% \vspace*{-1.3em}
\begin{center}

\begin{tabular}{|p{2.8cm}|p{3.0cm}|p{2.8cm}|}
\hline
\textbf{Product to be tested}  & \multicolumn{2}{c|}{\textbf{Test Cases To Be Implemented using the Proposed Approach}} \\
\textbf{} &\textbf{Single-Product Settings} & \textbf{Whole-Line Settings} \\
\hline

\SGM& 
3/99 (3\%)  & 
3/99 (3\%)  \\
   		
\DM& 
1/86 (1\%) & 
 1/108 (1\%) \\
 
\GAC&
1/92 (1\%)&
0/102 (0\%) \\

\VW&	
10/92 (11\%) & 
14/122 (11\%) \\
\hline

\end{tabular}

\end{center}
\label{table:exp:saving}
% \vspace*{-0.2em}
\end{table}

%\begin{table}
%\scriptsize
%\caption{Test Development Costs Savings}
% \vspace*{-1.3em}
%\begin{center}

%\begin{tabular}{|p{1.5cm}|p{2.7cm}|p{3.3cm}|p{3cm}|}
%\hline
%\textbf{Product to}  & \multicolumn{3}{c|}{\textbf{Test Cases To Be Implemented}} \\
%\textbf{be tested} &\textbf{Current Practice} & \textbf{Our approach in the Single-Product Settings} & \textbf{Our approach in the Whole-Line Settings} \\
%\hline
%
%\SGM& 86&
%3 & 
%3  \\
%   		
%\DM& 96 & 
%1 & 
% 1 \\
% 
%\GAC& 83 &
%1&
%0 \\
%
%\VW&	113 & 
%10 & 
%14 \\
%\hline
%
%\end{tabular}
%
%
%\end{center}
%\label{table:exp:saving}
% \vspace*{-0.2em}
%\end{table}

% !TEX root =  Main.tex
\begin{table*}
\scriptsize
\caption{Development Process Savings}
% \vspace*{-1.3em}
\begin{center}

\begin{tabular}{|p{1.2cm}|p{1.1cm}|p{1.9cm}|p{2.4cm}|p{2.4cm}|}
\hline
\textbf{Product}  &\textbf{Test Suite} &\textbf{Number of} & \multicolumn{2}{c|}{\textbf{Test cases to be Executed to \CHANGED{Cover all the Failing Test Cases}}} \\
\textbf{to be tested} &\textbf{Size}&\textbf{\CHANGED{Failing Test Cases}} & \textbf{Current Practice} & \textbf{Proposed Approach} \\
\hline

\SGM& 	86 &
39 &
	84 (97.67\%)		
	& 62 (72.09\%) \\		
	
\DM& 	96  &
27 &
 80 (83.33\%)	& 
 40 (41.66\%) \\
 
\GAC& 	83 &	
20 &
77 (92.77\%)		&
 43 (51.80\%) \\
 
\VW&	113  &	
14&
69 (61.06\%)    &
 30 (26.54\%)\\

\hline

\end{tabular}

\end{center}
\label{table:exp:savingsDevProc}
% \vspace*{-0.4em}
\end{table*}

%IN total we end up with more test cases, so more accurate test suite, but for now let's skip discussing it
%96+3
%107+1
%102+0
%108+14

%RQ6 aims to determine to what extent our approach can contribute to the reduction of system testing costs. %development costs.
Our test case classification and prioritization approach may reduce both (i) test case development costs (i.e., the number of test cases that need to be designed and implemented by engineers to test the software) and (ii) software development time (e.g., by \CHANGED{executing more failing test cases} at early stages of testing). 

As a surrogate metric to measure the savings, for each product of the STO product line, we report the number and percentage of test cases that can be reused when adopting the proposed approach (see Table~\ref{table:exp:saving}).
Columns \emphx{Single-Product Settings} and \emphx{Whole-Line Settings} 
report the results achieved by the approach when reusing only the test cases inherited from one previous product and from the test suites of all the previous products in the product line, respectively. 
As seen in these two columns, the effort required to implement test cases is very limited since,
with the proposed approach, engineers need to implement only the test cases required to cover new scenarios.
For instance, in the whole-line settings for product P4, engineers do not need to implement any test case at all. 
Instead, testing teams at IEE currently do not rely on approaches that support systematic reuse of test cases, a practice which often leads to re-implementing most of the test cases from scratch. 
Finally, one benefit provided by the whole-line settings is the identification of new scenarios, as discussed in Section~\ref{sec:rq2}; this is the case of product P5 where the whole-line configuration settings lead to the identification of four additional scenarios not identified with the single-product settings.

To evaluate the impact of our approach on software development time, we measured the percentage of test cases that need to be executed in order to \CHANGED{cover all the failing test cases in a product.}
%identify all the failures in a product. %Table~\ref{table:exp:savingsDevProc} shows the results. 
Column \emphx{Current Practice} in Table~\ref{table:exp:savingsDevProc} reports the percentage of test cases that need to be executed when considering the order followed by IEE engineers, which is based on domain knowledge. Column \emphx{Proposed Approach} reports the results we obtained. For all the products, our approach \CHANGED{covers all the failing test cases with less test cases than the current practice.}
%identifies all the failures with less test cases than the current practice. 
This is particularly true for product P5 where our approach requires the execution of less than half of the test cases prioritized by engineers. %test cases required by the current practice. %This is an important result for our industry partner. 
By using our approach, IEE can detect and fix failures earlier and thus speed up their software development.

\MREVISION{R4.14}{We evaluated the run-time performance of our approach with the whole-line settings. We executed the test case classification and prioritization five times for each new product. The execution times are shown in Tables \ref{table:classificationPerformance} and \ref{table:prioritizationPerformance}. Our tool was executed on an Intel quad-core i7 processor (2.40 GHz) with 6 MB Intel Smart Cache, and 8 GB of memory, running Windows 7.}

% !TEX root =  Main.tex
\begin{table}[t]
\scriptsize
\caption{Execution Times of Our Approach for Test Case Classification with Whole-line Settings (in seconds)}
% \vspace*{-1.3em}
\begin{center}
\begin{tabular}{|p{1.4cm}|p{1.40cm}||p{0.85cm}|p{0.90cm}|p{0.90cm}|p{0.90cm}|p{0.90cm}||p{0.90cm}|}

\hline

\textbf{Classified Test Suites}&\textbf{Product to be Tested}& \textbf{1st Run} & \textbf{2nd Run}& \textbf{3rd Run}& \textbf{4th Run}&\textbf{5th Run}&\textbf{Average}\\

\hline
\CH&	\SGM&			21	&	16	&	13	&12&13&15\\
\CH, \SGM&	\DM&			22	&	18	&	17	&17&18&18.4\\
\CH, \SGM, \DM&	\GAC&	27	&	16	&	17	&16&17&18.6\\
\CH, \SGM, \DM, \GAC&\VW&	29	&	16	&	17	&17&16&19\\
\hline

\end{tabular}
\end{center}
\label{table:classificationPerformance}
% \vspace*{-1.2em}
\end{table}

% !TEX root =  Main.tex
\begin{table}[t]
\scriptsize
\caption{Execution Times of Our Approach for Test Case Prioritization with Whole-line Settings (in seconds)}
% \vspace*{-1.3em}
\begin{center}
\begin{tabular}{|p{1.4cm}|p{1.40cm}||p{0.85cm}|p{0.90cm}|p{0.90cm}|p{0.90cm}|p{0.90cm}||p{0.90cm}|}

\hline

\textbf{Prioritized Test Suites}&\textbf{Product to be Tested}& \textbf{1st Run} & \textbf{2nd Run}& \textbf{3rd Run}& \textbf{4th Run}&\textbf{5th Run}&\textbf{Average}\\

\hline
\CH&	\SGM&			15	&	13	&	15	&13&12&13.6\\
\CH, \SGM&	\DM&			27	&	21	&	22	&21&21&22.4\\
\CH, \SGM, \DM&	\GAC&	38	&	30	&	33	&32&31&32.8\\
\CH, \SGM, \DM, \GAC&\VW&	49	&	40	&	41	&39&38&41.4\\
\hline

\end{tabular}
\end{center}
\label{table:prioritizationPerformance}
% \vspace*{-1.2em}
\end{table}

\MREVISION{R4.14}{According to the results, our approach requires less than 30 seconds to classify test cases in our case study, and less than 50 seconds to prioritize the same test cases. These results suggest that our selection and prioritization strategies are fast enough to be used in practical settings.}

\subsection{\MREVISION{R4.9, R4.15}{Reflections on Industrial Adoption}}

In addition to the research questions discussed above, we further reflect on the challenges for our approach to be widely transferable to industry. Based on our observations in the course of our efforts to get it adopted within IEE, we identified three challenges: \textit{modeling effort}, \textit{degree of automation}, and \textit{tool integration}.

\subsubsection{Modeling Effort}

In the current practice at IEE, like in many other environments, there is no systematic way to model variability information in use case specifications and diagrams. IEE engineers attach brief notes to use case specifications to indicate what may vary in the specification. They are reluctant to use feature models traced to
use case specifications because having feature models requires considerable additional modeling effort with manual assignment of traceability links at a very low level of granularity, e.g., sequences of use case steps. Therefore, in our approach, we employ the PL use case extensions presented in Section~\ref{subsec:elicitationVariability} that enable engineers to model variability directly in use cases without any feature modeling. In our discussions with IEE engineers, they stated that the effort required to apply the extensions for modeling variability was reasonable. They considered the extensions to be sufficiently simple to enable communication between engineers and customers. 

IEE engineers discuss variability with the customer to decide what to include in each product. In order to employ our test case classification and prioritisation approach in such an industrial setting, each customer should also be trained about the modeling method. IEE engineers mentioned that training customers about the modeling method may be more of a challenge since the company may need customers’ consent and effort in modeling variability the way we suggest.  

\subsubsection{Degree of Automation}

Based on our observations at IEE, we noticed that: (i) the current practice has no systematic way and automated tool support to decide which test cases from the previous products to execute and in which order for a new product; (ii) typically, multiple engineers from both the customer and supplier sides are involved in the decision-making process; (iii) engineers have to spend several days to manually review the entire set of system test cases from the previous products; and (iv) the intended updates of system test cases to cover new scenarios are manually identified and carried out by engineers. On the other hand, PUMConf supports 
test case classification and prioritisation activities in the context of product lines. The decision making process is automated in the sense that engineers are guided through the configuration decisions for classifying, prioritizing and modifying system test cases. Using PUMConf, for a new product, engineers select reusable and retestable system test cases to be run in the proposed order. 

At this current stage, our approach does not support the evolution of PL use case models. We still need to address and manage changes in variability aspects of PL use case specifications and diagrams, such as adding a new variation point in the PL use case diagram. Engineers need to be automatically guided to classify and prioritize system test cases for such changes. As future work, we plan to provide an automated regression test selection approach addressing changes in PL use cases. Our approach currently supports only one objective for test case prioritization, i.e., prioritizing test cases with higher failure likelihood. But multiple objectives may be required such as minimum execution time and maximum severity fault identification. Therefore, we also plan to extend our automated test case prioritization approach with multi-objective search.

\subsubsection{Tool Integration}

PUMConf is currently implemented as a plugin in IBM DOORS, in combination with commercial modeling tools used at IEE, i.e., IBM Rhapsody and Papyrus. PUMConf highly depends on the outputs of these tools. In another company, these tools might be replaced with other tools or the newer versions of the same tools. Future changes in the tool chain from one company to another will need to fulfill the following constraints: (i) a new requirements management tool for PL use cases should be extensible in such a way that we can implement the PL use case extensions, (ii) a new tool for establishing traceability links should be extensible in such a way that we can assign traceability links conforming to our traceability metamodel, and (iii) a new tool for system test cases should be extensible in such a way that our approach takes inputs from the other tools to classify and prioritize system test cases.

\subsection{Threats to Validity}

\emphx{Internal validity.} To limit threats to internal validity, we %considered an industrial case study. More precisely, %in our experiments 
considered the test cases developed by IEE engineers and the historical information collected over the years of system development. To avoid bias in the results, we considered the use case specifications written by IEE engineers and simply reformulated them according to the PUM methodology~\cite{Hajri2015, Hajri2016c}.

\MREVISION{R2.2}{We used all the PUM features in the PL use case diagram and specifications for STO, with the exception of the conflict relationship between use cases and the variant order group. These two features are not needed in STO, and they do not lead to worse results as long as the input PL use case models are correct and complete. Indeed, the effect that a conflict relationship has on the writing of use case specifications is that, when a conflicting use case is selected for a variation point, then another use case (i.e., the conflicting one) should be excluded from the use case specifications. This results in the removal of use case scenarios and the addition of new, untested use case scenarios. Concerning the classification of system test cases, based on the validation of our algorithm, conflict relationships should not introduce errors in our results, once again as long as the input models are correct and complete. Concerning test case prioritization, since the products considered in our evaluation differ with respect to previous products in terms of both removal and addition of scenarios, we do not expect deviations from our findings, even in the presence of conflict relations in PL models. Similar conclusions can be drawn for the case of variant order groups. All our use case models were confirmed by the IEE engineers to be correct and complete.} 

\emphx{External validity.} %The main threat to external validity regards the generalizability of the results. 
To mitigate the threat to generalizability, we considered a software product line that includes nontrivial use cases, with multiple customers and many sources of variability, in an application domain where product lines are the norm.  % in an application domain with various potential customers and numerous sources of variability. %is a good representative for the type of systems developed by our industry partner, IEE. 
The fact that STO has been installed on cars developed by major car manufacturers all over the world guarantees that the configuration decisions for STO cover a wide spectrum of possible configurations. 
%Fabrizio: 08.04.2020 previous versio
%and that the testing process put in place by IEE adheres to state-of-the-art quality standards. 

\MREVISION{R4.15}{In our experiments, we relied on test suites that exhaustively exercise the requirements for previous product versions.
The testing process put in place by IEE
aims to verify every use case scenario.
It therefore ensures that the software under test conforms with its use case specifications and thus adheres to expected quality standards, which is mandatory for safety-critical systems. 
For this reason, we believe the IEE test suites to be representative of what is typically found in other types of safety-critical systems.}
%Based on our knowledge of the automotive sector built over multiple collaborations with automotive companies, 
%Based on our experience built over the years with various automotive companies, we expect that the type of configuration decisions characterizing the STO product line and the type and number of test cases developed for STO are representative of other types of embedded automotive systems.

\MREVISION{R2.3, R4.15}{To achieve widespread applicability, we decided to rely on common requirements engineering practices (i.e., use case modeling and requirements traceability). Therefore, companies which already employ use cases for requirements elicitation only need to employ our extensions for product lines and restricted use case modeling. Based on our experience with various companies, we expect this transition to entail reasonable effort in safety-critical domains where these common practices are already in place to ensure compliance with standards. However, for organizations having less systematic and meticulous requirements and traceability  practices, we expect more effort to be required for adoption as this represents a fundamental change in practices and skills.}
%
% !TEX root =  Main.tex
\section{Conclusion}
\label{sec:conclusion}
This paper presents an automated test case classification and prioritization approach that supports use case-driven testing in product lines. For new products in a product line, it automatically classifies and prioritizes system test cases of previous product(s), and provides guidance in modifying existing system test cases to cover new use case scenarios that have not been tested in the product line before.

We improve the testing process in product lines by informing engineers about the impact of requirements changes on system test cases in a product family and by automatically and incrementally classifying and prioritizing system test cases. Such classification attempts to determine what test cases need to be rerun or modified, whereas prioritization helps ensure failing test cases are executed as soon as possible. 

Our test case classification and prioritization approach is built on top of our previous work (i.e., Product line Use case Modeling method and the Product line Use case Model Configurator), and supported by a tool integrated into IBM DOORS.
The key characteristics of our tool support are (1) the automated identification of the impact of requirements changes on existing system test cases, possibly leading to their selection or modification for a new product, (2) the automated identification of new use case scenarios in the new product that have not been tested in the product line, (3) the automated generation of guidance for modifying existing system test cases to cover those new scenarios, and (4) the automated prioritization of the selected system test cases for the new product.
We performed an industrial case study in the automotive domain, whose results suggest that our approach is practical and beneficial in industrial settings. More specifically, it provides an effective way to classify and prioritize system test cases in industrial product lines and to provide guidance for modifying existing system test cases for new products.

%This work is one of the last steps to achieve our long term objective~\cite{Hajri2016a}~\cite{Hajri2017b}, i.e., a change management approach in the context of use case-driven development and testing. Our plan is to support change impact analysis and regression test selection that help analysts properly manage changes in requirements and system test cases in a product family. 

%This work is one of the last steps to achieve our long term objective~\cite{Hajri2016a}~\cite{Hajri2017b}, i.e., the support for change impact analysis and regression test selection to help engineers manage changes in requirements and system test cases in a product family. 
%Our approach does not support the evolution of PL use cases. We still need to address and manage changes in variability aspects of PL use cases such as adding a new variation point in the PL use case diagram, and their impact on test cases in the context of test case selection and prioritization. 

\section*{Acknowledgments} We gratefully acknowledge funding from FNR and IEE S.A. Luxembourg, the grants FNR/P10/03 and FNR10045046, the Canada Research Chair program, and the European Research Council (ERC) under the European Union’s Horizon 2020 research and innovation programme (grant agreement No 694277).
%\sectopic{Acknowledgment.} Financial support was provided by IEE and FNR under grants FNR/P10/03.

%\bibliographystyle{IEEEtran}
%\bibliography{regressionTestSelection.bib}

\bibliographystyle{spmpsci}      % mathematics and physical sciences
\bibliography{Main.bib}   % name your BibTeX data base

\end{document}